%% file: main.tex
\documentclass[5p,times]{elsarticle}
\usepackage[utf8]{inputenc}
\usepackage{amsmath,amssymb,amsfonts}
\usepackage{filecontents}

\usepackage{graphicx}
\usepackage{textcomp}
\usepackage{xcolor}
\usepackage{stackengine}

\usepackage{wrapfig}

\usepackage{makecell}
\usepackage{float}
\usepackage{amsmath,amssymb}
\usepackage{comment}
\usepackage{setspace}
\usepackage{wrapfig}
\usepackage{vcell}
\usepackage{array}
\usepackage{bbm}
\usepackage{amsthm}
\usepackage{caption}
\usepackage{forest}

\usepackage{algorithm}
\usepackage{algpseudocode}
\usepackage{tabularx}
\DeclareMathOperator*{\argmax}{arg\,max}

\makeatletter
\newcommand{\multiline}[1]{%
  \begin{tabularx}{\dimexpr\linewidth-\ALG@thistlm}[t]{@{}X@{}}
    #1
  \end{tabularx}
}
\algnewcommand{\parState}[1]{\State%
  \parbox[t]{\dimexpr\linewidth-\algmargin}{\strut #1\strut}}
\makeatother

\algdef{SE}[VARIABLES]{Variables}{EndVariables}
   {\algorithmicvariables}
   {\algorithmicend\ \algorithmicvariables}
\algnewcommand{\algorithmicvariables}{\textbf{Global Variables}}

\usepackage{tikz}
\newcommand{\tikzmark}[2]{\tikz[remember picture,baseline=(#1.base)] \node (#1) {#2};}
\usetikzlibrary{shapes.geometric}
\usetikzlibrary{shapes}
\definecolor{royalblue}{cmyk}{1,0.50,0,0}
\definecolor{lavander}{cmyk}{0,0.48,0,0}
\definecolor{violet}{cmyk}{0.79,0.88,0,0}

\tikzstyle{nblue}=[thin, text=blue]
\tikzstyle{nred}=[thin, text=purple]

\tikzstyle{cyanbox}=[line width=0.5mm, draw=cyan, rounded corners=.2cm, text=cyan]
\tikzstyle{bluebox}=[line width=0.5mm, draw=blue, rounded corners=.2cm, text=blue]
\tikzstyle{greenbox}=[line width=0.5mm, draw=olive, rounded corners=.2cm, text=olive]
\tikzstyle{redbox}=[line width=0.5mm, draw=purple, rounded corners=.2cm, text=purple]

\tikzstyle{cblue}=[circle, draw, thin,fill=black!20, scale=0.8]
\tikzstyle{qgre}=[rectangle, draw, thin,fill=white!20, scale=0.8]
\tikzstyle{rpath}=[ultra thick, red, dashed, opacity=0.4]
\tikzstyle{legend_isps}=[rectangle, rounded corners, thin, 
                       fill=gray!20, text=blue, draw]
                        
\tikzstyle{legend_overlay}=[rectangle, rounded corners, thin,
                           top color= white,bottom color=green!25,
                           minimum width=2.5cm, minimum height=0.8cm,
                           pinegreen]
\tikzstyle{legend_phytop}=[rectangle, rounded corners, thin,
                          top color= white,bottom color=cyan!25,
                          minimum width=2.5cm, minimum height=0.8cm,
                          royalblue]
\tikzstyle{legend_general}=[rectangle, rounded corners, thin,
                          top color= white,bottom color=lavander!25,
                          minimum width=2.5cm, minimum height=0.8cm,
                          violet]
\usepackage{natbib}

\tikzset{
itria/.style={
  draw,dashed,shape border uses incircle,
  isosceles triangle,shape border rotate=90,yshift=-1.45cm},
rtria/.style={
  draw,dashed,shape border uses incircle,
  isosceles triangle,isosceles triangle apex angle=90,
  shape border rotate=-45,yshift=0.2cm,xshift=0.5cm},
ritria/.style={
  draw,dashed,shape border uses incircle,
  isosceles triangle,isosceles triangle apex angle=110,
  shape border rotate=-55,yshift=0.1cm},
letria/.style={
  draw,dashed,shape border uses incircle,
  isosceles triangle,isosceles triangle apex angle=110,
  shape border rotate=235,yshift=0.1cm}
}

\pgfdeclareshape{left triangle}{
    \nodeparts{}
    \anchor{center}{\pgfpoint{0cm}{0cm}}
    \behindbackgroundpath{
    \path [draw,dashed] (0,0) -- (0,-2) -- (-1,-2) -- cycle;
   }
}
\pgfdeclareshape{right triangle}{
    \nodeparts{}
    \anchor{center}{\pgfpoint{0cm}{0cm}}
    \behindbackgroundpath{
    \path [draw,dashed] (0,0) -- (0,-2) -- (1,-2) -- cycle;
   }
}
\newcommand{\isrevision}{1}
\usepackage{graphicx}
\graphicspath{ {./images/} }
\usepackage{amsthm}
\usepackage{amssymb} 
\usepackage{amsmath} 

\usepackage{xspace}
\usepackage{multicol}

\usepackage{lipsum}
\usepackage{mathtools}
\usepackage{cuted}

\usepackage{tabularx}
\usepackage{standalone}

\usepackage[normalem]{ulem} 

\ifnum\isrevision=1
 	\usepackage[colorinlistoftodos,textsize=scriptsize]{todonotes} 
    
    \newcommand{\stkout}[1]{\ifmmode\text{\sout{\ensuremath{#1}}}\else\sout{#1}\fi}
    \newcommand\rev[3]{\textcolor{red}{\begin{scriptsize}{#1}\end{scriptsize}\stkout{#2}}\textcolor{blue}{#3}}
\else
    \usepackage[colorinlistoftodos,textsize=scriptsize, disable]{todonotes}
    
	\newcommand\rev[3]{\ignorespaces#3\ignorespaces\unskip}
\fi

\usepackage{etoolbox}
\AtBeginEnvironment{tabular}{\scriptsize}

\date{November 2021}

\newtheorem*{nrpa_c}{Proposition 1}
\newtheorem*{nepa_c}{Proposition 2}
\newtheorem*{nrpa_mem}{Proposition 3}

\newcommand\xrowht[2][0]{\addstackgap[.5\dimexpr#2\relax]{\vphantom{#1}}}
\usepackage{geometry}

\begin{document}
\begin{frontmatter}

\author[1]{Maxime Elkael}
\author[2]{Massinissa Ait aba}
\author[1]{Andrea Araldo}
\author[1]{Hind Castel-Taleb}
\author[1]{Badii Jouaber}

\address[1]{Telecom SudParis, SAMOVAR, IP-Paris, \textit{\{Maxime.Elkael, Hind.Castel, Badii.Jouaber, Andrea.Araldo\}@telecom.sudparis.eu}}
\address[2]{Davidson consulting, \textit {Massinissa.ait-aba@davidson.fr}}

\title{Monkey Business: Reinforcement learning meets neighborhood search for Virtual Network Embedding}
\begin{abstract}
    In this article, we consider the Virtual Network Embedding (VNE) problem for 5G networks slicing. This  problem requires to allocate multiple Virtual Networks (VN) on a substrate virtualized physical network while maximizing among others, resource utilization, maximum number of placed VNs and network operator's benefit. We solve the online version of the problem where slices arrive over time. Inspired by the Nested Rollout Policy Adaptation (NRPA) algorithm, a variant of the well known Monte Carlo Tree Search (MCTS) that learns how to perform good simulations over time, we propose a new algorithm that we call Neighborhood Enhanced Policy Adaptation (NEPA). The key feature of our algorithm is to observe NRPA cannot exploit knowledge acquired in one branch of the state tree for another one which starts differently. NEPA learns by combining NRPA with Neighbordhood Search in a frugal manner which improves only promising solutions while keeping the running time low. We call this technique a monkey business because it comes down to jumping from one interesting branch to the other, similar to how monkeys jump from tree to tree instead of going down everytime. NEPA achieves better results in terms of acceptance ratio and revenue-to-cost ratio compared to other state-of-the-art algorithms, both on real and synthetic topologies. 
\end{abstract}
\end{frontmatter}


\section{Introduction}

The fifth-generation (5G) communications system is envisioned to serve a variety of novel services and industries, such as, autonomous vehicles, Virtual Reality (VR), Augmented Reality (AR) and remote healthcare, each requiring different Quality-of-Service (QoS). 
In this context, network slicing is a new way to manage telecommunication networks in a similar manner to what is done with cloud computing, relying on virtualization. The idea is that an operator owns a physical network (analogous to a {data center} for cloud computing) that can host multiple virtual networks (or slices). The operator can instantiate slices on the fly to a third party service provider. Each slice provides resources in an isolated, adaptable and dynamic manner. Service providers can have demands for slices with specific QoS/security constraints, topologies and resource requirements. Slices would be implemented using Network Function Virtualization (NFV) and software defined networking (SDN). The former allows to instantiate Network Functions (NFs - entities managing networking blocks such as authentication, gateways, etc). The latter enables packet routing to be modified by a centralized controller, enabling efficient and adaptable configuration of links ("virtual links") of the slice.
The interest of such an approach mainly lies in its flexibility:
one could envision supporting various use cases such as autonomous vehicles (requiring ultra low latency and ultra high reliability), virtual reality (requiring high throughput) or sensor networks (requiring an enormous amount of connections on a small area)\cite{AT18}. In the current "one size fits all" paradigm, supporting such use cases would imply building a new physical network for each one, which would hardly be economically viable.

In this context, an important question is how to place slices on such a network: clients give the operator slice (virtual network) requests in the form of interconnected NFs (e.g. a graph), and the operator tries to embed them onto the physical infrastructure (\textit{e.g.} to provide enough CPU for each virtual node and enough bandwidth for each virtual link between those nodes), by accepting as many slices as possible, so as to maximize the operator's gain.
This problem is known as Virtual Network Embedding\cite{OMAR} (VNE) problem which has been extensively studied in the recent years\cite{vineYard}\cite{nodeRank}\cite{survey}. The VNE being  NP-complete and inapproximable \cite{NP}, running an exact algorithm is not an option in most cases. Various methods have been studied for this problem, among which many are heuristic algorithms based on Linear programming\cite{vineYard}\cite{MASSI21}, ranking algorithms\cite{nodeRank} or reinforcement learning (RL) methods \cite{HA17}\cite{DL}. We are particularly interested in the latter ones, as these approaches enable to construct heuristics in an autonomous manner, based on experience and learning while solving the on-line version of the problem, where slices arrive and leave the system over time. Research on RL methods is still lacking, as current neural networks based methods either have hard constraints on the network topologies \cite{deepVine} or require very large amount of computing resources for training \cite{DL}. On the other hand, Monte Carlo based methods such as \cite{HA17} still have large rooms for improvement, as we show in this work. These shortcomings of other RL approaches are further developped in section 2. \\
In this paper, our main contribution is to provide a new state-of-the-art algorithm called Neighborhood Enhanced Policy Adaptation for this problem which combines reinforcement learning techniques with neighborhood search.
This article builds on our previous work from \cite{LKL21}, since it improves the proposed Nested Rollout Policy Adaptation (NRPA)\cite{NRPA} algorithm with a neighborhood search technique. To our knowledge this is the first time the Monte-Carlo Tree Search (MCTS) based NRPA algorithm is complemented with such neighborhood search technique for any problem, enabling it to beat several state-of-the-art algorithms. 
The MCTS approach (called Maven-S) from \cite{HA17} uses UCT (Upper Confidence bound for Trees)\cite{survMCTS}, which is adapted to stochastic problems. On the other hand, NRPA is specifically adapted to deterministic optimization problems. In our case, we do not know the slices in advance making the arrival process stochastic; however, once a slice arrives it is fully observable, and so is the physical network, making NRPA more adapted to tackle the placement of slices, given the current state of the network. 
NRPA learns through exploring the NF placement possibilities of the slice at random while learning weights for biasing future explorations, which enables it to focus on regions of the search space that have been the most rewarding so far while still maintaining a good level of exploration. These characteristics make it a very efficient algorithm for solving the VNE. However we show it can be further enhanced when combining it with neighborhood search. They key idea is that NRPA bases its search on the tree structure of the search space, which is good for quickly finding good solutions. However it can limit exploration of new, better branch once the algorithm has converged. Neighborhood search enables us to exploit knowledge of those good solutions for jumping to better branches of the search tree (similar to how monkeys jump from branch to branch) and continue the search from there, which enables better future exploration. 
We also propose a heuristic for initialization of the weights, and we show our numerical results, showing an improvement in the slice acceptance probability on real and synthetic networks compared to other methods, and therefore an increase in financial gains for an operator. 
Our contributions in this paper are then the following :
\begin{itemize}
    \item We combine NRPA with neighborhood search and our heuristic weight initialization, deriving the Neighborhood Enhanced Policy Adaptation (NEPA) algorithm for the virtual network embedding problem which outperforms state-of-the-art methods in both acceptance and revenue-to-cost ratio on all tested instances, including both synthetic and real topologies. Our approach is particularly effective on real topologies, on which it can even triple the number of accepted slices compared to some of the previous algorithms. We also investigate the topological features of those real topologies and explain how our algorithm can exploit them. Note that NRPA had never been used for the VNE problem.
    \item We publish a large set of testing scenarios for the community to experiment with, patching a lack of publicly available instances for quicker experimentation and comparisons (126 instances).
    \item We publish our implementations of several algorithms publicly (including NRPA, NEPA and algorithms from \cite{UEPSO}\cite{HA17}), since during this work, we found most algorithms lacked a well-documented  implementation. 
    \item To our knowledge, we are the first to explore the combination of NRPA with neighborhood search for any problem. We believe the idea can be exploited in other application where NRPA has been successful and where good neighborhood search algorithms are known such as the  Travelling Salesman Problem (TSP)\cite{TSP_NRPA} or the Vehicle Routing Problem (VRP)\cite{nrpaD}
    \item Finally, we assess wether the results of NEPA for the VNE can be improved by utilizing the reward function described in \cite{EvFunc} (see Appendix B).
\end{itemize}

The paper is organized as follows :
Section I introduced our work, section II presents our litterature review of the VNE, then section III presents our model. We describe NEPA in section IV, section V presents our numerical experiments,   {and in section VI  we summarize our work and we propose extensions and future perspectives}

\section{Litterature review}

Several methods have already been proposed for the VNE problem. 
\subsection{Mathematical programming} Notoriously, some work has been done for exact VNE using Mathematical programming. In \cite{OPT}, the authors propose an ILP formulation. This has the advantage to give guaranteed optimal solutions. However, since the VNE is NP-hard\cite{Rost-hard}, such an approach would not be able to cope with even medium slices with a reasonable execution time. Hence, a lot of work in the literature focus on heuristic algorithms. In \cite{vineYard}, two heuristics based on linear programming and rounding (either randomized or deterministic) are derived. These give good results in terms of acceptance and revenue-to-cost ratio, although most other approaches manage to beat them (\cite{HA17}\cite{UEPSO}\cite{DL}). These two rounding heuristics also sometimes suffer from relatively high runtimes, as \cite{DL} shows they run up to 13 times slower than the approach from \cite{HA17} for worst results, and that for some cases the approaches are even unable to run due to a lack of computational resources. In \cite{MASSI21}, an ILP heuristic is derived by reducing the number of candidate paths to a small amount, which enables the solver to find a solution quicker. However since it is ILP-based the algorithm is still non-polynomial. Our approach addresses these issues by proposing a solution which both runs quickly (sub-second runtime) and provides high quality (state-of-the art) embeddings.

\subsection{Graph neural networks} Some recent papers \cite{graphVine}\cite{neuroVine} process the problem with a deep neural network for performing the embedding (note that in this section we do not consider approaches using neural networks in conjunction with RL). In \cite{graphVine}, the graph is clustered with a graph neural network, which then helps guide the embedding procedure. On the other hand \cite{neuroVine} pre-processes the network in order to reduce the state-space, making the problem more manageable for other algorithms. Overall, \cite{neuroVine} addresses a slightly different problem than we do, since the paper is concerned with feeding a VNE algorithm (such as ours) with hints for solving the VNE, and both could be used in conjunction. On the other hand, \cite{graphVine} is concerned with the VNE, and although it has good results, the runtime is a significant problem as it is exponential in the number of nodes. The authors patch this issue with the use of a GPU. However, our experiments show that although the runtime is manageable it is still higher than all other algorithms we tested (in the order of ...).
\subsection{Heuristics and meta-heuristic techniques}
There is also a wealth of meta-heuristic algorithms for the VNE. This includes genetic algorithms \cite{GENETIC} and ant colony optimization \cite{VNE-AC}. However the most popular class of meta-heuristic approach for the VNE is particle swarm optimization (PSO), with several well performing algorithms such as \cite{UEPSO}, \cite{RWPSO} or \cite{EPSO}. These PSO approaches work by initializing "particles" as a swarm of random solutions which move in the space of candidate solutions. They find new solutions by sarcastically combining the best solutions found so far with current solutions.

Regarding heuristics, in \cite{GRC}, authors propose metric for evaluating a nodes' resource capacity/demand and then match highly demanding virtual nodes to highly available physical nodes. A similar idea is used in \cite{nodeRank} where it is combined with the Pagerank algorithm for ranking nodes.

These heuristic and meta-heuristic approaches show relatively good performances that we aim at beating in this article. Especially, to our knowledge, none of them exploits the fact that solutions can be improved by keeping virtual nodes close to one another. In that regard, our work could inspire enhanced versions of the cited algorithms. 

\subsection{Reinforcement learning approaches}
The family of approaches that interests us the most is reinforcement learning. First of all \cite{HA17} showed how to use the Monte Carlo Tree Search algorithm (MCTS) \cite{survMCTS} for the VNE problem. MCTS intelligently explores the space of possible placement solutions in order to find the best, but its exploration is based on multi-armed bandit theory, which assumes stochastic rewards. Instead, the outcome of a given embedding is deterministic and our method more efficaciously exploits determinism. 
Both can be considered \emph{online} methods, since they can immediately take decisions on any slice arrivals.

By contrast, \emph{offline methods} accumulate knowledge during an extensive learning (training) stage, which is then reused for a near-instantaneous high-quality embedding. Recently, DeepVine \cite{deepVine} used a deep neural network in order to learn embedding. This approach learns from graphs that are turned into images, enabling easy use of convolutional neural network (CNN) architectures. Although successful, this method makes strong assumptions about the input graphs: CNNs rely on the networks to be grid-shaped.
Another method is \cite{DL} where the neural neural network is fed directly with graphs. In this article, they use the A3C (Asynchronous Advantage Actor-Critic) algorithm for learning, which has been successful for other RL tasks. 
These approaches rely on function approximators (namely, neural networks) coupled with model-free RL techniques. This use of neural networks enables them to deal with big state-spaces, but comes at the cost of having no convergence guarantees to an optimal embedding or even an approximation. On the other hand, online methods like ours can be tweaked to guarantee that given enough time, they could converge to the optimal solution.
They are also able to handle similar state-spaces compared to neural-network based methods.

The huge computation needed to perform a very costly a-priori training (\textit{e.g.}, training for 72h on 24 for parallel instances of the problem~\cite{DL}) may make these offline methods~\cite{DL}\cite{deepVine} infeasible in practical situations. In particular when applying  embedding on different scenarios (or with different conditions or constraints), the huge offline learning phase must start from scratch. It is also an open question whether or not in a real world scenario we will have enough samples in order to enable such algorithms to learn. The advantage of online methods is instead their ability to immediately adapt and take decisions on new instances of the problem. 

For these reasons we improve upon the state-of-the-art online methods~\cite{HA17} \cite{LKL21}, providing convergence at regime toward the optimal embedding, sample efficiency and better empirical performance.
\section{VNE with a MDP approach}
\begin{table}
\center
\begin{small}
\begin{tabular}{ | l | l |}
   \hline
   Notation  &  Description\\
   \hline
   VNE & Virtual Network Embedding\\
   \hline
    VNR & Virtual Network Request/Slice \\
    \hline
   $G(V,E)$ & Physical network with nodes V and links E\\
   \hline    
   \makecell{$H^x(V^x,$\\$E^x,t^x_a,t^x_d)$}& $x^{th}$ slice with nodes $V^x$, links $E^x$, arrival and departure dates $t^x_a$ and $t^x_d$\\
   \hline
   $CPU_{v_i}$ & CPU capacity of physical node $v_i$\\
   \hline
   $BW_{v_i, v_j}$ & BW capacity of physical link $(v_i,v_j)$\\
   \hline
   $CPU^o_{v_i}$ & Occupied CPU of physical node $v_i$ \\
   \hline
   ${BW^o_{v_i,v_j}}$ & Occupied BW of physical link $(v_i,v_j)$\\
   \hline
   $CPU^d_{v_j^x}$ & CPU demanded by virtual node $v_j^x$\\ 
   \hline
   $BW^d_{v_i^x,v_j^x}$ & BW demanded by virtual link $(v_i^x,v_j^x)$\\
   \hline
   $\bar{BW}_{v_i, v_j}^x$ & Bandwidth used by slice $x$ on physical link $(v_i, v_j)$\\
   \hline
   $\bar{CPU}_{v_j}^x$ & CPU used by slice $x$ on physical node $v_j$\\
   \hline 
   MDP & Markov Decision Process\\
   \hline
   $\mathcal{A}$ & Set of possible actions in MDP\\
   \hline
   $s(k)$ & State of MDP at step $k$\\
   \hline
   $a_k$ & Action chosen in MDP at step $k$\\
   \hline
   NRPA & Nested rollout policy adaptation \\
   \hline    
   MCTS & Monte Carlo Tree Search \\
   \hline 
   $P$ & Policy function (associates a State-action couple with its weight\\
   \hline
   $\mathcal{P}$ & Link-mapping function (associates virtual links with physical paths)\\
   \hline
\end{tabular}
\end{small}\\
\caption{Notation and Acronyms}
\end{table}
The physical network belongs to an operator. At any point in time, the operator has a full knowledge of the state of the network, e.g. the amount of resources available, the slices it hosts and the resources they use. The operator receives slice requests from its clients over time. These requests are descriptions of a virtual network they would like to embed on the network, including resources required and topology. The goal of the operator is to place the incoming slices on its network in order to maximize a given objective (slice acceptance rate in our case).
\subsection{Graph theoretic notation}
The VNE problem can be formally described as a graph embedding problem: 
\begin{itemize}
    \item the physical network is represented as an undirected graph $G(V,E)$, where  $V$ is the set of $n$ physical nodes, $v_1, \ldots,\;v_i,$ $ \ldots,\; v_n$,
    that represent several
physical machines where virtual network functions can be hosted, and $E$ is the set of the physical edges between the nodes.  So we have: 
\begin{itemize}
    \item Each physical node $v_i$ is characterized by a  
    
    
{CPU} capacity,  $CPU_{v_i}$ and an occupied {CPU} quantity, ${CPU^o_{v_i}}$. One could extend this model with other resource types (RAM, HDD, ...) without loss of generality. 
\item On the other hand, each physical edge $(v_i, v_j) \in E$ is weighted by a maximum bandwidth amount, $BW_{v_i, v_j}$ and an occupied bandwidth amount ${BW^o_{v_i, v_j}}$. In case $BW_{v_i, v_j} = 0$, then we consider that there is no edge between $v_i$ and $v_j$.
\end{itemize}
\item We denote by $H^x(V^x, E^x, t_a^x, t_d^x)$ the undirected graph describing  the $x^{th}$ slice with the resources needed: each virtual node of the slice, $v_i^x \in V^x$ carries a CPU demand, $CPU^{d}_{v_i^x}$ and each virtual link $(v_i^x, v_j^x) \in E^x$ carries a bandwidth demand, $BW^{d}_{v_i^x, v_j^x}$. Since we are in a dynamical system, each slice also has a time of arrival $t_a^x$ and a time of departure $t_d^x$. Observe that as slices are placed or leaving, the physical occupied resources, ${CPU^o_{v_i}}$ and ${BW^o_{v_i, v_j}}$ change over time. The problem is to map each virtual node on a physical node and each virtual link on a physical path between the two host of its extremities, taking into account the available resources.
\end{itemize}
\subsection{Problem constraints}
If at a certain instant time instant the  $x^{th}$ slice request arrives, placement decisions must satisfy the following constraints:
\begin{itemize}
    \item Each placed virtual node should have enough CPU, e.g. if we choose $v_i$ hosts virtual node $v_j^x$ we should have $CPU_{v_j^x}^d\leq CPU_{v_i} - {CPU^o_{v_i}}$ (where $CPU_{v_i} - {CPU^o_{v_i}}$ represents available CPU on node $v_i$)
       \item For virtual link $(v_m^x, v_p^x)$ all physical links $(v_i, v_j)$ it uses should be chosen so $BW^d_{v_m^x, v_p^x} \leq BW_{v_i, v_j} - {BW^o_{v_i, v_j}}$ (where $BW_{v_i,v_j} - {BW^o_{v_i,v_j}}$ represents the available bandwidth between nodes $v_i$ and $v_j$) such that these links form a path between the physical nodes hosting $v_m^x$ and $v_p^x)$. 
    \item If two virtual nodes belong to the same slice, they can't be placed on the same physical node. This constraint is present in most previous works on the VNE\cite{HA17,vineYard,nodeRank}. It ensures reliability by preventing a significant portion of a slice from going off if a single physical node is down. To our knowledge, the optimal trade-off between sharing physical nodes (thus economizing bandwidth) and redundancy has not been well studied. Our approach, as most of the others cited, could easily be adapted to a relaxation of this constraint.
\end{itemize}

\begin{figure}
\centering
\begin{tikzpicture}[auto, thick]
  \foreach \place/\x in {{(-2.5,0.3)/1}, {(-1.75,-0.55)/2},{(-1.2,0.55)/3},
    {(-0.75,-0.7)/4}, {(-0.25,0)/5}, 
    {(1.5,0)/8}}
  \node[cblue] (a\x) at \place {};
  
  \node[cblue, label=above left:19] (a6) at (0.25,0.7) {};
  \node[cblue, label=below:7] (a7) at (0.75,-0.3) {};
  \node[cblue, label=below:8] (a9) at (2.5,0.4) {};
  
  \path[thin] (a1) edge (a3);
  \path[thin] (a2) edge (a3);
  \path[thin] (a3) edge (a6);
  \path[thin] (a2) edge (a4);
  \path[thin] (a5) edge (a6);
  \path[thin] (a5) edge (a4);
  \path[thin] (a5) edge (a2);
  \path[thin] (a5) edge (a7);
  \path[thick] (a6) edge[color=red] node [color=black, thin, draw=black, near start, yshift=2ex, below=12pt]  {12} (a7);
  \path[thick] (a6) edge[color=red] node [color=black, thin, draw=black, midway, yshift=2ex, below=-4pt]  {13}(a9);
  \path[thin] (a6) edge (a8) ;
  \path[thick] (a8) edge[color=red] node [color=black, thin, draw=black, midway, yshift=2ex, below=12pt]  {4} (a9);
  \path[thick] (a7) edge[color=red] node [thin, color=black, draw=black, midway, yshift=2ex, right=34pt, below=10pt]  {7} (a8);
 
    \node[qgre, label=10] (b6) at (0.25,2.7) {};
    \node[qgre, label=above right:6] (b7) at (0.75,1.7) {};
    \node[qgre, label=8] (b9) at (2.5,2.4) {};
 
  \path[thin] (b9) edge node [draw=black, midway, yshift=2ex, below=10pt]  {4}  (b7);
  \path[thin] (b7) edge node [draw=black, midway, yshift=2ex, below=8pt]  {7} (b6);
  \path[thin] (b9) edge node [draw=black, midway, yshift=2ex, right]  {11} (b6);
 
  \foreach \i in {9, 7, 6}
    \path[rpath] (a\i) edge (b\i);
 
\end{tikzpicture}   
    \caption{Slice (white nodes) embedded on physical network (gray nodes). Link demands and remaining capacities are boxed, used physical links are in red. CPU demands and capacities are non-boxed.}
    \label{fig:placement}
\end{figure}
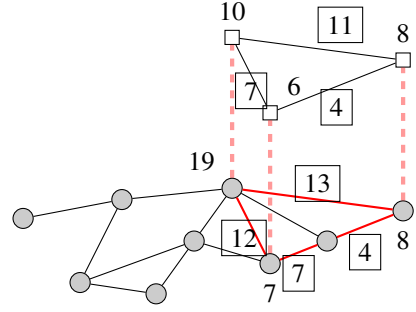%

\begin{figure*}
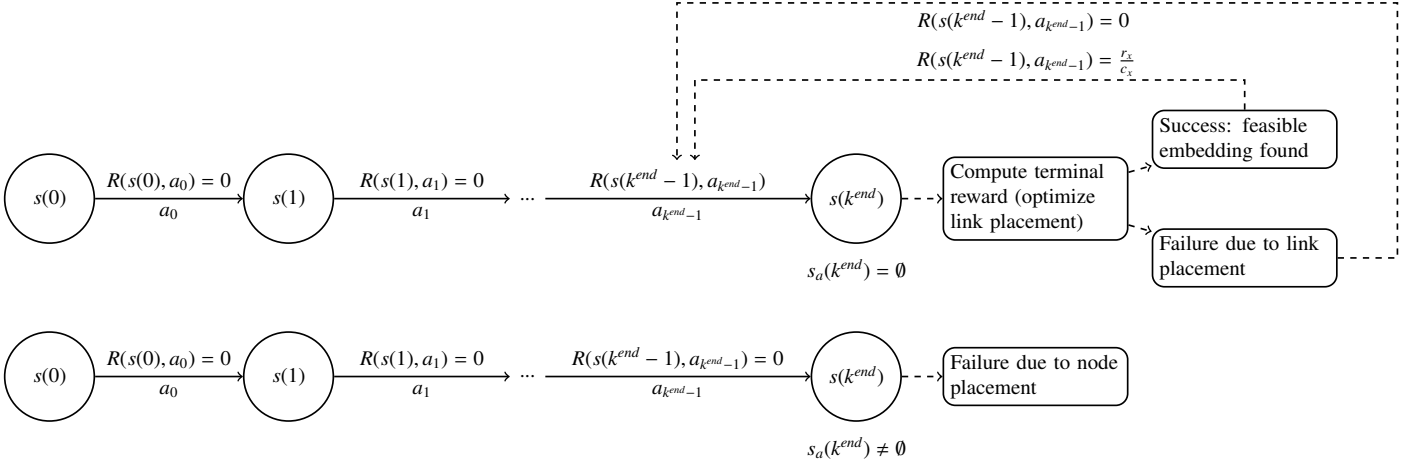

  \includestandalone[width=1\textwidth]{fig_mdp}
    \caption{Example sequences of actions in the MDP. Dashedy arrows are transitions not occurring in the MDP (no action choice).}
    \label{fig:diag}
\end{figure*}

\subsection{Online VNE description}
An example of slice is shown in Figure \ref{fig:placement}.
We solve the VNE online :
\begin{itemize}
    \item when a slice $x$ arrives at time $t^x_a$, we directly try to embed it.
If a feasible solution is found, the slice is placed on the physical network, consuming the corresponding CPU and bandwidth resources, \textit{i.e.} updating the corresponding $CPU^o_{v_i}$ and $BW^o_{v_i, v_j}$. If no solution is found, the slice leaves the system and is dropped.
\item 
When time $t^x_d$ is reached, the slice leaves the physical network and resources are freed. 
\end{itemize}
The full system time is continuous and gives us the arrival and departure dates for slices ($t^x_a$ and $t^x_d$ refer to this scale) we assume  the VNE  is instantaneous : 
in the same instant the slice request arrives, the corresponding MDP (described in the next subsection) is solved, instantaneously, and the slice is either placed or discarded.
For each virtual node to be placed, we select a physical node via RL (Reinforcement learning). In order to learn an optimized sequence of decisions for virtual resource placement  via RL, one needs to frame the VNE problem as a Markov Decision Process (MDP)\cite{Sutton}.

\subsection{MDP description}



A MDP is a system made up of two elements: the agent
{ (the network operator in our case)} and the environment
{ (the description of the slice to place and of the state of the physical network in our case \textit{i.e.} the amount of resources available and occupied)}. 

{Observe that our MDP works as a sequence of steps, each step corresponding to the decision of placing a virtual node onto a physical node. Note that these steps do not have any time-dimension, they can be considered to be all taken instantaneously. Also observe that our MDP is fully deterministic: all transitions and all rewards (which we will define later) are deterministic and computable in advance.}

In our particular setting, we consider the optimization problem where we have to place a single slice at a time. This means that as soon as one slice requests arrives, an MDP  is initialized in order to decide the embedding of each virtual node and link it demands. 

~\\
We assume that the agent only decides where to place each virtual node. After all virtual nodes
{ of a certain slice} have been placed, we calculate 
{Link placement} with a shortest path heuristic (see algorithm \ref{algo:BFS}). Therefore, we adopt MDP only for virtual node placement. 
~\\

\subsubsection{Elements of the MDP }

Let $s(k)=(s_a(k), s_b(k))$ be a state of the MDP, it is composed of two components :
\begin{itemize}
    \item $s_a(k)$ is the set of virtual nodes yet to be embedded at
    {step} $k$.  
    \item  $s_b(k)$ represents, at step $k$, the occupation of the physical nodes by the virtual nodes. It is a vector with $|V|$ elements where $s_b(k)[i] = j$ if virtual node $v_j^x$ from slice $x$ is hosted on physical node $v_i$. If $v_i$ hosts no node from the current slice, $s_b(k)[i] = 0$ (we assume indexes of virtual nodes are strictly positive integers).
\end{itemize}
{For the incoming slice $x$, we consider the virtual nodes $v^x_j\in V^x$ one by one
\footnote{{The order in which we iterate through virtual nodes can be chosen arbitrarily}} and we take an action $i$ which corresponds to placing it on a physical node $v_i$.} Therefore, the set of 
{possible} actions $\mathcal{A}=\{1,\ldots,n\}$ corresponds to the physical nodes of $V$. Choosing action $i$ would mean placing the current 
{virtual} node on 
$v_i$. 
We also consider $\mathcal{A}(s(k))$
{$\subseteq\mathcal{A}$}  the set of legal actions from 
state $s(k)$, which will be specified later.
\subsection{System's evolution}

The main steps of the system evolution are described as follows:
\begin{enumerate}
    \item At 
    {step} $0$, $s(0)=(V^x, u)$, where $u$ is a vector of $|V|$ components all equal to 0. 
    \item At 
    {step} $k\geq 0$, from the state $s(k)$
    {, let} $v^x_l$
    { be} the first virtual node of $s_a(k)$. Then 
    $\mathcal{A}(s(k))$ is the set of actions $j \in \mathcal{A}$ such that $CPU^d_{v_l^x}${$\le$} $CPU_{v_j} - {CPU^o_{v_j}}$ and $s_b(k)[j] = 0$. 
    \\
    Assume the chosen action from $\mathcal{A}(s(k))$ is $a_k=i$.
    Then the virtual node $v^x_l$ is embedded on physical node $v_{i}$ and we have a  transition to the state $s(k+1)=(s_a(k)-\{v^x_l\}, s_b(k) + b_{i})$ where $b_{i}$ is a  vector with the $i^{th}$ component equal to index l of virtual node $v_l^x$ and all other components equal to 0.
\end{enumerate}
The embedding process continues at each 
{step} until we reach 
the final state 
{at a certain step $k^\text{end}$}, where $\mathcal{A}(s(k^{end})) = \emptyset$. At this point, two situations can occur:
\begin{itemize}
    \item Either the node embedding is a success, so the set of virtual nodes is 
    {$s_a(k^\text{end})=\emptyset$}. The second part of the state holds a vector
    { $s_b(k^\text{end})$} indicating which physical nodes are used by each virtual node of the slice. 
    So the final state is $(\emptyset,u')$, where $u'[i]=l$ if virtual node $v_l^x$ is hosted by physical node $v_i$ .
    \item Or the embedding fails, which means that for a virtual node, there is no suitable physical node to host it e.g. $s_a(k^{end}) \neq \emptyset$. In this case, the entire slice is rejected. 
\end{itemize}
If node embedding is successful, the Link embedding is calculated using algorithm \ref{algo:BFS} which is a shortest path heuristic. Then, if link embedding is successful too, we need to update the physical network to acknowledge for the used resources, \textit{e.g.} update $CPU^o_{v_i}$ and $BW^o_{v_i, v_j}$ for all physical nodes $v_i$ and physical links $v_i, v_j$ used by the slice. On the other hand, if one of the two phases fails, the slice is discarded.
~\\

\subsubsection{Reward Function}
We now define the reward obtained by the agent over the course of its actions.
Let us first  define the revenue of the operator $r^x$ (representing the revenue gained thanks to a client paying for slice $x$)  and the cost $c^x$ (the cost induced by operating the physical resources allocated to host the slice)  
{ for a successfully placed slice} $x$ as: 
\begin{align} 
r^x 
&= \sum\limits_{\forall v_i^x, v_j^x \in V^x} BW_{v_i^x, v_j^x}^{d} + \sum\limits_{\forall v_m^x \in V^x} CPU_{v_m^x}^d
\\
c^x
&= \sum\limits_{\forall (v_i, v_j) \in {E}} {\bar{BW}}_{v_i, v_j}^x + \sum\limits_{\forall v_i \in V} \bar{{CPU}}_{v_i}^{x},
\end{align}
where for slice $x$, ${\bar{BW}}_{v_i, v_j}^x$ is the bandwidth used  on physical link $(v_i, v_j)$ and $\bar{{CPU}}_{v_i}^{x}$ the CPU used  on physical node $v_i$. In other words, service providers pay proportionally to the resource demands by their slices. The cost of operation of a slice is proportional to the physical resources consumed.
We define the immediate reward function of our MDP as:
\begin{equation}
R(s(k), a_k) = \left\{
    \begin{array}{ll}
        \frac{r^x}{c^x} \mbox{~~if } s_a(k+1) = \emptyset \mbox{\vbox{\noindent~and node and link}}\\ \mbox{~~~~~mapping are successful}\\
        0 \mbox{~~~ otherwise}
    \end{array}
\right.
\end{equation}
Examples of sequences of actions in the MDP are shown in figure \ref{fig:diag}, in which the circular states correspond to the states of the MDP, where the decisions are taken by the agent. 
The sequence at the top diagram corresponds to a successful embedding (after node and link placement), while the bottom one returns a  failure. Note that since rewards happens during transitions, the last reward is $R(s(k^{end}-1), a_{k^{end}-1})$ as it happens during the last transition, from $s(k^{end} - 1)$ to $s(k^{end})$.
\subsubsection{Objective function}
From the initial state $s(0)$, we consider  a sequence of $k^{end}$ actions :  $seq = a_0$, $a_1$, $\ldots$,$a_{k^{end}-1}$. 
We define the total reward from the state $s(0)$ for $seq$ as follows :
\begin{equation}
R^{seq}(s(0))= \sum\limits_{k=0}^{k^{end}-1} R(s(k), a_k)
\end{equation}

Then the  objective function is:
\begin{equation}
\max\limits_{seq}  R^{seq}(s(0))
\end{equation}
And the agent seeks to find the  best 
sequence of actions: 
\begin{equation}
    seq^* = \argmax\limits_{seq} R^{seq}(s(0))
\end{equation}
and the corresponding reward:
\begin{equation}
R^*(s(0))=R^{seq^*}(s(0))
\end{equation}

Notice that in practice, all rewards except the last one are equal to 0 due to equation (3). 
\sloppy
{With this definition of reward, the agent, i.e., the network operator always tries to choose valid embeddings,} since any valid embedding has a non-zero revenue-to-cost ratio. It also favors embeddings that use the least possible amount of resources, since 
{the reward increases} as $\sum\limits_{\forall(v_i, v_j) \in {E}} \bar{BW}_{v_i, v_j}^x$ decreases. An intuitive way to frame this is that the reward encourages the choice of embeddings that lead to placing virtual links on short physical paths, effectively trying to place the slice on a cluster of physical nodes. We do this {based} on the idea that if a slice uses the least possible amount of resources, then it will leave more resources available for future slices, thus enabling us to improve the acceptance ratio on the full scenario.
Note that, at best, each virtual link is mapped on a physical link of length 1. Note also that for a successfully embedded slice, $\sum\limits_{v_i \in V} \bar{CPU^x_{v_{i}}} = \sum\limits_{v_m^x \in V^x} CPU^d_{v^{x}_{m}}$, hence the best achievable reward is 1 and, the closer the reward is to 0, the worst the embedding is in terms of resource usage (with 0 being the worst reward, reserved for failed embeddings). 
{Therefore, this reward function quantifies} the quality of an embedding regardless of the size of the slice.
{This has clear advantages over} the reward function used in \cite{HA17} which is $r'_x - c_x$ with $r'_x = \alpha \sum\limits_{\forall v_i^x, v_j^x \in V^x} BW_{v_i^x, v_j^x}^{d} + \beta \sum\limits_{\forall v_m^x \in V^x} CPU_{v_m^x}^d$, where $\alpha, \beta$ are weight parameters which have to be tweaked. In \cite{HA17} they use parameters of $
1$ which provides an upper bound of 0 and no lower bound, making it harder to compare the quality of embeddings for different slice sizes. In the general case they do not provide any bound. This is particularly unfortunate for the MCTS algorithm they use, as it is based on the upper confidence bounds algorithm UCB-1, which provides its theoretical guarantees only for a reward bounded between 0 and 1.
\subsection{Characteristics of the MDP and implications on resolution}
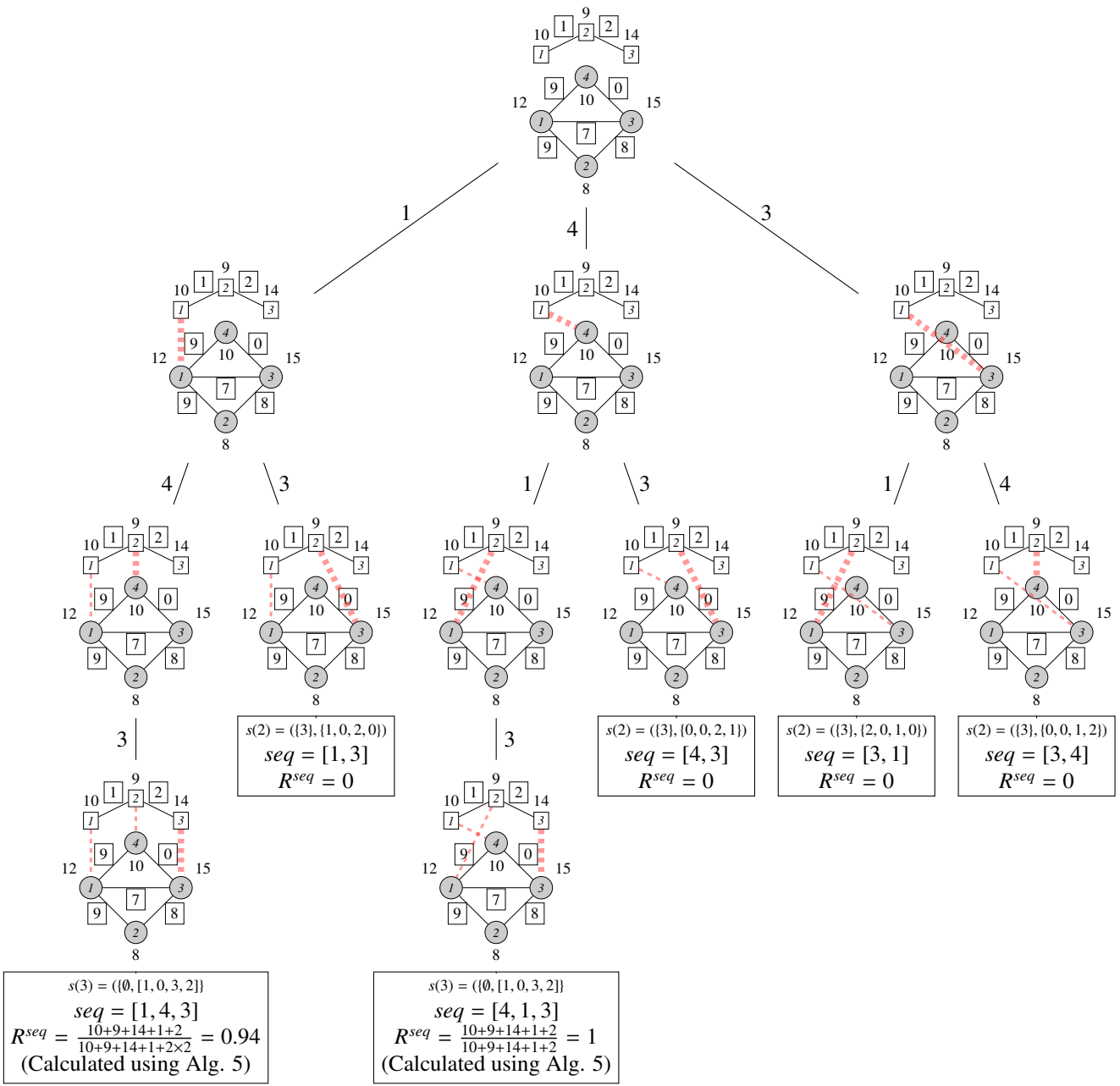
\begin{figure*}
\begin{tikzpicture}[level distance=4cm,
  level 1/.style={sibling distance=5.6cm},
  level 2/.style={sibling distance=2.8cm},
  level 4/.style={distance=1cm}]
  \node {\includestandalone[scale=0.7]{root_tikz}}
    child {node {\includestandalone[scale=0.7]{l1_1}}
      child {node {\includestandalone[scale=0.7]{l2_1}}
        child {node {\includestandalone[scale=0.7]{l3_1}}
          child {node[above=0.6cm, align=center, rectangle, draw] {
          \scriptsize
          $s(3) = (\{\emptyset,[1,0,3,2]\}$\\
          $seq=[1,4,3]$ \\$R^{seq}=\frac{10+9+14+1+2}{10+9+14+1+2 \times 2} = 0.94$\\(Calculated using Alg. 5)}}
          edge from parent  node[left] {$3$}
        }
        edge from parent  node[left] {$4$}
      }
      child {node {\includestandalone[scale=0.7]{l2_2}}
        child {node[above=1.1cm, align=center, rectangle, draw] {\scriptsize
        $s(2) = (\{3\}, \{1,0,2,0\})$\\
        $seq=[1,3]$\\
        $R^{seq}=0$}}
        edge from parent  node[right] {$3$}
      }
      edge from parent  node[left,above] {$1$}
    }
    child {node {\includestandalone[scale=0.7]{l1_2}}
    child {node {\includestandalone[scale=0.7]{l2_3}}
      child {node {\includestandalone[scale=0.7]{l3_2}}
          child {node[above=0.6cm, align=center, rectangle, draw] {
          \scriptsize
          $s(3) = (\{\emptyset,[1,0,3,2]\}$\\
          $seq=[4,1,3]$ \\$R^{seq}=\frac{10+9+14+1+2}{10+9+14+1+2} = 1$\\(Calculated using Alg. 5)}}
          edge from parent  node[right] {$3$}
        }
        edge from parent  node[left] {$1$}
    }
      child {node {\includestandalone[scale=0.7]{l2_4}}
        child {node[above=1.1cm, align=center, rectangle,draw] {
        \scriptsize
        $s(2) = (\{3\}, \{0,0,2,1\})$\\
        $seq=[4,3]$\\
        $R^{seq}=0$}}
        edge from parent  node[right] {$3$}
      }
      edge from parent  node[left] {$4$}
    }
    child {node {\includestandalone[scale=0.7]{l1_3}}
    child {node {\includestandalone[scale=0.7]{l2_5}}
      child {node[above=1.1cm, align=center, rectangle,draw] {
      \scriptsize
        $s(2) = (\{3\}, \{2,0,1,0\})$\\
        $seq=[3,1]$\\
        $R^{seq}=0$}}
        edge from parent  node[left] {$1$}
    }
      child {node {\includestandalone[scale=0.7]{l2_6}}
        child {node[above=1.1cm, align=center, rectangle,draw] {
        \scriptsize
        $s(2) = (\{3\}, \{0,0,1,2\})$\\
        $seq=[3,4]$\\
        $R^{seq}=0$}}
        edge from parent  node[right] {$4$}
        }
        edge from parent  node[right,above] {$3$}
    }
    ;
\end{tikzpicture}
\caption{Example MDP for a toy example. Observe transitions are deterministic and MDP has a tree topology.}
\label{fig:tree_full}
\end{figure*}

Since the MDP transition model for a given slice is completely known in advance and deterministic, one could be tempted to use a method such as dynamic programming to solve the problem. However, it would be unrealistic due to the number of states: there are $\frac{|V|!}{(|V|-|V^x|)!}$ final states (which corresponds to the number of possible repetition-free permutations of $|V|$ physical nodes of size $|V^x|$), each requiring to calculate link placement. For a slice of size 12 placed on a 50 nodes network , we have $5 \times 10^{19}$ possible terminal states.\\ 
Also note virtual nodes are taken in an arbitrary order, hence a given final placement is reachable only using a single sequence of actions. This implies the MDP has a tree topology (see Figure \ref{fig:tree_full} which illustrates the full tree of states for a toy example placement). We argue our algorithm should take this structure into account for exploration and exploitation. Particularly, we will see that existing MCTS methods (MaVEN-S from \cite{HA17} and NRPA) are interesting since they take advantage of the MDP's tree structure for finding good solutions. However this can lead to local optima once the algorithm has converged. The main motivation of our work is to escape these optima by "jumping" to unexplored branches of the tree that we can guarantee are better than the best solutions found. We will show this can be done by getting around the tree topology and sometimes exploring the solution space in a different manner.
\\
Next, we present our online learning algorithm (NEPA) which improves slice acceptance ratio, with reduced computation time by implementing this idea. As our result section will show, we only need to explore a few hundred complete sequences of actions for our algorithm.
\begin{figure*}
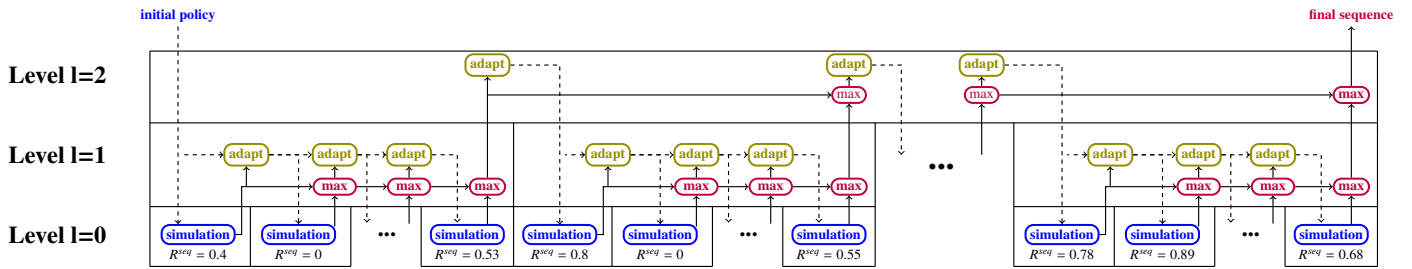

    \includestandalone[width=1\textwidth]{fig_nrpa}
    \caption{Example execution of NRPA for l=2. Dashed arrows represent the policy going from one function to the other, while plain arrows represent sequences returned between functions. A function needs to get values from all its predecessor before executing.}
    \label{fig:nrpa_diagram}
\end{figure*}

\section{Proposed algorithm}
The algorithm we propose in this paper is called NEPA. It is based on NRPA, adding weight initialization and neighborhood-search based refinement. For the sake of clarity, instead of directly presenting NEPA, we first present NRPA and weight initialization.
\subsection{{Review of} Nested Rollout Policy Adaptation (NRPA)}
The NRPA\cite{NRPA} algorithm is a Monte Carlo Search algorithm that aims at finding near-optimal solutions in deterministic environments. It is perfectly suited for our problem as in our model 
 a given action from a certain state always leads 
{deterministically} to the same state. This setup is similar to the puzzle games which NRPA solves remarkably well (with a world record for Morpion Solitaire) \cite{NRPA}. We describe NRPA in \textbf{Algorithm \ref{algo:NRPA}}. The idea of this algorithm is to consider the MDP as a tree that we have to explore ("search tree"). This is coherent with our model because we treat virtual nodes to place in an ordered manner, hence there is only a single way to reach a given final or partial state. NRPA explores the search tree with recursive calls to the search function. This search function is defined as such:
\begin{itemize}
    \item At level 0 a search call does a random simulation of legal actions in the MDP. It returns the reward obtained during that run of the MDP along with the sequence of actions used and the virtual link placement solution $\mathcal{P}$, calculated using \textbf{Algorithm \ref{algo:BFS}}. If we refer to figure \ref{fig:tree_full}, we can see a random simulation as a complete descent from the root of the tree to a leaf (\textit{e.g.} final) state. The return values from that descent are the corresponding $seq$ and $R^{seq}$.
    \item At level $l \neq 0$ the algorithm makes $N$ NRPA calls of level $l-1$. It then returns the best sequence returned by these "children" calls to its caller function, which is either a level $l+1$ NRPA call or the main function. In the latter case the returned sequence is the best sequence found over every simulations tried so far (called $seq^{best}$) and the NRPA algorithm terminates.
    \item Then, the control flow returns to the main function (algorithm \ref{algo:MAIN}), which updates the resources. 
\end{itemize}
The NRPA search function is combined with a policy learning procedure (see \textbf{Algorithm \ref{algo:adapt}} : Adapt procedure for NRPA). 
\paragraph{Policy improvement}
The principle is that during each simulation, we choose the sequence of actions $seq$ in a biased manner that leads to final states close to the best final state found so far, $R^{seq^{best}}$, which has been reached through sequence of actions $seq^{best}$. This random sampling enables us to focus on sequences of actions that resemble $seq^{best}$.

We shall now give the details through which we learn and then bias the simulations :

\begin{itemize}
    \item We define a policy matrix, $P$ which associates each possible tuple ($s(k), a_k$) with a real weight $P[s(k),a_k]$ from which probabilities are calculated during simulation.
\item  Given a certain initial state $s(0)=(s_a(0), s_b(0))$ and a policy matrix $P$, the algorithm will try a sequence of random actions dictated by the probabilities calculated from $P$.
\end{itemize}
After each NRPA call, the weights of actions of the best sequence found $seq^{best}$ are incremented with respect to the state where they should be chosen, \textit{e.g.} $P[s(i),a_i]$ is incremented for all $a_i \in seq^{best}$ (see \textbf{Algorithm  \ref{algo:adapt}}). Then, during  the simulation, when we are in state $s(k)$ and need to select the action $a_k = i$ randomly, we draw using Gibbs sampling, e.g. with probability $\frac{\exp P[s(k),i]}{\exp \sum\limits_{1 \leq j \leq |\mathcal{A}|} P[s(k), j]}$. A visualization of those steps is depicted in figure \ref{fig:nrpa_diagram}, where the recursive nature of the algorithm is particularly noticeable.

\subsection{Virtual links placement}

\textbf{Algorithm \ref{algo:BFS}}, is used for placing virtual links after the node placement is decided. It is used during each call to the simulation procedure (\textbf{Algorithm \ref{algo:simulation})}. The idea is to treat virtual links one by one by descending bandwidth demands, embedding them on the shortest path (in terms of hops) that has enough bandwidth. Note this is not an exact algorithm and it could replaced with other methods of link embedding. We do not use an exact method because the underlying problem of placing virtual links is an instance of the unsplittable flow problem which is itself NP-Hard \cite{ref_UFP}. One alternative could be to relax the problem and allow "path-splitting", making the problem solvable by linear programming \cite{PS}. It might be of interest and has been used for the VNE (see for example \cite{HA17}), with the relaxed version consistently improving performance metrics at the cost of a larger computation time (in the order of 40 times for their small cases). However it is unclear whether such an algorithm would be implementable in practice, due to scalability issues as well as the need to reorder packets on arrival, incurring potential additional delay and CPU processing times. For these reasons, the case of path-splitting is outside the scope of this article. To conclude with $NRPA$, we give in \textbf{Algorithm.\ref{algo:MAIN}} the main procedure which describes the calls of the different algorithms related to NRPA for a slice placement. 
\subsection{Heuristic weight initialization}
\begin{figure}
    \begin{minipage}{0.2\linewidth}
             \begin{tikzpicture}[auto, thick, scale=0.7]
  \foreach \place/\x in {{(-0.25,0)/1}, 
    {(1.5,0)/2}}
  \node[cblue] (a\x) at \place {\small \x};
  
  \node[cblue] (a3) at (0.25,0.7) {\small 3};
  \node[cblue] (a4) at (0.75,-0.3) {\small 4};
  \node[cblue] (a5) at (2.5,0.4) {\small 5};
  
  \path[thin] (a1) edge (a4);
  \path[thick] (a3) edge[color=black] node [color=black, thin, near start, yshift=2ex, below=12pt]  {} (a4);
  \path[thick] (a3) edge[color=black] node [color=black, thin, midway, yshift=2ex, below=-4pt]  {}(a5);
  \path[thin] (a3) edge (a2) ;
  \path[thick] (a2) edge[color=black] node [color=black, thin, midway, yshift=2ex, below=12pt]  {} (a5);
  \path[thick] (a4) edge[color=black] node [thin, color=black, midway, yshift=2ex, right=34pt, below=10pt]  {} (a2);
 
    \node[qgre] (b3) at (0.25,2.7) {\small 1};
    \node[qgre] (b4) at (0.75,1.7) {\small 3};
    \node[qgre] (b5) at (2.5,2.4) {\small 2};
 
  \path[thin] (b3) edge node [midway, yshift=2ex, below=10pt]  {}  (b4);
  \path[thin] (b5) edge node [midway, yshift=2ex, below=8pt]  {} (b4);
  \path[thin] (b5) edge node [midway, yshift=2ex, right]  {} (b3);
 
  \foreach \i in {3, 5}
    \path[rpath] (a\i) edge (b\i);
\end{tikzpicture}

\end{minipage}
 \hspace{0.05\textwidth}%
\begin{minipage}{0.6\linewidth}

      \begin{tabular}{|c|c|c|}
      \hline
          \makecell{Next\\action} & \makecell{Potential next\\state} & \makecell{Initial weight\\value} \\[0.1cm]
          \hline
          $a_k = 4$ & (\{\}, [0, 0, 1, \textbf{3}, 2]) &  $-\frac{d(3,4) + d(5,4)}{2} = -1.5$ \\[0.1cm]
          $a_k = 2$ & (\{\}, [0, \textbf{3}, 1, 0, 2]) & $ -\frac{d(3,2) + d(5,2)}{2} = - 1$\\[0.1cm]
          $a_k = 1$ & (\{\}, [\textbf{3}, 0, 1, 0, 2]) & $-\frac{d(3,1) + d(5,1)}{2} = -2.5$\\
          & & \\
          \hline
          
      \end{tabular}
\end{minipage}
    \caption{Example of state weight heuristic initialization when first choosing action from state (\{3\}, [0, 0, 1, 0, 2])}
    \label{fig:weight_init}
\end{figure}
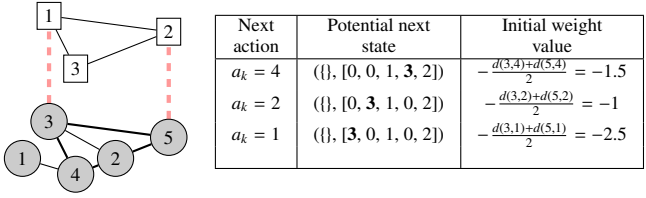

In standard NRPA, when one encounters an unseen state $s(k)$, all its  potential following states $s'(k+1)$, reached from $s(k)$ by choosing action $a_k$ are initialized with a weight $P[s(k), a_k] = 0$. However, this leads to exploring completely at random without exploiting knowledge of the problem. We propose to bias the weight initialization towards more interesting actions, drawing inspiration from \cite{nrpaD}. Our heuristic for weight initialization assumes that good embeddings tend to cluster virtual nodes, \textit{i.e.} to place virtual nodes of the same slice in close-by physical nodes, which reduces the mean length of the virtual links. When a new state-action couple $(s(k), a_k)$ is encountered, we initialize its weight with:
\begin{equation}
\label{eq:weights}
P[s(k),a_k]=\left\{
    \begin{array}{ll}
        -\sum\limits_{1 \leq i \leq n} \frac{d(i,a_k) \times \mathbbm{1}(s_b(k)[i])}{\sum\limits_{1 \leq j \leq n}\mathbbm{1}(s_b(k)[j])}, ~~~ if s_b(k) \neq \Vec{0}\\ 
        \frac{1}{n} ~~~~~~~~~~~~~~~~~~~~~~~~~~~~~~~~~~~~~~~~ otherwise
    \end{array}
\right.
\end{equation}
 where $\mathbbm{1}(s_b(k)[i])$ is equal to 1 if $s_b(k)[i]$ is non-zero (meaning that some virtual node from current slice is associated to physical node $v_i$) and zero otherwise. The function $d(i, j)$ returns the distance (in terms of hops) between physical node $i$ and $j$. Note this distance does not take bandwidth into account, making the function computable in advance before starting NRPA, which makes the complexity of weight initialization negligible. In other words we penalise the physical nodes that are far from the ones used up to the current state to embed the current slice.
 Figure \ref{fig:weight_init} gives an example of such an initialization when the NRPA algorithm first encounters the state (\{3\}, [0, 0, 1, 0, 2])\}. Note that candidate physical nodes that are close to previously placed virtual nodes get a higher weight, since we assume they tend to be more interesting choices. We show an example of such a weight initialization in Figure \ref{fig:weight_init}. Notice that the highest weight corresponds to placing virtual node 3 on physical node 2, which leads to using the least resources.

\begin{algorithm}[H]
\caption{MAIN placement procedure} \label{algo:MAIN}
\setstretch{0.96}
\begin{algorithmic}[1]
\Require $G(V, E)$ :  Physical network, $H^x(V^x, E^x)$ : slice to place 
\Ensure $G(V, E)$ : Physical network, 
$seq^{best}$ : best node placement,
$\mathcal{P}^{best}$: link mapping corresponding to $seq^{best}$
\small
\State  Choose level parameter $l$ (and level $l'$ for NEPA) and number of iterations per level $N$ 
\State  Initialize policy $P$ {as all 0s}
\State  Derive initial state s(0) from $G$ and $H^x$
\If{we call NRPA}
\State $R^{seq^{best}}, seq^{best}, \mathcal{P}^{best} \gets$ \textbf{NRPA}($l$, $N$, P, s(0), $E^x, G$)
\EndIf
\If{we call NEPA}
\State $R^{seq^{best}}, seq^{best}, \mathcal{P}^{best} \gets$ \textbf{NEPA}($l$, $N$, P, s(0), $E^x, G$, l')
\EndIf
\If{$R^{seq^{best}} \geq 0$} 
\State Update occupied resources of $G$ with the placed slice
\EndIf
\State return $G$, $seq^{best}, \mathcal{P}^{best}$
\end{algorithmic}
\end{algorithm}
\begin{figure}
\begin{algorithm}[H]
\caption{NRPA Algorithm}\label{algo:NRPA}
\setstretch{0.96}
\begin{algorithmic}[1]
\Require $l$: Search level , $N$: max iterations, $P$ : Policy, $s(0)$: Initial state, $E^x$: Virtual links, $G(V,E)$: Physical network 
\Ensure $R^{seq^{best}}$: Best score, $seq^{best}$:  best sequence of actions to achieve it, $\mathcal{P}^{best}$: link mapping corresponding to $seq$ 
\small
\If{$l = 0$}
    \State\Return \textbf{SIMULATION}($s(0)$, $P$, $E^x$, $G$)
\EndIf
\State $R^{seq^{best}} \gets - \infty$
\State $seq^{best} \gets \emptyset$
\State $\mathcal{P}^{best} \gets \emptyset$ 
\For{N iterations}
    \State $R^{seq}, seq, \mathcal{P} \gets$ \textbf{NRPA}($l-1, N, P, s(0), E^x, G)$
    \If{$R^{seq^{best}} \leq R^{seq}$}
        \State $R^{seq^{best}} \gets R^{seq}$
        \State $seq^{best} \gets seq$
        \State $\mathcal{P}^{best} \gets \mathcal{P}$
    \EndIf
    \State $P \gets$ \textbf{ADAPT}($P$, $seq^{best}$)
\EndFor
\State \Return $R^{seq^{best}}, seq^{best}, \mathcal{P}^{best}$
\end{algorithmic}
\end{algorithm}
\begin{algorithm}[H]
\caption{ADAPT procedure for NRPA}
\setstretch{0.96}
 \label{algo:adapt}
\begin{algorithmic}[1]
\Require $P$: Policy matrix, $seq$ : sequence of actions 
\Ensure Update of $P$ biased towards drawing actions from $seq$
\small
\State $P_{new} \gets P$
\For{$k = \{0,...,|seq|\}$}
    \State $v_l^x \gets $ first node of $s_a(k)$
    \State $P_{new}[s(k), a_k]$ $ \mathrel{+}= 1$ ~~~~~~~~~~~~// $a_k$ is the $k^{th}$ action of $seq$
    \For {$m \in \mathcal{A}(s(k))$}
        \State $P_{new}$[s(k), m] $\mathrel{-}$= exp($\frac{P[s(k), m]}{\sum\limits_{j \in \mathcal{A}(s(k))}\exp (P[s(k), j])}$)
    \EndFor
    \State $s$(k+1) $\gets$ $(s_a(k) - \{v_l^x\}, s_b(k)+b_{a_k})$
\EndFor
\State return $P_{new}$
\end{algorithmic}
\end{algorithm}
\begin{algorithm}[H]
\setstretch{0.96}
\caption{SIMULATION procedure}
 \label{algo:simulation}
\begin{algorithmic}[1]
\Require $s(0)$: Initial State, $P$: Policy matrix, $E^x$: Set of virtual links, $G$: Physical network 
\Ensure $seq$: sequence of actions, $R^{seq}$:  the reward it yielded, $\mathcal{P}$: path mapping found by \textit{Alg.} \ref{algo:BFS} for $seq$
\small
\State $seq \gets \emptyset,~ k \gets -1$
\While{$\mathcal{A}(s(k)) \neq \emptyset$}
    \State k++
    \State $v_l^x \gets$ first node of $s_a(k)$
    \State Deduce $\mathcal{A}(s(k))$ for $v_l^x$
    \State \multiline{$a_k$ $\gets$ random-choice($\mathcal{A}(s(k))$)\\// \textit{draws action from $\mathcal{A}_(s(k))$ with probability {$\frac{\exp(P[s(k), a_k])}{\sum\limits_{j \in \mathcal{A}(s(k))} \exp(P[s(k), j])}$}}}
    \State $b \gets \Vec{0}$
    \State $b[a_k] \gets l$ 
    \State $s(k+1) \gets (s_a(k) - \{v_{l}^x\}, s_b(k)+b)$
    \State $seq \gets seq \bigcup a_k$
\EndWhile
\State $R^{seq}$, $\mathcal{P}$ $\gets \textbf{VLINK}(E^x, seq, G)$
\State \Return $R^{seq}$, $seq$, $\mathcal{P}$
\vspace{-0.125cm}
\end{algorithmic}
\end{algorithm}
\end{figure}
\begin{algorithm}
\setstretch{0.96}
 \caption{VLINK (virtual link placement) procedure}
 \label{algo:BFS}
\begin{algorithmic}[1]
\Require Set of virtual links $E^x$, sequence of actions $seq$, Physical network $G$
\Ensure $R^{seq}$ : Reward yielded by $seq$, $\mathcal{P} = \{\mathcal{P}_{v_i^x,v^x_j}, \forall (v_i^x, v_j^x) \in E^x \}$ : set of physical path used by each virtual link

\While{$E^x \neq \emptyset$}
\State Pick $(v^x_i, v^x_j) \in E^x$, the most demanding link.
\State \multiline{Find the shortest path $\mathcal{P}_{v_i^x,v^x_j}$ between the physical nodes hosting $v_i^x$ and $v^x_j$, minding only physical links with available bandwidth (at least equal to $BW^d_{v^x_i, v^x_j}$)}
\If{$ \mathcal{P} \neq \emptyset$}
\State Update the physical links   occupied by $\mathcal{P}_{v_i^x,v^x_j}$
\State $\mathcal{P} \gets \mathcal{P} \bigcup \mathcal{P}_{v_i^x,v_j^x}$
\Else
\State return $0, \emptyset$
\EndIf
\EndWhile
\State compute $R^{seq} $
\State return $R^{seq}, \mathcal{P}$
\end{algorithmic}
\end{algorithm}
\begin{figure*}[t]
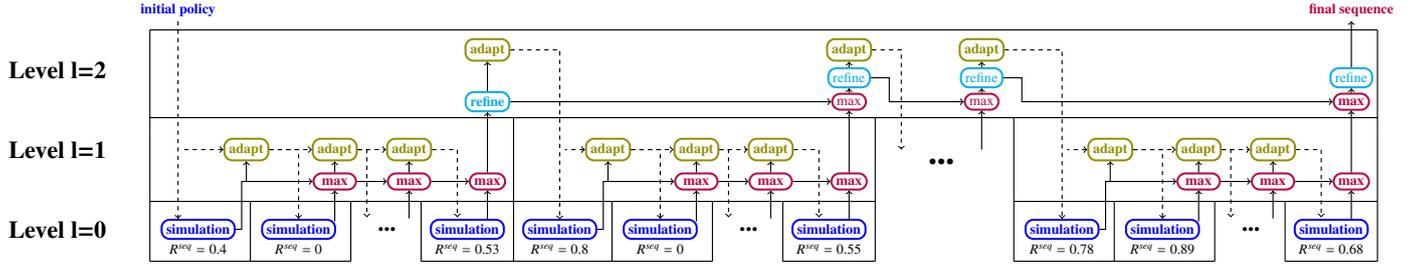

    \includestandalone[width=1\textwidth]{fig_nepa}
    \caption{Example execution of NEPA for l=2, l'=2. Dashed arrows represent the policy going from one function to the other, while plain arrows represent sequences returned between functions. A function needs to get values from all its predecessor before executing.}
    \label{fig:nepa_diagram}
\end{figure*}
\subsection{Neighborhood Enhanced Policy Adaptation (NEPA)}
Observe that once NRPA has found a reasonably good $seq^{best}$, the weights of the partial state leading to it (\textit{e.g.} the weights on the path from the root of the tree to the best final state found) will start increasing. This means if a random simulation deviates from $seq^{best}$ early on, it will then draw actions from a state which is considered almost unexplored, and where the knowledge of $seq^{best}$ is not used. For example, in Fig. \ref{fig:tree_full}, if the best sequence found so far is [1, 4, 3] and at first step the chosen action is 4, it would be desirable to exploit the knowledge that in the best sequence found, virtual node 3 goes on physical node 3, as it is true in [1, 4, 3]. In our toy example, this would imply that the algorithm would have a higher chance to find the optimal sequence, [4, 1, 3]. However, with NRPA this is not how things go: if we descend to an unexplored part of the tree, there is no way to reuse the information gained from the known best sequence. This is the reason we introduce NEPA, which we call a \textit{monkey business} algorithm: our goal is to improve NRPA by enabling it to use its knowledge of $seq^{best}$ for finding better branches, similar to how monkeys jump from one branch to another. When monkeys explore the jungle, they swing from branch to branch, they go faster than if they went back down each time they want to move. Similarly, NEPA swings from branch to branch to explore the MDP while NRPA has to go down every time it wants to explore a new zone of the search space.


One possible solution for this could be to increase weights of all states resembling those found by executing $seq^{best}$. However, this would incur significant cost due to the factorial number of total states, potentially requiring to design approximate methods or workarounds that are outside of the scope of this paper.

Instead, we observe NRPA often finds good embeddings that would be very easy to improve upon by changing the placement of only a small subset of nodes. We also observe that improving such embeddings could be interesting for discovering better sequences of actions that resemble $seq^{best}$ but would not necessarily be discoverable through NRPA's weight mechanism (such as in our example above). Our key idea is that we can improve upon $seq^{best}$ through neighborhood search, finding a new, better sequence without taking the tree structure of the MDP into account. Then, once this improved sequence has been found, we reinject it into NRPA to use as its new $seq^{best}$, which it can further improve. 

In this section, we devise our method for discovering such sequences while keeping the computational complexity reasonable. We call the resulting algorithm Neighborhood Enhanced Policy Adaptation (NEPA) as it combines NRPA with neighborhood search for improving good solutions. The idea of NEPA is to choose a level $l'$ of search at which the solutions should be improved. Then, when the NEPA search reaches level $l'$, each solution found (which correspond to the best solution of each level $l' - 1$ call) is refined through the neighborhood search procedure described in Algorithm \ref{algo:refine}. 

We define the neighborhood of a final state $s(k^{end})$ (corresponding to  an embedding solution of the virtual nodes on the physical network), as the set of final obtained by moving a virtual node to another physical nodes. 
\begin{algorithm}[ht]
\setstretch{0.96}
\caption{REFINE procedure}
 \label{algo:refine}
\begin{algorithmic}[1]
\Require $K$: Max number of physical nodes candidates, $X$:Number of iterations, $G$: Physical network, $H^x$: Slice, $seq$: node placement sequence, $\mathcal{P}$: link mapping for $seq$, $R^{seq}$: reward yielded by $seq$ and $\mathcal{P}$
\Ensure $R^{seq^{ref}}$: Refined solution reward, $seq^{ref}$: refined node mapping, $\mathcal{P}^{ref}$: refined mapping
\State $seq^{ref} \gets seq$
\State $R^{seq^{ref}} \gets R^{seq}$ 
\State $\mathcal{P}^{ref} \gets \mathcal{P}$
\For{X iterations}
\State $previous\_R \gets R^{seq^{ref}}$

\State  \multiline{Compute most promising virtual node  to move in placement given by $seq^{ref}$, $\mathcal{P}^{ref}$ using eq.(9)}
\State \multiline{Build set of K best physical nodes (ranked using eq.(\ref{eq:weights})) $V_{r}\subset V$ suitable for hosting the virtual node}
\For{$v \in V_{r}$}
\State \multiline{Put virtual node  on  physical node $v$, \textit{i.e.}:} 
    \State\hspace{0.2cm}\multiline{\textbullet \hspace{0.05cm} Compute updated version of $seq^{ref}$  $seq$}
    \State\hspace{0.2cm}\multiline{\textbullet \hspace{0.05cm}  Compute new shortest paths $\mathcal{P}$ considering the new position of the virtual node }
    \State\hspace{0.2cm}\multiline{\textbullet \hspace{0.05cm} Update resources used on the physical node and links newly used.} 
\State Compute $R^{{seq}}$ using $seq$ and $\mathcal{P}$ 
\If{$R^{seq}$ $>$ $R^{seq^{ref}}$}
                \State $R^{seq^{ref}}$ $\gets$ $R^{seq}$
                \State $seq^{ref}$ $\gets seq$
                \State $\mathcal{P}^{ref} \gets \mathcal{P}$
\Else
    \State \multiline{Restore resources used on physical node and link to values matching $seq^{ref}, \mathcal{P}^{ref}$}
\EndIf
\EndFor
    \If{previous\_R = $R^{seq^{ref}}$}
        \State break
    \EndIf
    \EndFor
    \State \Return $R^{seq^{ref}}, seq^{ref}, \mathcal{P}^{ref}$
\end{algorithmic}
\end{algorithm}
\subsubsection{Main steps of NEPA}
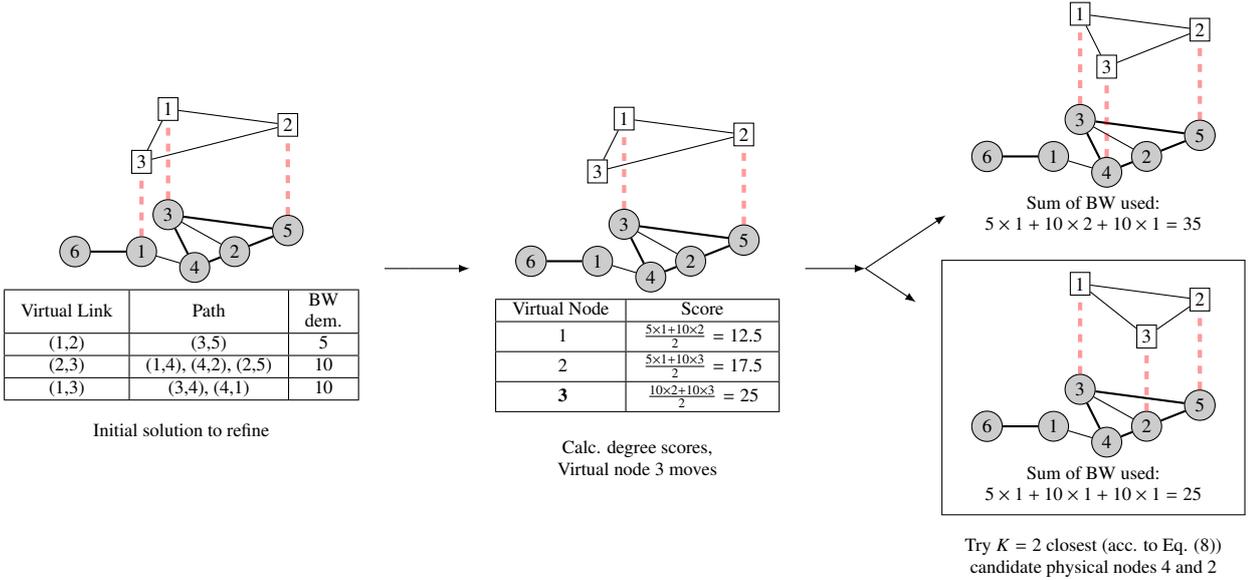
\begin{figure*}
\begin{tikzpicture}[level distance=30mm,
  sibling distance=40mm,
  arrow/.style={edge from parent/.style={draw,-latex}}
]

\node {
    \begin{tabular}{c}
        \includestandalone[]{fig5_1}\\
        \begin{tabular}{|c|c|c|}
            \hline
             Virtual Link & Path & \makecell{BW\\ dem.} \\
             \hline
             (1,2) & (3,5) & 5\\
             \hline
             (2,3) & (1,4), (4,2), (2,5) & 10\\
             \hline
             (1,3) & (3,4), (4,1) & 10 \\
             \hline
        \end{tabular}\\
        \makecell{\\Initial solution to refine}
    \end{tabular}
}
child[grow=right] {
 child[arrow] {node{
 \begin{tabular}{c}
    \includestandalone[]{fig5_2}\\
    \begin{tabular}{|c|c|}
        \hline
         Virtual Node & Score  \\
         \hline\xrowht[()]{5pt}
         1 & $\frac{5 \times 1 + 10 \times 2}{2} = 12.5$ \\
         \hline\xrowht[()]{5pt}
         2 & $\frac{5 \times 1 + 10 \times 3}{2} = 17.5$ \\
         \hline\xrowht[()]{5pt}
         \textbf{3} & $\frac{10 \times 2 + 10 \times 3}{2} = 25$  \\
         \hline
    \end{tabular} \\
    \makecell{Calc. degree scores, \\\vspace{-1cm}Virtual node 3 moves} 
\end{tabular}
 }
 child[grow=right] {
    child {node{
    \begin{tabular}{c}
    \fbox{
    \begin{tabular}{c}
         \includestandalone{fig5_4}\\
         \makecell{Sum of BW used: \\
        $5 \times 1 + 10 \times 1 + 10 \times 1 = 25$
        } 
    \end{tabular}
    }\\
    \makecell{
        \\
        Try $K=2$ closest (acc. to Eq. (\ref{eq:weights}))\\ candidate physical nodes 4 and 2
    }
    \end{tabular}
    }}
    child {node{
    \begin{tabular}{c}
        \includestandalone[]{fig5_3}  \\
        \makecell{Sum of BW used: \\
        $5 \times 1 + 10 \times 2 + 10 \times 1 = 35$
        } 
    \end{tabular}
    }}
 }
 }
};

\end{tikzpicture}
\caption{Example iteration of Refine with K=2. The placement chosen after the iteration is in the box. We assume all nodes have enough CPU and all links have enough BW. The chosen solution has the lowest amount of bandwidth used as reward only depends on it (CPU demands and uses are the same for a given slice)}
\label{fig:refine}
\end{figure*}

We describe in \textbf{Algorithm \ref {algo:NEPA}} the main steps of NEPA algorithm. 
It is similar to the NRPA algorithm  except that if we reach a level $l=l'$, then we choose to refine the solution by searching a  neighboring solution as described in the following algorithm: 
\begin{enumerate}
\item \textbf{Algorithm \ref{algo:refine}} first finds the nodes with the largest potential improvement among the already placed virtual nodes (\textit{e.g.} we choose a single node $v^x_B$ to move). We find it by calculating: 
\begin{equation}
    score(v_m^x)=\frac{\sum\limits_{v_p\in V^x}BW^d_{v_m^x,v_p^x} \cdot d(v_m^x, v_p^x)}{deg(v_m^x)}
\end{equation}
    for each virtual node $v^x_m$, where $l(v_m^x, v_p^x)$ is a function returning the length of the physical path used by virtual link $(v_m^x, v_p^x)$ and $deg(v_m^x)$ is the degree of virtual node $v_m^x$. The virtual node which maximizes this metric is considered as the most promising for improvement, since it is the one which consumes the most bandwidth compared to its number of neighbors.
\item The refining procedure is then to try several candidate physical nodes that could be better suited to host the selected virtual node $v^x_{B}$, in terms of reducing resource consumption. For each candidate, we remap the virtual node on them (which corresponds to flipping values in the state vector), then remap its adjacent virtual links. After all candidate physical nodes have been tried, the new placement of $v^x_{B}$ is then the one that leads to a maximum reward (see eq (3)).
\end{enumerate}
To control the execution time of the algorithm, we introduce two parameters: 
\begin{itemize}
\item $X$ : is the number of times that the  process is  repeated, note that the process is also stopped  if a full trial does not lead to any improvement. 
Typically a criterion can be to do no more than $|V^x|$ tries.  This ensures the runtime is reasonable while spending more time on larger slices, since they tend to be harder to place. 
\item $K$ is the number of candidate physical nodes. For choosing candidates, the simplest thing would be to try all possible physical nodes. However this would lead to poor scalability. Instead, we use our weight initialisation function and define our $K$ candidates as the $K$ nodes with the highest distance score. For example, in Fig. \ref{fig:refine}, which shows a refinement iteration with $K=2$, the two candidates for hosting virtual node 3 are physical nodes 4 and 2 because they are the two nodes that are the closest to physical nodes 3 and 5, which host the other virtual nodes.

\end{itemize}
After the refinement, the resulting placement is treated like a normal state by NEPA, \textit{e.g.} if it is the best found so far, its weight is incremented. In this sense, NEPA (Alg. \ref{algo:NEPA}) maintains the structure of NRPA (\textbf{Algorithm \ref{algo:NRPA}}). Note that in practice, NRPA could potentially have found any sequence of actions (\textit{e.g.} node placement) found by NEPA. However, the virtual link embedding corresponding could be different since NRPA places only using \textbf{Algorithm \ref{algo:BFS}} while in the case of refinements, NEPA uses \textbf{Algorithm \ref{algo:refine}} which can find a different link embedding for the sequence than what \textbf{Algorithm \ref{algo:BFS}} could have found. This is particularly important since in practice, we observe that some sequences found by \textbf{Algorithm \ref{algo:refine}} have no valid solution if using only \textbf{Algorithm \ref{algo:BFS}}. Hence, when using NEPA, it is necessary to save not only the best sequence of actions, but also the link embedding result in case it needs to be restored after execution for future use. This is typically done at the end of \textbf{Algorithm \ref{algo:refine}} by saving the link embedding solution in the global data structure $S_{emb}$ 

NEPA requires only very little modifications to NRPA (see algorithm \ref{algo:NEPA}), which is particularly noticeable in figure \ref{fig:nepa_diagram}, as it illustrates the way NEPA makes its function calls recursively. Also note how few refine calls there are compared to the number of max operations, since those we chose $l' = 2$.

Our method enables us to discover better solutions that would not be easy to find once standard NRPA has converged: once NRPA has found a local optimum, the probabilities of choosing the states of the best sequence found in NRPA will go towards 1, meaning exploration could become poor while there is no point exploiting the same region anymore. With NEPA, since we change the best sequence found so far, we open the opportunity of exploring completely new, but better parts of the search space. A single change in the first few actions can lead to discovering a whole new part of the state-space where most states are undiscovered, leading to a highly explorative phase with a very good sequence as a starting point (which we newly found through the refinement procedure). Hence NEPA exploits its neighborhood search mechanism in order to help NRPA escape local optima.
\begin{algorithm}[H]
\setstretch{0.96}
 \caption{NEPA Algorithm}
 \label{algo:NEPA}
\begin{algorithmic}[1]
\Require Search level $l$, Number of iterations $N$, Policy matrix $P$, Initial state $s(0)$, Set of virtual links $E^x$, Physical network $G$, \color{blue}Refinement level $l'$\color{black}
\Ensure Best score achieved and best sequence of actions to achieve it
\small
\If{$l = 0$}
    \State \Return \textbf{SIMULATION}($s(0)$, $P$, $E^x$, $G$)
\EndIf
\State $R^{seq^{best}} \gets -\infty$
\State $seq^{best} \gets \emptyset$
\State $\mathcal{P}^{best} \gets \emptyset$
\For{N iterations}
    \State $R^{seq}$, $seq, \mathcal{P} \gets$ \textbf{NEPA}($l-1, N, P, s(0), E^x, G, l'$)
    \If{$R^{seq^{best}}$ $\leq$ $R^{seq}$}
        \State $R^{seq^{best}} \gets$ $R^{seq}$
        \State $seq^{best}$ $\gets$ $seq$
        \State $\mathcal{P}^{best} \gets \mathcal{P}$
    \EndIf
    \color{blue}
    \If{l = l' and $R^{seq^{best}} \neq 0$}
        \State \multiline{$R^{seq^{best}}$, $seq^{best}$, $\mathcal{P}^{best}$ = \textbf{REFINE}(K, X, $G$, $H^x$, $seq^{best}$, $\mathcal{P}^{best}$, $R^{seq^{best}}$)}
    \EndIf
    \color{black}
    \State $P \gets$ \textbf{ADAPT}($P$, $seq$)
\EndFor
\State \Return $R^{seq^{best}}$, $seq^{best}$, \color{blue} $\mathcal{P}^{best}$ \color{black}
\end{algorithmic}
\end{algorithm}
\subsection{Complexity}
In this section, we outline complexity results for both NRPA and NEPA.
We first start by calculating their computational complexity, then memory complexity is discussed.
\subsubsection{Computational Complexity}
\begin{nrpa_c}
The NRPA algorithm has a computational complexity of $O(|V| \times N^l)$ for sparse physical and virtual graphs.
\end{nrpa_c}
Let $T(N, l)$ be the function associating the algorithms' parameters (number of iterations per level N and search level l) with the number of simulations executed (\textit{e.g.} the number of calls to Alg. 3). We shall prove $T(N, l) = N^{l}$ by induction on $l$:
For $l=0$ the relationship is verified.
We now assume that our hypothesis is verified, \textit{e.g.} $T(N, l) = N^{l}$. We will now show that this implies $T(N, l+1) = N^{l+1}$.
\begin{equation} \label{eq1}
\begin{split}
T(N, l+1)& = N \times T(N, l) =  N \times N^{l} = N^{l+1}
\end{split}
\end{equation}
This proves $T(N, l) = N^{l}$. By the same argument, one could show that the same NRPA search would perform $N^l$ adaptations of its policy (\textit{e.g.} $N^l$ calls to Alg. 2).

Algorithms \ref{algo:adapt} and \ref{algo:simulation} are really similar and treat a sequence of length $|V^x|$. For each element of the sequence, both algorithms loop through the list of legal moves. At worst, at each step, all physical nodes that have not already been chosen are legal. In such a case, the complexity of both nested loops would be $O(|V^x|\times|V|)$. 
In order to compute the rewards in the simulation procedure, we place virtual links of the embedding found. This is done using algorithm \ref{algo:BFS}. A breadth-first search (BFS), used in Alg. 4 for finding shortest path, has a complexity of $O(|V| + |E|)$ which we perform $E^x$ times in Alg. 4. The complexity of the simulation procedure (Alg. 3) is then $O((|V| + |E|)\times|E^x|)$. Furthermore, at worst we have $|E| = \frac{|V|(|V|-1)}{2}$ and $|E^x| = \frac{|V^x|(|V^x|-1)}{2}$, so the complexity of the link embedding phase is $O(|V|^2 \times |V^x|^2)$.

It can then be concluded that the complexity of the NRPA algorithm (Alg. 1) is $O(|E^x| \times |E| \times N^{l}) = O(|V^x|^2 \times |V|^2 \times N^{l})$. Note that we used the simplest possible embedding function for links. If one swaps it out for a more elaborate function, such as an exact method\cite{UNSPL}, one taking congestion\cite{CONG}, delays\cite{BENHAM} or survivability\cite{SURV} into account, the complexity would typically increase. We expect such a costlier function to be used in a more realistic setting. Also note that in a typical scenario, both physical network and virtual networks are sparse, \textit{e.g.} $|V|^2 >> |E|$ and $|V^x|^2 >> |E^x|$ hence the complexity of BFS can be assumed to be reduced to $O(|V|)$ and the complexity of NRPA to be $O(|V^x| \times |V| * N^l)$. Furthermore in most cases $|V| >> |V^x|$ and it further reduces to $O(|V| \times N^l)$.

\begin{nepa_c}
The NEPA algorithm has a computational complexity of $O(|V| \times (N^l + N^{l'} \times K \times X))$ when physical and virtual graphs are sparse, where K is the number of candidates per refinement, X is the maximum number of times we try to refine the solution and l' is the level where refinements are performed. 
\end{nepa_c}
First, note that the number of simulations does not change compared to NRPA and is still $N^l$. It follows that the total number of operations performed by the simulation part of the algorithm is $O(|V| \times N^l$ as in NRPA. \\
The complexity of NEPA is then $O(|V| \times N^l + Z)$ where Z is the number of operations incurred by all the refinement steps.
At each refinement step, we perform $X \times K$ BFS searches of complexity $|E|$. The total number of operations performed by refinements is then $Z = O(N^{l'} \times K \times X \times |E|)$. In the case of a sparse graph, this number is $Z = O(N^{l'} \times K \times X \times |V|)$. The total computational cost of NEPA is then 
\begin{equation}
O(|V| \times N^l + N^{l'} \times K \times X \times |V|) = O(|V| \times (N^l + N^{l'} \times K \times X))
\end{equation}
Overall, NEPA has a greater theoretical complexity than NRPA. However, numerical results from Appendix B show that NEPA is far more effective than NRPA when they are given equal time.

\subsubsection{Memory Complexity}

\begin{nrpa_mem}
The NRPA and NEPA algorithms have a memory complexity of $O(|V^x| \times N^l)$
\end{nrpa_mem}

We will now prove proposition 3. First, in the worst case, each simulation procedure call can lead to finding $|V^x|$ new unexplored states, each of which requires to store a float representing its weight in the policy. If every state found in every simulation call is seen only once, we have to store $N^l \times |V^x|$ floats since as seen in the previous proofs, we call the simulation procedure $N^l$ times. The other source of memory consumption in NRPA is the storage of the sequences, $seq$ and $seq^{best}$. Those are both of length $|V^x|$. Since NRPA calls itself recursively, we also have to count the sequences stored by its infant calls. There are at worst l such infants since the recursive call depth is of l and there is only one call of a given level active at the same time, and once a call returns it frees the memory. Hence the memory consumption of the stored sequences is $O(l \times |V^x|)$. The total memory complexity of NRPA is $O(|V^x| \times N^l + l \times |V^x|) = O(|V^x| \times N^l)$.
For NEPA the memory complexity remains the same as NRPA because the refinement procedure does not incur a significant memory usage, as it only requires memory to store the best solution (a virtual network which requires $O(|V^x|$ memory in case of a sparse graph).
      
\subsection{Dimensionality reduction and pre-treatment}

In practice, we make a slight modification to the MDP model in order to make the NEPA search more effective. First, we note that for a given couple of virtual node $v_i^x$ and physical node $v_j$, if the maximum amount of bandwidth required by links adjacent to $v_i^x$ exceeds the maximum available bandwidth of links adjacent to $v_j$, then we know one of the adjacent links of $v_i^x$ would be impossible to embed if $v_i^x$ was placed on $v_j$. Hence, we reduce the size of the action space by removing such actions before running NRPA. Similarly, if the sum of the bandwidth adjacent to $v_i^x$ exceeds the sum of bandwidths available on links adjacent to $v_j$, then we know it would not be possible to place all virtual links if $v_i^x$ was placed on $v_j$, hence we remove this action from the set of possible actions for placing $v_i^x$. 
Finally, before the placement, we sort the nodes according to the number of physical nodes that could host them. This draws on the idea that if a node has only few possibilities for placement, we should treat it first, otherwise there would be a high chance of blocking the possible host with another virtual node placed before. By doing this, we avoid exploring some unfeasible placements.

\section{Numerical Results}
In this section we extensively compare NEPA with several other methods from the state-of-the-art, demonstrating the superiority of of its performance consistently on various scenarios. We first compare on synthetic physical networks generated randomly. Then, algorithms are tested with real physical networks from the topologyZoo dataset. Finally, in order to assess the performance of each tested algorithm {on} large problems {against the theoretical optimum}, we compare on a set of Perfectly Solvable Scenarios \cite{PhDFischer}, which are constructed so the optimal is known but is very hard to achieve. This step is often overlooked in the literature but we argue it is of key importance in order to assess the quality of each algorithm. Note we make sure the range of CPU and Bandwidth capacities fit reality: for CPUs, a typical server CPU (such as intel Xeon) would have between 8 and 56 cores (for example Xeon Platinum 9282). Also note that some server motherboards can host 2 CPUs (for example ASUS WS C621E). For ethernet links, it is common to find bandwidths in the order of 50-100 Gbps, see for example \cite{ETH}.

\subsection{Compared methods}
\begin{figure}
    \begin{center}
    \includegraphics[width=0.45\textwidth]{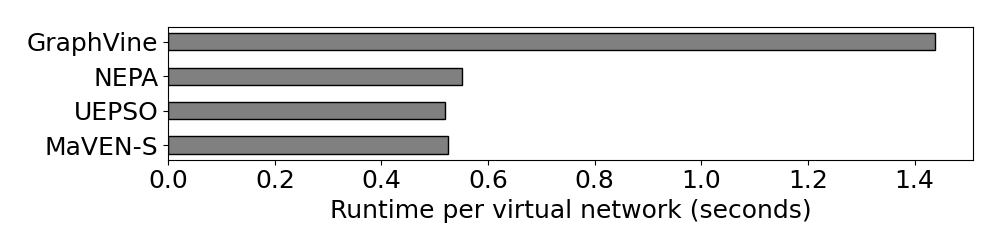}
    \caption{Mean runtime per slice for each algorithm (calulated by averaging runtime per slice on all runs of varying $\lambda$ scenarios) 
    }
    \label{fig:runtime}
    \end{center}
\end{figure}

All our experiments are ran with an Ubuntu machine with a 16-core Intel Xeon Gold 5222s machine with 32 GB RAM, excepted for GraphVine which requires to be ran on another machine equipped with a GPU (see below).
We compare our proposed method NEPA with the following methods:
\begin{itemize}
    \item MaVEN-S \cite{HA17} is a Monte Carlo Tree Search based algorithm which uses a model equivalent to ours for modeling the VNE and the same shortest path algorithm for final reward calculation. It makes sense to compare it with our method as it is similarly based on randomly simulating node placements but uses a different exploration strategy. This strategy, called Upper Confidence Bound for Trees explores the MDP as a tree of states (rooted in the initial state).
    It chooses where to descend in the tree by balancing exploration of new states and exploitation of known states, with the objective to minimize the regret of exploring new states given the expectated reward yielded by known states.  
    
    \item UEPSO \cite{UEPSO} is a particle swarm optimization (PSO) based meta-heuristic algorithm that shows good performance for the VNE. 
    \item GraphVine \cite{graphVine} is a recently proposed method that exploits graph neural networks for selecting the physical nodes on which to place the virtual nodes. Note that GraphVine{, like us,} learns online, {different from other neural network} approaches such as \cite{DL}{,} which requires an extensive offline training first, tied to the physical network. {For this reason, a comparison with these other approaches would not be fair. This is why we prefer to compare with~\cite{graphVine} instead.}
\end{itemize}
{For the sake of clarity, we keep in this section only the comparison with the state-of-the-art methods (mentioned above). W}e postpone the ablation study of NEPA (and its improvement over NRPA) to Appendix B, which shows the benefits brought to NEPA by weight initialization and neighborhood-based refinements.
{We implemented a}ll these methods {in} the Julia programming language and made the code available as open source\cite{github}, except for GraphVine for which we use the publicly available Python/Pytorch implementation.
In order to compare in the fairest manner possible, we run the following experiments:
\begin{itemize}
    \item We run NEPA with parameters $N = 5$ and $l = 3$.
    \item MaVEN-S is ran with a computational budget (\textit{e.g.} the total number of link placement attemps it executes per slices) of 445 link placements per slice. Note we tried to run it for longer times (up to 670 iterations per slice) without a significant improvement of results.
    \item Since UEPSO is a non-recursive algorithm, it is easier to stop it at any moment and get a valid placement. Hence here, we simply stop UEPSO after a certain amount of time equal to the mean time taken by NEPA.
    \item Finally we run GraphVine with the default implementation, as it is a quite different algorithm which does not rely on repeated simulations and since it can exploit a GPU. However with our original machine, we note that it is the slowest to run. As shown in \cite{graphVine}, the algorithm is better suited for using a GPU. For that reason, we run it on another computer which has a GPU (as it gave the best runtime). This machine uses an nvidia A3000 and an Intel i7-11850H CPU. 
\end{itemize}
Runtimes are depicted in Figure \ref{fig:runtime}. We run each of the described experiments 10 times with different random seeds, excepted for GraphVine for which we run it only once due to the high computational cost. Note that we compute 99\% confidence intervals of acceptance and revenue-to-cost ratio for MaVEN-S, UEPSO and NEPA. Some figures do not display them because they are too narrow to be visible on figures. (\textit{e.g.} confidence interval in the order of less than $\pm$ 0.01 for acceptance and revenue-to-cost ratio)

\subsection{Results on synthetic physical topologies}
We start our experiments with a sensitivity analysis. For this part, we generate scenarios with default parameters and we vary each of these one by one in order to assess the results on a representative set of cases. Default parameters are reported in Table \ref{fig:def_params}. We choose to generate our slices and virtual networks with the Waxman generation algorithm as it is commonly used {in} the VNE litterature \cite{HA17}\cite{survey}. We choose to generate 500 slices per scenario as we validated experimentally this gave enough time for the system to stabilize in terms of acceptance ratio.
We then vary parameters in the following ways :
\begin{itemize}
    \item We generate slices with varying Poisson arrival rates between $\lambda=0.02$ and $\lambda = 0.08$ arrivals per second. (results in Fig. 9.1/9.5)
    \item We generate slices with sizes (number of virtual nodes) with minimum size $7 + i$ and maximum size $13 + i$ for $i \in [0, 9]$. (results in Fig. 9.2/9.6)
    \item We modify the physical network from the default scenario by removing bandwidth and {CPU} capacities in increments of 5 from {links and nodes of} the physical network, making resources scarcer. Since initially the resource capacities (CPU and BW) are chosen uniformly at random between 50 and 100, their mean value is about 75. Since we remove from all nodes and links, the mean number of resources for the different scenarios is 70, 65, 60, down to 45. (results in Fig. 9.3/9.7)
    \item We generate 10 different physical networks and slice sets for each physical network size {of} 50, 60, 70, 80, 90, 100{ nodes}. In those scenarios, we use the default parameters, but with $\lambda = 0.04$ and slice sizes as {specified} in {Table }\ref{table_var_sizes}. We scale the size of slices {with respect to the physical network size} since our early experiments showed that if the slice sizes were the same for all physical networks tested, it resulted in too easy scenarios for larger {physical} networks, where most algorithms reached performances close to {100\% acceptance rate}, making the comparison pointless. (results in Fig. 9.4/9.8)
\end{itemize}
\begin{table}
    \centering
    \begin{tabular}{|l|l|}
    \hline
         Parameter & Default Value \\
    \hline
         Slice arrival rate $\lambda$ & 0.02 \\
    \hline
         Slice departure rate $\mu$ & 0.005 \\
    \hline
         Slice generator & Waxman ($\alpha = 0.5, \beta = 0.2$)\\
    \hline
         Number of slices & 500 \\
    \hline
         Min $|V^x|$ & 7 \\
    \hline
        Max $|V^x|$ & 13 \\
    \hline
        $|V|$ & 75 \\
    \hline
        $|E|$ & 273 \\
    \hline
        CPU demands (number of cores) & 1 - 50 \\
    \hline
        BW demands (Gbps) & 1 - 50 \\
    \hline
        Physical CPU capacities (number of cores per node) & 50 - 100 \\
    \hline
        Physical BW capacities (Gbps per link)& 50 - 100 \\
    \hline
    \end{tabular}
    \caption{Default scenario generation parameters}
    \label{fig:def_params}
\end{table}
\begin{table}
    \centering
    \begin{tabular}{|l|l|}
    \hline
        Number of physical nodes of  & Number of virtual nodes \\
    \hline
        50 & 7-13 \\
    \hline
        60 & 8-14 \\
    \hline
    70 & 9-15 \\
    \hline
    80 & 10-16 \\
    \hline
    90 & 11-17 \\
    \hline
    100 & 12-18 \\
    \hline
    \end{tabular}
    \caption{Mean number of nodes of virtual networks for each size of physical network tested}
    \label{table_var_sizes}
\end{table}


\begin{figure*}
\begin{minipage}{.24\textwidth}
  \includegraphics[width=\textwidth]{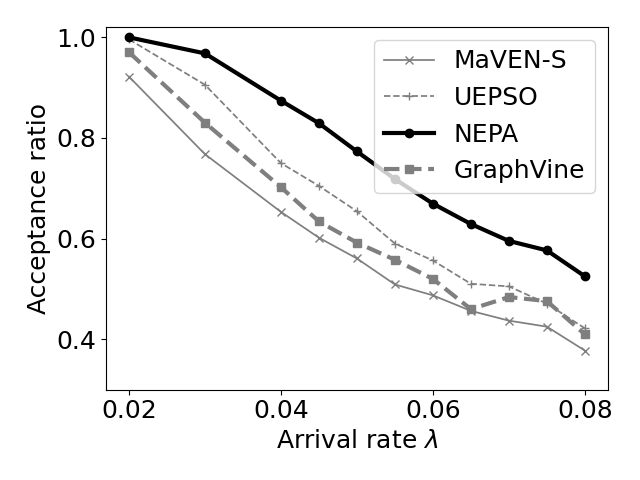}
  \centering \small 9.1 Acceptances for varying arrival rate ($\lambda$)
  \includegraphics[width=\textwidth]{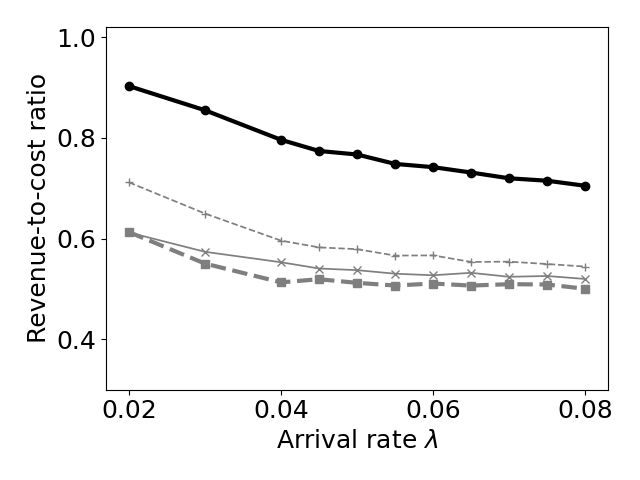}
  \centering \small 9.5 Revenue-to-cost ratio for\\ varying $\lambda$
\end{minipage}
\begin{minipage}{.24\textwidth}
  \includegraphics[width=\textwidth]{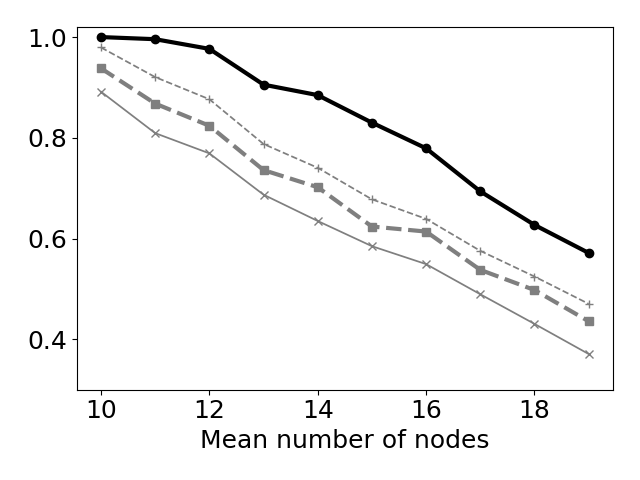}
  \centering \small 9.2 Acceptances for varying slice size
  \includegraphics[width=\textwidth]{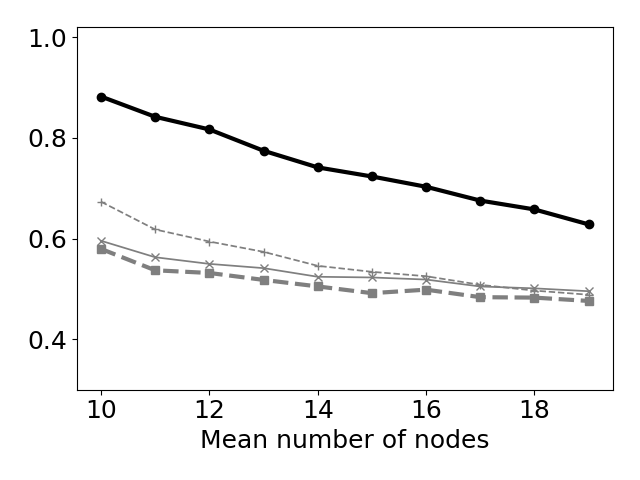}
  \centering \small 9.6 Revenue-to-cost ratio for varying slice size
\end{minipage}
\begin{minipage}{.24\textwidth}
  \includegraphics[width=\textwidth]{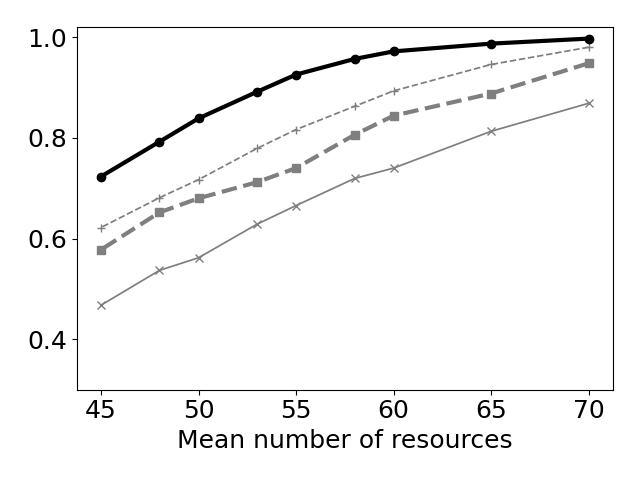}
  \centering \small 9.3 Acceptances (varying CPU \& BW capacities)
  \includegraphics[width=\textwidth]{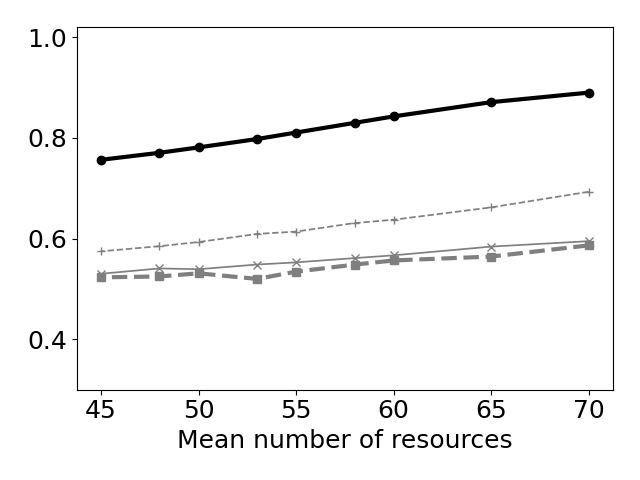}
  \centering \small 9.7 Revenue-to-cost ratio (varying CPU \& BW capacities)
\end{minipage}
\begin{minipage}{.24\textwidth}
  \includegraphics[width=\textwidth]{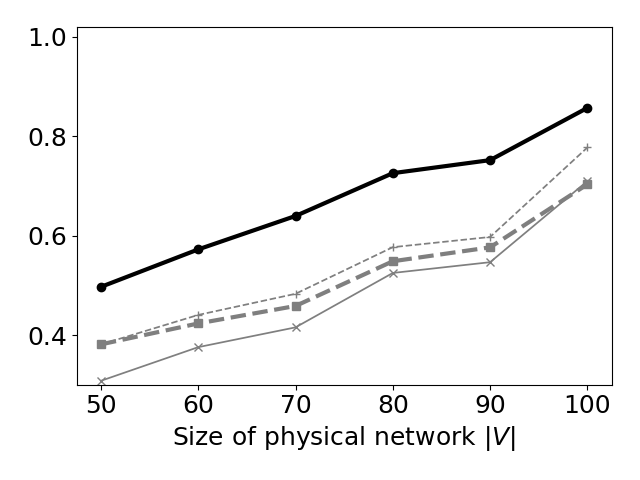}
  \centering \small 9.4 Mean acceptance ratios for varying physical network sizes
  \includegraphics[width=\textwidth]{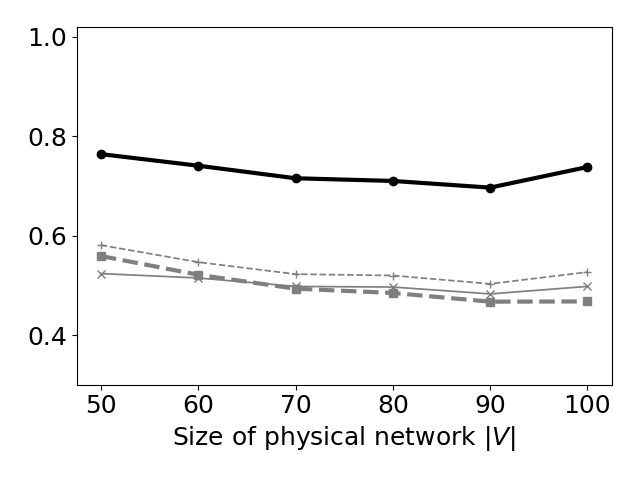}
  \centering \small 9.8 Revenue-to-cost ratio (varying network sizes)
\end{minipage}
    \caption{Results for sensitivity analysis experiments}
    \label{fig:results_sensitivity}
\end{figure*}

Figure \ref{fig:results_sensitivity} shows that on every tested scenario, NEPA beats all other algorithms by a large margin, consistently beating MCTS of around 50\% of acceptance and the best of other contenders (which are close to each other, above MCTS) by 15\%, regardless of the case. 
In terms of revenue-to-cost ratio, it is striking to note that NEPA beats other algorithms by an even larger margin than for acceptance. This means NEPA tends to use less physical resources
, which is the reason why it achieves a better acceptance. This suggests that reducing the overall consumption of each slice enables us to leave more resources for future incoming slices, making it possible to place them.
~\\
For variable size physical networks (figure \ref{fig:results_sensitivity}.4), we observe again that NEPA beats other contenders by a large margin, since it accepts up to 60\% more than MaVEN-S, and consistently beats it by 20 points of acceptance.
It is remarkable to note how regular the patterns are in the acceptance plots, especially given that results are averaged for different topologies (recall that in the variable size experiment, for each seed, we generate a different random topology). The difference between algorithms is almost always the same regardless of sizes and difficulty of the instance, with NEPA as a clear winner.  
Regarding revenue-to-cost ratios, we note that all algorithms excepted NEPA have average ratios between 0.5-0.6. NEPA beats them by a large margin, since it is the only one to consistently reach 0.7 to 0.75 of revenue-to-cost ratio, showing again the effectiveness of the neighborhood based refinement in increasing the quality of the solutions found. 

\subsection{Real Topologies}

We try all algorithms with real topologies from the TopologyZoo \cite{Zoo} dataset as physical networks
. We choose to use topologies that have between 60 and 200 nodes and are connected. This leaves us with 26 topologies with $|V|$ between 60 and 197. We use bandwidth capacities chosen randomly between 250 and 300 Gbps and CPU capacities between 50 and 100 cores. Slices are generated with our standard scheme but with $\lambda = 0.04$ arrivals per second.

\begin{figure*}
    \centering
    \centerline{
    \includegraphics[scale=0.4]{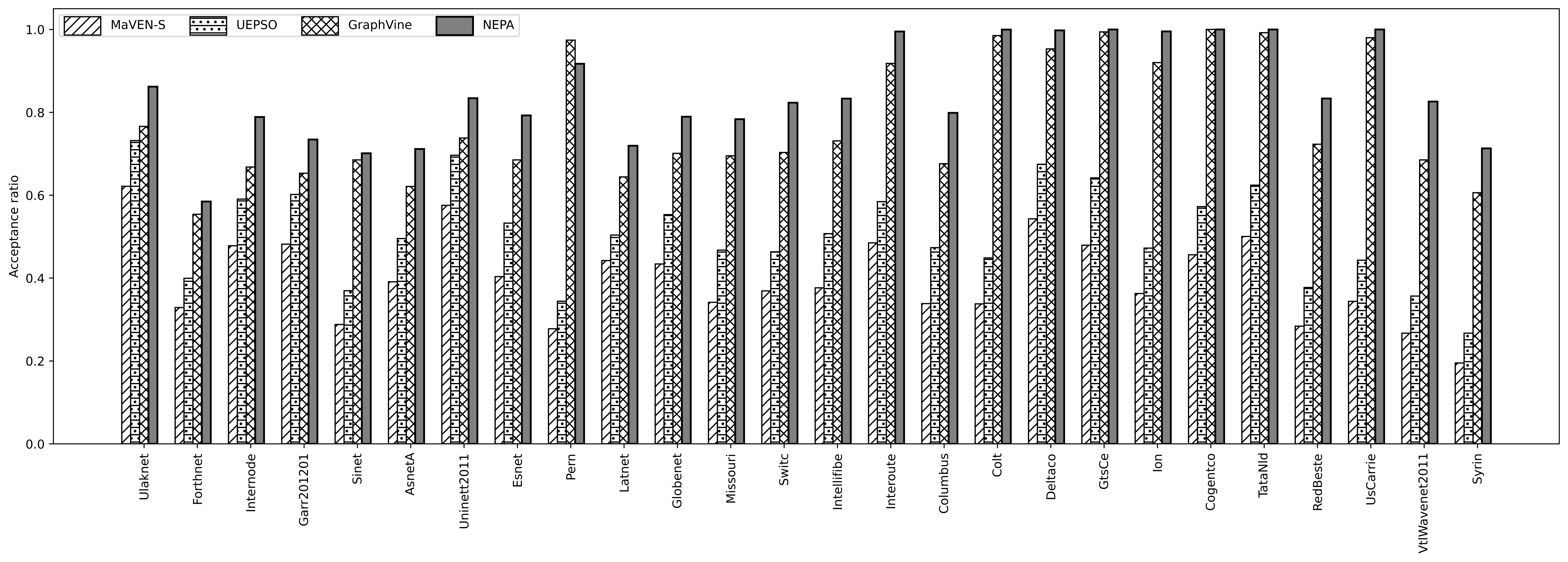}
    }
    \caption{Acceptance on real physical networks. Results are ordered by increasing shortest-path length variance.}
    \label{fig:acc_real}
\end{figure*}
\begin{figure*}
    \centering
    \centerline{
    \includegraphics[scale=0.40]{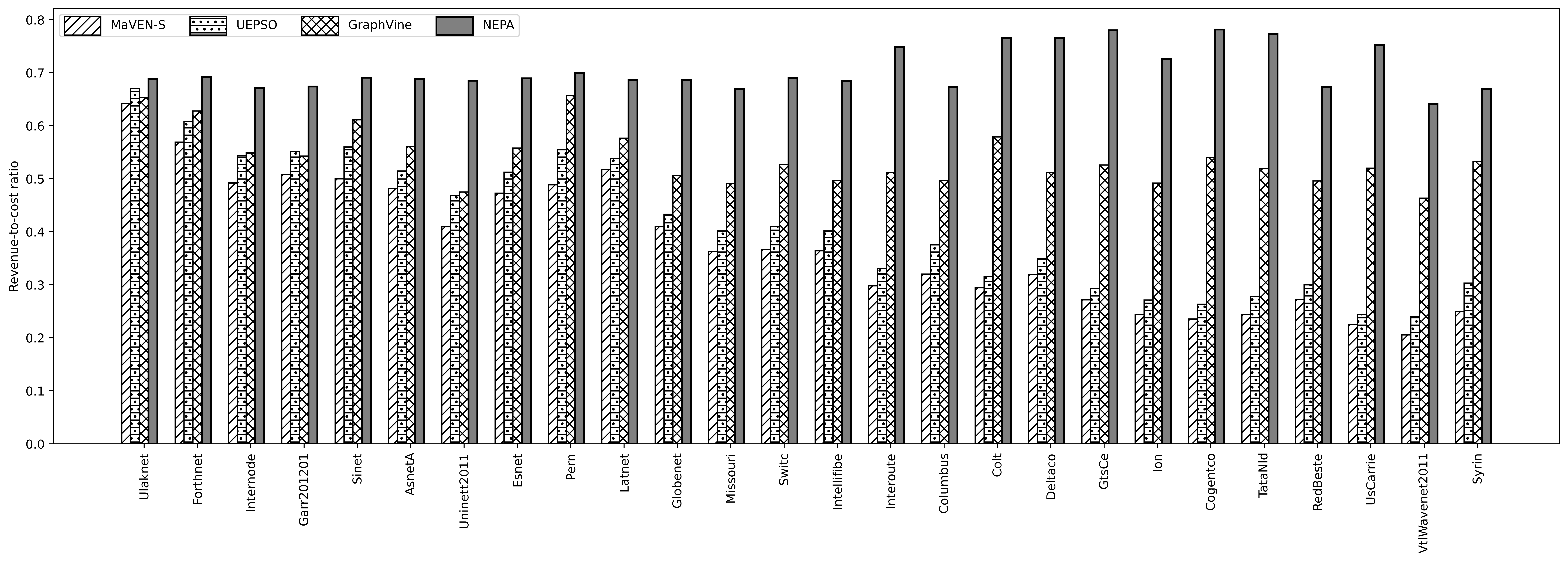}
    }
    \caption{Revenue-to-cost ratios on real physical networks}
    \label{fig:rtc_real}
\end{figure*}

The results depicted in Figures~\ref{fig:acc_real} and \ref{fig:rtc_real} show that NEPA is a lot more effective than UEPSO and MCTS. We achieve improvements of at least one order of magnitude in terms of acceptance compared to these algorithms with our best result being to more-than-triple their acceptance ratio on the Syrin topology by using NEPA. 
GraphVine interestingly performs much better on those topologies than on random ones, however NEPA still is the best in terms of acceptance rate with only few experiments where GraphVine manages to reach a similar acceptance as NEPA, an only 2 where it beats our algorithm by a thin margin. In terms of revenue-to-cost ratios, results are on par with acceptance, since again NEPA beats other algorithms (GraphVine aside) by an order of magnitude. We note that on some instances, GraphVine has a worst revenue-to-cost ratio than NEPA but still matches it in terms of acceptance (CogentCo, GtsCe, Pern, ...). This observation implies that although improving revenue-to-cost ratio is a key factor in order to reach a higher acceptance, it is not the only parameter to look for, since an approach can have a worst revenue-to-cost ratio but a better long term acceptance ratio.

Overall, our results suggest that NEPA is the best suited method compared to state of the art algorithms when it comes to placing slices on real-world networks. We note however that although the GraphVine method struggled on random topologies, it is competitive when it comes to real networks, although not as good as NEPA overall. We think it would be a great area of future research to try to combine both methods, as NEPA might be able to leverage the addition of GraphVine's neural network for reusing information learned accross experiences.
Results also suggests that the topology of the physical network has a great influence on the performances of each algorithms. 

\begin{table*}
    \centering
    \begin{tabular}{|c|c|c|c|c|}
         \hline
         & Mean distance & Diameter & \makecell{Standard deviation of shortest path length} & \makecell{Clustering coefficient}  \\
         \hline
        Correlation & 0.72  & 0.65 & 0.71 & -0.17\\
        \hline
        p-value & $3\times10^{-5}$ & $3\times10^{-4}$ & $5\times10^{-5}$ & 0.38\\
        \hline
    \end{tabular}
    \caption{Correlation between graph topological statistics and improvement ratio from NRPA to NEPA for real topologies.}
    \label{fig:correlation}
\end{table*}

Starting from that observation, we investigate the key topological features that enable NEPA to peform so much better in those cases. Our data exploration reveals that real topologies tend to have a larger diameter and mean shortest path length (\textit{e.g} overall longer paths) than generated ones. They also tend to have a lower link density (\textit{e.g.} they have "less" edges). We depict statistics for these topologies compared to generated ones in \ref{fig:stats_paths_sim}, in Appendix C. The difference between real and synthetic topologies questions the appropriateness of the widely used in the litterature Waxman Generator for VNE studies.

Furthermore, there is often a larger standard deviation in the length of shortest paths (\textit{e.g.} distances) in real topologies than in synthetic ones. We notice that the differences between NEPA (which uses distance information a lot) and other algorithms is the most important for real topologies where the standard deviation in shortest path length is the largest. For example with Syrin, where acceptance rises from 0.18 - 0.23 for MaVEN-S and UEPSO to 0.77 with NEPA, and where the shortest path length standard deviation is $6.77$. We observe the same pattern with VtlWavenet2011, UsCarrie, RedBeste, Cogentco or TataNld. On the other hand, when standard deviation is low (Ulaknet, Internode, Sinet, Forthnet, ...), we notice that the differences between algorithms are much lower, as the information to be leveraged from distances is less important, since choosing a "bad" placement would result in a smaller augmentation of the cost. Note however that NEPA still beats all other algorithms in those cases, although it is by a thinner margin. We quantify the advantage NEPA gets from exploiting distance information (\textit{e.g.} using weight initialization and refinement) by calculating the augmentation ratio between the acceptance of NEPA and the acceptance of NRPA-W (which is depicted in Appendix B) for each real-topology scenario. We choose to compare against NRPA-W as it is the same algorithm, but with no help from distance-based information during node placement. We then calculate the correlation between the augmentation ratio and different topological measures for results on the real-world topologies.

Those correlation results (obtained using Pearson correlation coefficient) are depicted in Table \ref{fig:correlation}. We find strong positive correlations of 0.65, 0.71 and 0.72 respectively for diameter, standard deviation of distances and mean distance, meaning distance information is particularly important to exploit when the physical network has a high standard deviation in the distribution of distances, such as in many of the real networks studied. This explains why our algorithm can perform so much better on these instances. This is relatively intuitive to understand: these cases correspond to instances where there are a lot of chances to make "high-cost mistakes", \textit{e.g.} where a single virtual link could incur à lot of cost by being placed on two physical nodes that are far from one another. Our distance-based techniques explicitly mitigate this by ensuring virtual nodes are placed close to one another, which results in an even greater performance boost on those cases. Also notice that in this paragraph our analysis was focused on standard deviation but applies to the other distance related metrics, as mean distance, diameter and standard deviation all have a correlation between one another of 0.99, according to our measurements.

\subsection{Specific case: Perfectly solvable scenarios}
Third, we evaluate each algorithm on perfectly solvable scenarios (PSS). A PSS is a kind of scenario proposed by Fischer \cite{PhDFischer}
that is generated such that there are only slice arrivals and no departure, and such that it is possible to place all slices. The scenario is generated so the only solutions where all slices are placed leave 0 remaining resources. Hence it is a very hard, but theoretically feasible scenario (\textit{e.g.} 100\% of acceptance is reachable). 

We argue evaluating algorithms on such scenarios is an important but often overlooked practice in the literature. Indeed, it is generally infeasible to evaluate the suboptimality gap as computing the exact placement would be computationally too expensive. 
We generate 10 PSS scenarios, using the additive algorithm from \cite{PhDFischer}
. The generation is done by first generating slices, then "adding" them in order to form the physical network. The "addition" step is done by treating each virtual node iteratively, either reusing an already created physical node or creating a new one for the current virtual node (the choice is made probabilistically). Then, once all physical nodes have been created, they are linked so that if two nodes host neighboring virtual nodes, bandwidth is added to the link between them equal to the requirement of the corresponding virtual link.

Each scenario $PSS i$ is generated from a batch of 100 random slices of random size $7+i$ to $10+i$ and a probability of reusing existing nodes of 0.93. This parameter was chosen empirically as it enabled us to generate graphs of sizes similar to those we experimented with in the previous section, as shown in Table \ref{fig:sizes_PSS}.
\begin{table*}
    \centering
    \begin{tabular}{|l|c|c|c|c|c|c|c|c|c|c|}
        \hline
        Instance & PSS0 & PSS1 & PSS2 & PSS3 & PSS4 & PSS5 & PSS6 & PSS7 & PSS8 & PSS9\\
        \hline
        $|V|$  & 67 &  86 & 73 & 94  & 111 & 104 & 124 & 121 & 135 & 150 \\
        $|V^x|$ & 7-10 & 8-11 & 9-12 & 10-13 & 11-14 & 12-15 & 13-16 & 14-17 & 15-18 & 16-19 \\ 
        \hline
    \end{tabular}
    \caption{Number of nodes for physical and virtual networks of PSS scenarios}
    \label{fig:sizes_PSS}
\end{table*}
\begin{figure*}
    \includegraphics[scale=0.365]{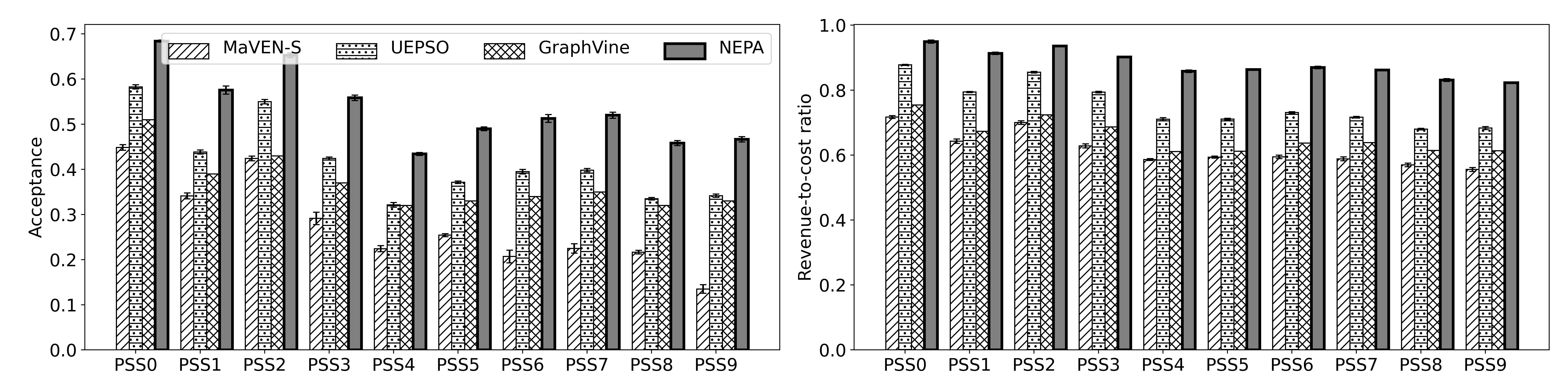}
    \caption{Acceptance and revenue-to-cost ratios for perfectly solvable scenarios}
    \label{fig:pss}
\end{figure*}
Our experiments show the same kind of results (shown in figure \ref{fig:pss}) as for the previous part, \textit{e.g.} that NEPA outperforms all other methods by an order of magnitude (consistently beating the second best method, UEPSO by accepting up to 35\% more slices in PSS9), both in terms of acceptance and revenue-to-cost ratio. However, it is striking to note that we never achieve a result of 100\% of acceptance (the best one is NEPA on PSS0 - the smallest case - with 69\%), even though the revenue-to-cost ratio gets really close to 1 (up to 0.965 in the first scenario). This means that although we achieve an almost perfect online optimization objective, resources on the physical network are badly used, leaving a lot of "holes" which are unusable. We believe this shows the need for the VNE community to investigate better reward functions which could assess the quality of a solution with other metrics than pure resource usage (with the goal to define whether a virtual network "fits" its embedding or not). In that regard, a recent article \cite{EvFunc}
made a first step in that direction by proposing to enrich the reward function with degree information, which slightly helps improving acceptance depending on the algorithm used. However, the results shown in Appendix A demonstrate this reward function has no significant impact on the results of the NEPA placement, suggesting it is ineffective when the algorithm is already very good. Another possibility would be to place virtual networks in batches, which might enable us to combine placements better, at the cost of a higher computational complexity.

\section{Discussion}

Our results illustrate that the widely adopted idea of optimizing placement for reduced bandwidth \cite{HA17} \cite{DL} \cite{GRC} \cite{graphVine} consumption in order to let more resources for future virtual networks good. We show that pushing this logic a step further by explicitly reducing the consumption of the found solution enables our algorithm to reach even better results. The main hurdle with the refinement step is the computational cost, which we overcome by selecting promising solutions to refine instead of trying to refine any solution. The NRPA algorithm is easily adaptable into NEPA due to its recursive nature. It is an open question whether other algorithms such as UEPSO could be modified in order to similarly select promising states to be refined, which would enable them to keep the computational cost low while finding better embeddings.

We shall now focus on the differences between MaVEN-S (which we call a mean-based approach) and NRPA and NEPA (which we call max-based approaches), in an effort to try to explain why max-based approaches perform so much better than the MCTS-based MaVEN-S algorithm (refer to Appendix B which shows the ablation study of NEPA, also demonstrating that NRPA without the improvements brought by NEPA outperforms MaVEN-S), while both types of algorithms are Monte Carlo Search algorithms that try to balance exploration and exploitation of the tree formed by the MDP underlying our embedding problem. 

\begin{figure}[H]
    \centering
    \includegraphics[width=0.5\textwidth]{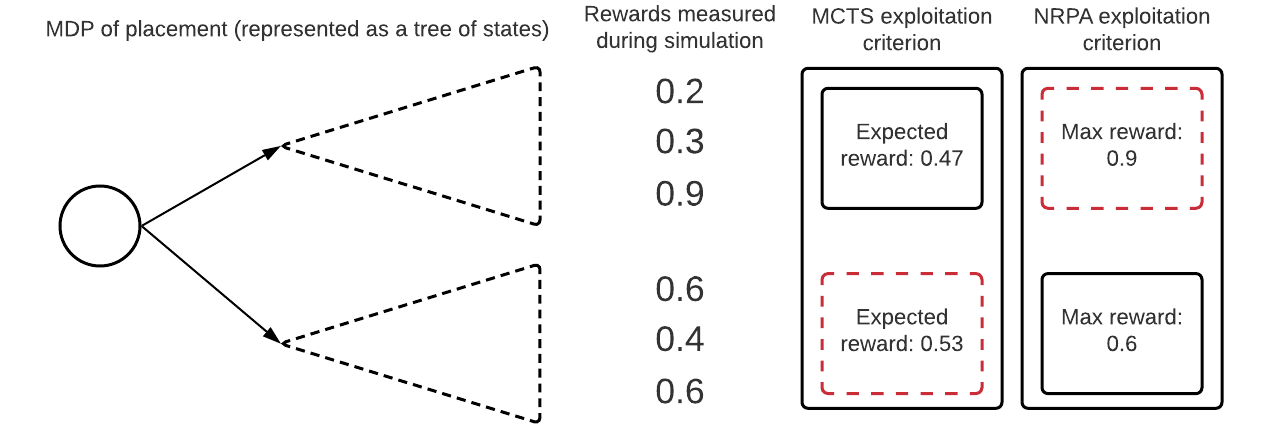}
\caption{Toy example of MDP exploration choices}
    \label{fig:tree}
\end{figure}

In figure \ref{fig:tree}, we illustrate with a toy example the potential results obtained after executing 6 random simulations (with a policy that could either be given by NRPA/NEPA's policy matrix or by MaVEN-S's tree). The tree represents the MDP, with the final values obtained through simulations at its leaves. This tree will serve us to illustrate the key difference between algorithms: max-based approaches assume that the best solutions lie near the single best solution found so far, hence they will explore regions of the search tree even with low expected value as long as they contain the best solution found so far. On the other hand, MCTS is designed to explore the states with the best expected (mean) value. Hence, on the tree from figure \ref{fig:tree}, MaVEN-S, would exploit more the states of the bottom sub-tree, since their mean value would be of 0.53, while max-based approaches would go for the top sub-tree since the maximum known value is of 0.9, even though the mean value would only be of 0.47.

We argue this is desirable for the VNE problem we solve, because optimizing for the mean expected reward is typically suited for problems where there is uncertainty, \textit{e.g.} where taking one action from a given state can yield to several different states. This is not the case for the VNE, where a choice of action from a given state always yields to the same state: the model is a deterministic MDP. Hence it makes more sense to choose actions only according to the best sequence found so far and not according to the best mean value. This is what max-based approaches do since they optimize considering the best sequence found, as opposed to mean-based MaVEN-S, which partly explains why MaVEN-S is outperformed.
\section{Conclusion}
We have proposed a new approach for the virtual network embedding problem, which consists in placing virtualized networks ("slices") on a physical infrastructure (typically a 5G network). Our algorithm builds upon the NRPA (nested rollout policy adaptation) algorithm by combining its recursive reinforcement learning procedure with neighborhood search and by initializing the weights it uses for learning based on distances in the physical network. The resulting algorithm, called NEPA (Neighborhood Enhanced Policy Adaptation with Distance-weights) shows state-of-the-art results on commonly used synthetic benchmarks as well as on real network topologies, while keeping the running time comparable to earlier algorithms. Experiments on real topologies show that since it exploits distances between placed nodes, NEPA brings acceptance and revenue-to-cost improvements of an order of magnitude compared other meta-heuristic and Monte Carlo search algorithms. It also beats a state-of-the-art graph neural network based approach (GraphVine). On random topologies, which are the most explored in the VNE litterature, we also showed that NEPA is the most robust of the tested approaches since it always reaches the best acceptance and revenue-to-cost ratio. These results will help in solving the resource allocations problems in future 5G networks, but also help the VNE community better evaluate its algorithms, since we characterized how topological features can induce enormous differences between results from different methods. In the future, we would like to demonstrate how to use NEPA on other combinatorial problems where good neighborhood search policies are also available, such as the TSP and the VRP. We also plan on incorporating offline learning by reusing the learned weights from past NEPA runs in order to learn better how to initialize future weights as currently these datas are not used once the placement is decided. In that regard, a combination with GraphVine would be particularly appealing. Implementing NEPA on a real 5G network is also planned in the near future. Finally, we contribute to the VNE community by making our set of instances and of implementations available online \cite{github}.
\bibliography{refs}

\clearpage

\section*{Appendix A: Comparison of results for NEPA with and without the Alternative Reward Function Based on Degrees}

In this Appendix, we expose our simulation results on all tested instances when we use the AFBD (alternative function based on degrees) proposed in \cite{EvFunc} in the reward function. This function uses a combination of the sum of Bandwidth used and the degrees of the used nodes as a cost function.
The formula for the AFBD cost function for virtual network $H^x$ placed on $G$ is the following :
$$ AFBD(G, H^x) = \sum\limits_{\forall (v_i, v_j) \in {E}} {\bar{BW}}_{v_i, v_j}^x + \sum\limits_{v_i^x \in V^x}deg(host(v_i^x)) - deg(v_i^x) $$
{Where $deg(v_i)$ is the degree of node $v_i$ and $host(v_i^x)$ is the physical node hosting virtual node $v_i^x$.}
~\\
The idea behind that choice is to keep minimizing the length of the used paths, but while preserving resources on high degree nodes when possible. This revolves around the intuition that higher degree nodes tend to offer more link embedding possibilities, hence, if a virtual network can be placed by using more constraining physical nodes, it should be done, since some future virtual networks might require less constrained ones in order to be placed. In \cite{EvFunc}, the authors claim to achieve improvements (in the order of a several percents of acceptance) that could be transferred to other meta-heuristic algorithms.
~\\
We try to find out if this is the case with NEPA, since after our investigations it is the best performing algorithm at our disposal. Note that the reference meta-heuristic used in \cite{EvFunc} is Harmony Search, which has comparable performances to UEPSO, as shown by the same authors in \cite{CompHS} (in that article UEPSO is referred to as PSOI and shows very close acceptance ratio with Harmony Search based methods).

\begin{figure}[H]
    \centering
    \includegraphics[scale=0.4]{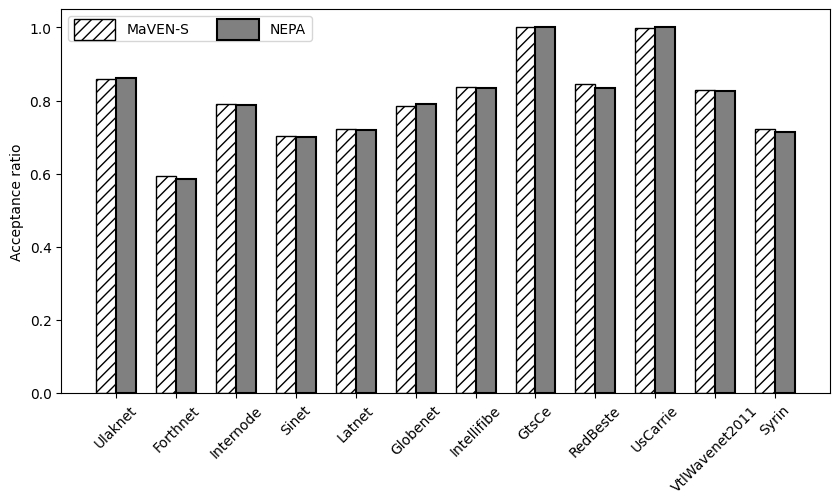}
    \caption{Acceptance for several real topologies}
    \label{fig:acc_real_afbd}
\end{figure}

\begin{figure}[H]
\begin{minipage}{.24\textwidth}
  \includegraphics[width=\textwidth]{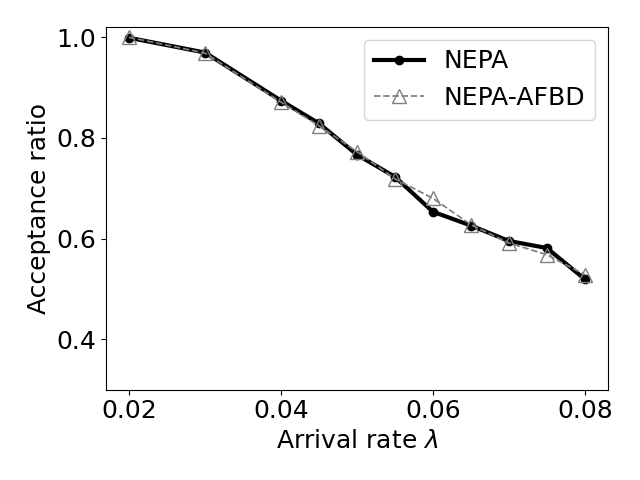}
  \centering \small 15.1 Acceptances for varying arrival rate ($\lambda$)
  \includegraphics[width=\textwidth]{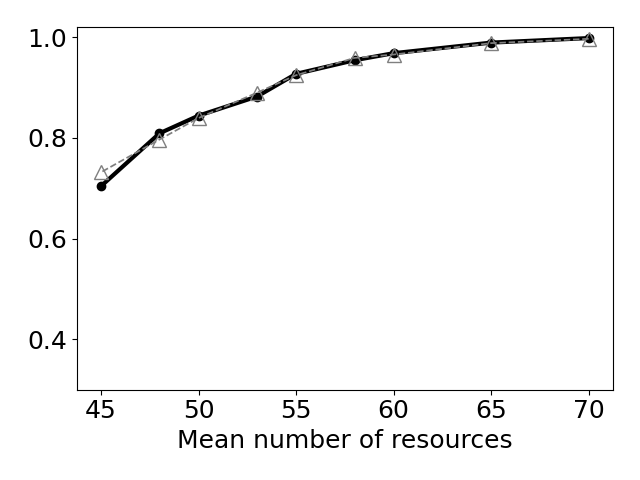}
  \centering \small 15.2 Acceptances for varying CPU/BW capacities
  \includegraphics[width=\textwidth]{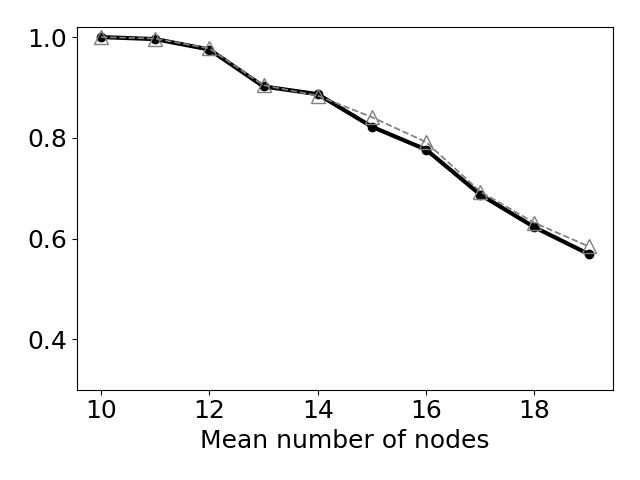}
  \centering \small 15.2 Acceptances for varying slice size
  \includegraphics[width=\textwidth]{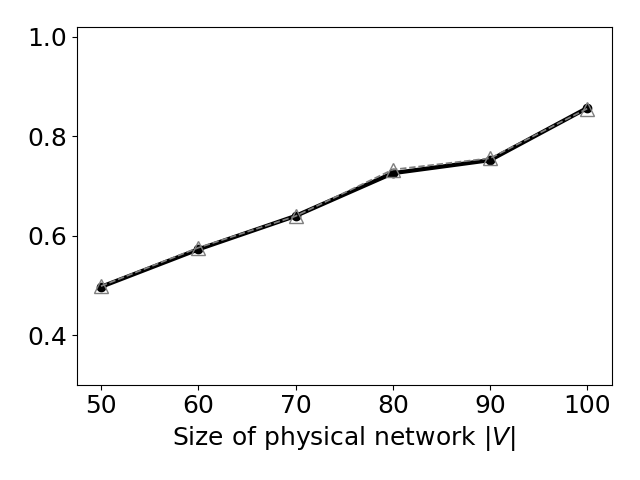}
  \centering \small 15.4 Mean acceptance ratios for varying physical network sizes
\end{minipage}
\begin{minipage}{.24\textwidth}
  \includegraphics[width=\textwidth]{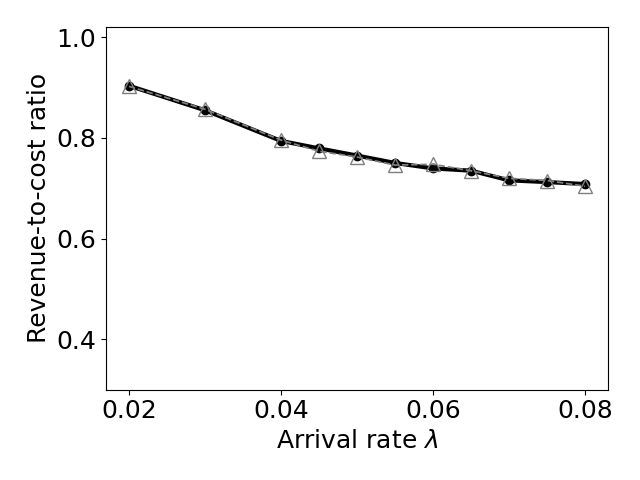}
  \centering \small 15.5 Revenue-to-cost ratio for\\ varying $\lambda$
  \includegraphics[width=\textwidth]{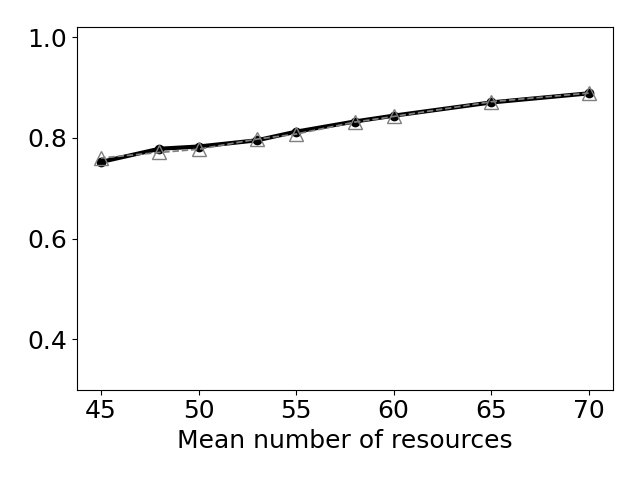}
  \centering \small 15.6 Revenue-to-cost ratio for varying CPU/BW capacities
  \includegraphics[width=\textwidth]{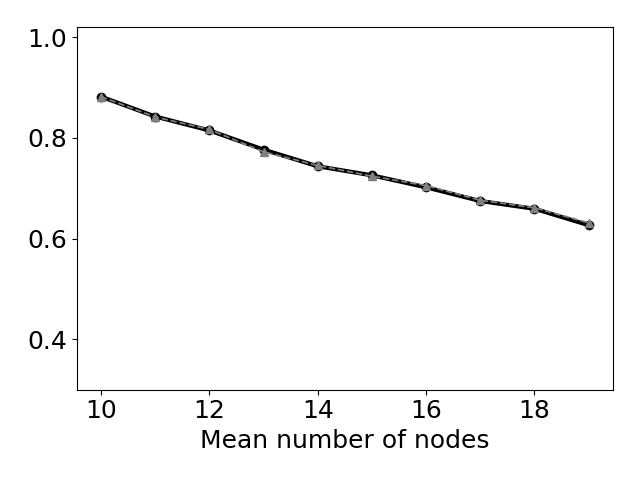}
  \centering \small 15.7 Revenue-to-cost ratio for varying slice size
  \includegraphics[width=\textwidth]{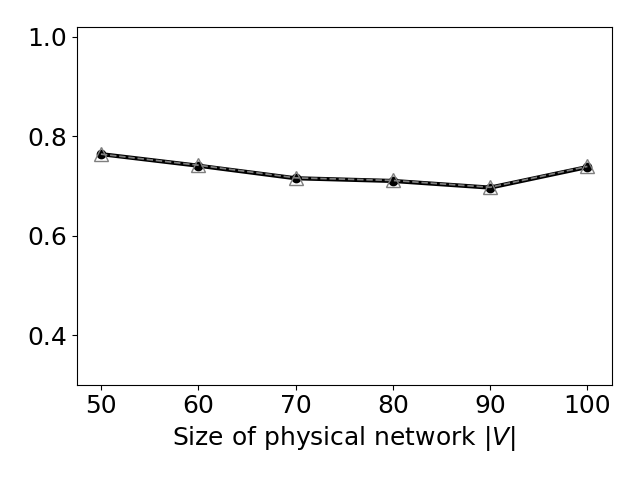}
  \centering \small 15.8 Revenue-to-cost ratio (varying network sizes)
\end{minipage}
    \caption{Results for sensitivity analysis experiments}
    \label{fig:results_sensitivity_afbd}
\end{figure}

\begin{figure}[H]
    \centering
    \includegraphics[scale=0.4]{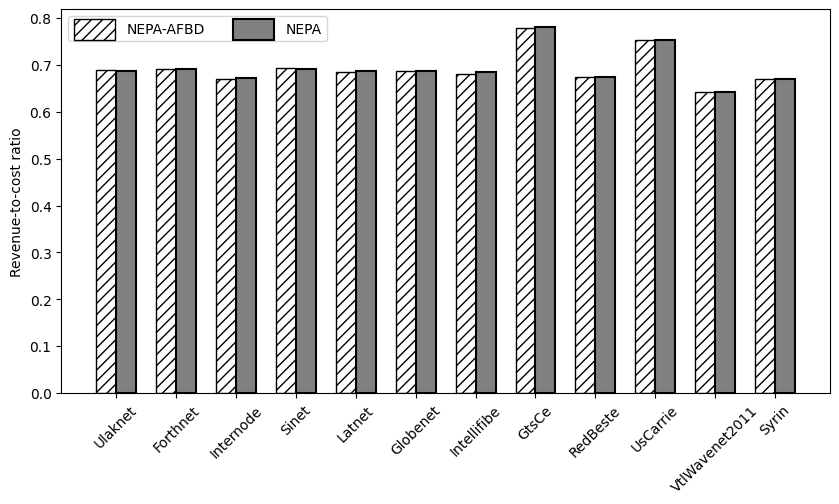}
    \caption{Revenue-to-cost ratio for several real topologies}
    \label{fig:rtc_real_afbd}
\end{figure}

Since NEPA tries to maximize a reward function, and the AFBD function is a cost function, which should be minimized, we take $\frac{1}{AFBD(G, H^x)}$ as the reward function in this appendix. We run NEPA on all the cases of section 5.2 (synthetic networks) and section 5.3 (real topologies), but only depict a subset of those instances due to space constraints (the rest can be found online in \cite{github}). Our results (figs. \ref{fig:acc_real_afbd}\ref{fig:results_sensitivity_afbd}\ref{fig:rtc_real_afbd}) show there is no significative difference when using AFBD with NEPA.
\vspace*{0.6cm}

\clearpage

\clearpage
\section*{Appendix B: Ablation study of NEPA components}
In this appendix, we compare NEPA and NRPA with their counterpart that do not use weight initialization. The presented results were ran using the parameters and instance generation described in the main article. However, due to space constraints, we only present a representative subset (10 topologies) of the ablation study experiments for real networks. The rest can be found on \cite{github}. We call NRPA-W and NEPA-W the versions of our algorithms that do not use our weight initialization function (but initialize all weights to 0)

Note that we chose parameters $l=3$, $N=7$ for NRPA and NRPA-W, as we observed that this resulted in the same runtime for NEPA/NEPA-W (with parameters $l=3, N=5$) and these approches.

\begin{figure}[H]
    \centering
    \includegraphics[scale=0.4]{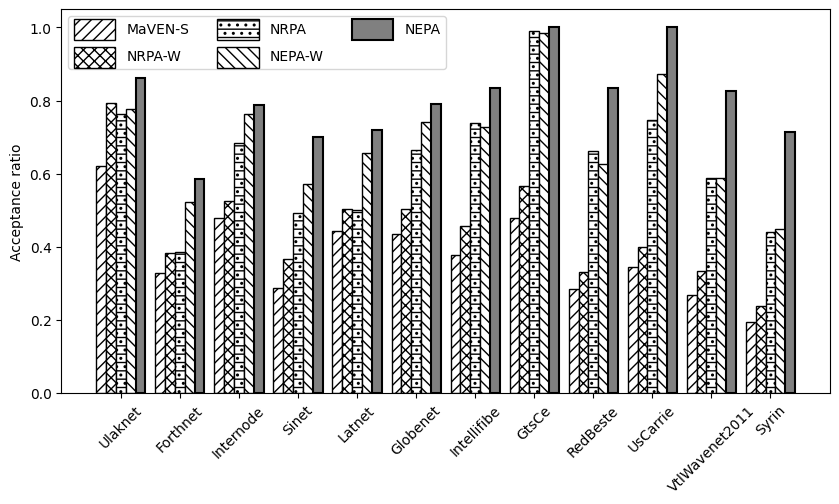}
    \caption{Acceptance for several real topologies}
    \label{fig:acc_real_apdx}
\end{figure}

\begin{figure}[H]
\begin{minipage}{.24\textwidth}
  \includegraphics[width=\textwidth]{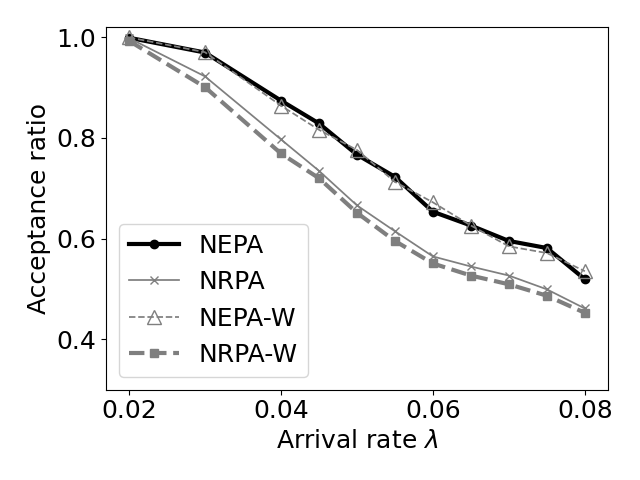}
  \centering \small 18.1 Acceptances for varying arrival rate ($\lambda$)
  \includegraphics[width=\textwidth]{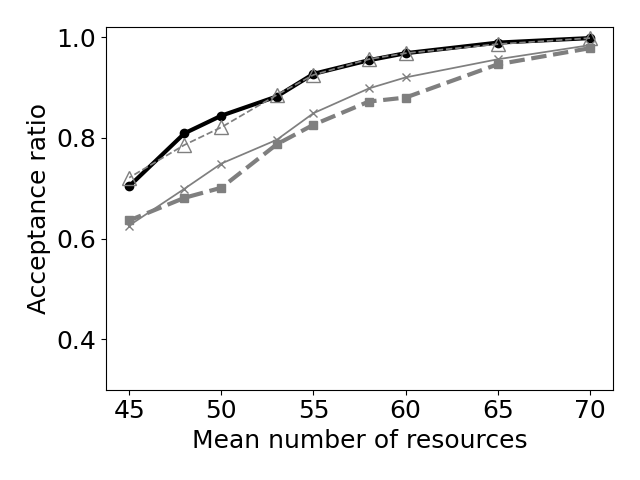}
  \centering \small 18.2 Acceptances for varying CPU/BW capacities
  \includegraphics[width=\textwidth]{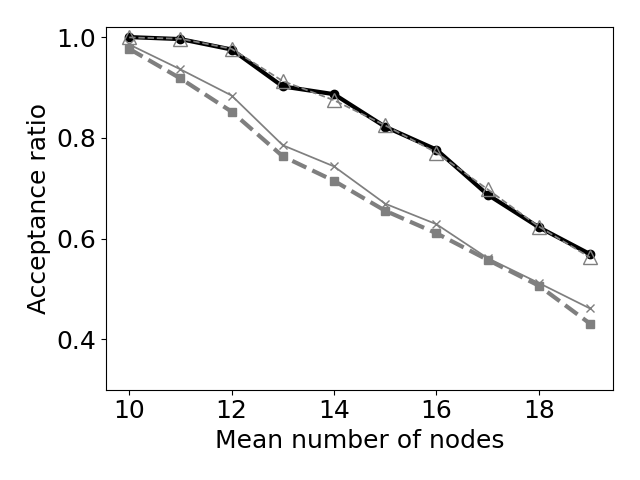}
  \centering \small 18.2 Acceptances for varying slice size
  \includegraphics[width=\textwidth]{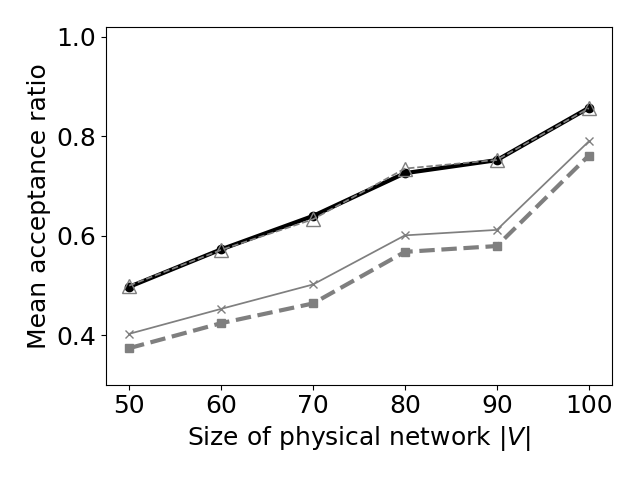}
  \centering \small 18.4 Mean acceptance ratios for varying physical network sizes
\end{minipage}
\begin{minipage}{.24\textwidth}
  \includegraphics[width=\textwidth]{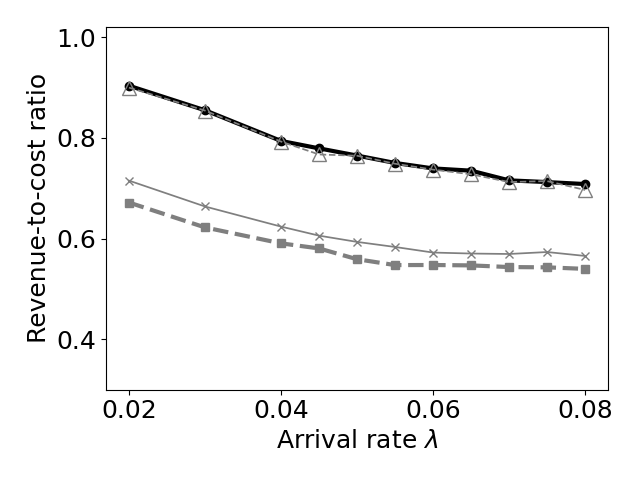}
  \centering \small 18.5 Revenue-to-cost ratio for\\ varying $\lambda$
  \includegraphics[width=\textwidth]{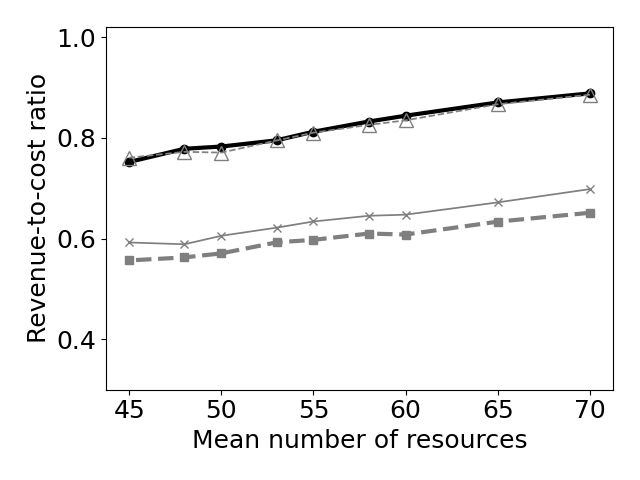}
  \centering \small 18.6 Revenue-to-cost ratio for varying CPU/BW capacities
  \includegraphics[width=\textwidth]{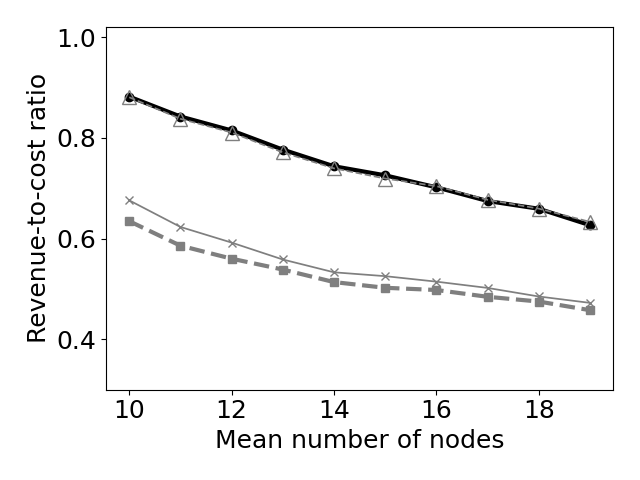}
  \centering \small 18.7 Revenue-to-cost ratio for varying slice size
  \includegraphics[width=\textwidth]{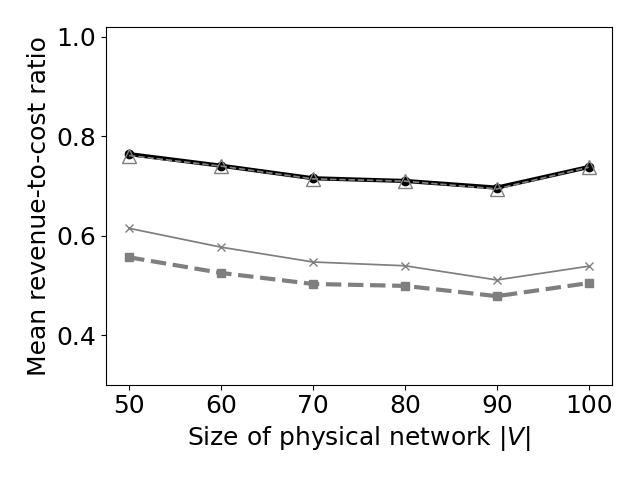}
  \centering \small 18.8 Revenue-to-cost ratio (varying network sizes)
\end{minipage}
    \caption{Results for sensitivity analysis experiments}
    \label{fig:results_sensitivity_apdx}
\end{figure}

\begin{figure}[H]
    \centering
    \includegraphics[scale=0.4]{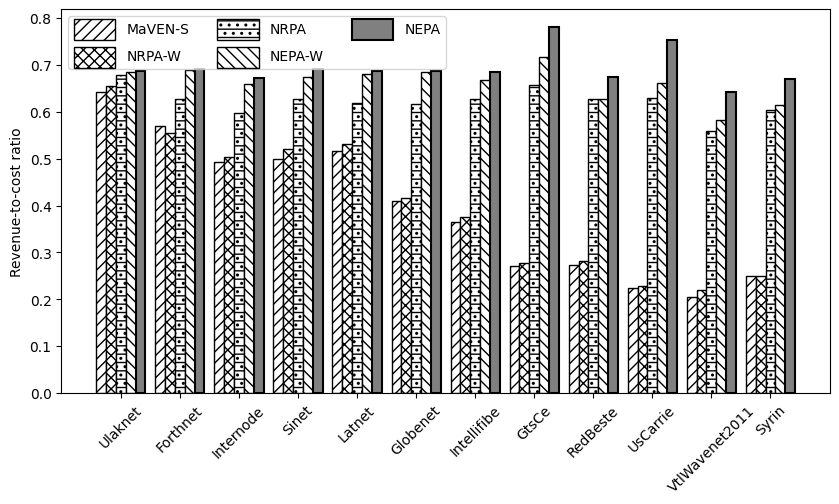}
    \caption{Revenue-to-cost ratio for several real topologies}
    \label{fig:rtc_real_apdx}
\end{figure}

Our results from figure \ref{fig:results_sensitivity_apdx} show that NRPA outperforms standard NRPA-W by a thin margin on random topologies, hinting that the weight heuristic helps slightly for improving the results. For NEPA-W, there is no significative difference in results for randomly generated topologies compared to NEPA, meaning that the refining operation already exploits well the shortest path information. However, on real topologies (figure \ref{fig:acc_real_apdx}), the difference is more significative. NEPA outperforms NEPA-W by a large acceptance margin on several topologies, such as Pern and UsCarrie. In terms of revenue-to-cost ratios (figure \ref{fig:rtc_real_apdx}, NEPA also outperforms NEPA-W on real topologies. It means on some real topologies, the extra exploitation of shortest path information is of importance. This makes sense intuitively as, as we have seen before in the paper, real topologies typically have larger diameters, mean distances and distance standard deviation, meaning an error of placement related to distances can be much more costly that on randomly generated topologies.

When we compare NRPA and NRPA-W on real topology, we also observe that on most of the cases, NRPA beats NRPA-W both in terms of acceptance and revenue-to-cost ratio by a larger margin than what was observed with random topologies. Similarly here, we believe this is largely due to less errors caused by choosing distant nodes when other less costly solutions existed.

Finally, we observe that on all cases, NEPA significantly outperforms NRPA, meaning that neigborhood-based enhancement is effective at finding better solutions.
\clearpage
\begin{figure*}
\section*{Appendix C: statistics on network topologies}
\end{figure*}
\begin{table*}
\center
    \begin{tabular}{|c|c|c|c|c|}
        \hline
        Instance & \makecell{Mean distance} & Diameter & \makecell{Distance Standard Deviation} & Clustering Coefficient \\ \hline
        Intellifibe & 6.33 & 15 & 2.99 & 0.088 \\ \hline
        Latnet & 3.96 & 12 & 2.17 & 0.058 \\ \hline
        VtlWavenet2011 & 13.07 & 31 & 6.50 & 0.000 \\ \hline
        Syrin & 11.95 & 31 & 6.77 & 0.000 \\ \hline
        Globenet & 5.23 & 15 & 2.46 & 0.081 \\ \hline
        Forthnet & 3.28 & 7 & 1.06 & 0.016 \\ \hline
        GtsCe & 9.83 & 21 & 4.18 & 0.082 \\ \hline
        Ulaknet & 2.44 & 4 & 0.59 & 0.000 \\ \hline
        Sinet & 3.93 & 7 & 1.41 & 0.000 \\ \hline
        Internode & 3.60 & 6 & 1.14 & 0.013 \\ \hline
        UsCarrie & 12.09 & 35 & 6.46 & 0.058 \\ \hline
        RedBeste & 10.59 & 28 & 5.66 & 0.009 \\ \hline
        Missouri & 6.23 & 14 & 2.73 & 0.025 \\ \hline
        Interoute & 7.62 & 17 & 3.39 & 0.105 \\ \hline
        Columbus & 7.24 & 18 & 3.62 & 0.045 \\ \hline
        Garr201201 & 3.62 & 8 & 1.28 & 0.054 \\ \hline
        Cogentco & 10.51 & 28 & 5.10 & 0.012 \\ \hline
        Deltaco & 7.16 & 23 & 3.80 & 0.107 \\ \hline
        AsnetA & 3.78 & 8 & 1.43 & 0.127 \\ \hline
        Switc & 6.09 & 13 & 2.80 & 0.110 \\ \hline
        Uninett2011 & 4.25 & 9 & 1.62 & 0.018 \\ \hline
        Ion & 10.14 & 25 & 4.79 & 0.011 \\ \hline
        Pern & 4.55 & 8 & 1.89 & 0.004 \\ \hline
        Esnet & 4.32 & 9 & 1.68 & 0.034 \\ \hline
        Colt & 9.35 & 20 & 3.70 & 0.040 \\ \hline
        TataNld & 9.85 & 28 & 5.17 & 0.065 \\ \hline
        
    \end{tabular}
    \caption{Statistics on each real topology}
    \label{fig:stats_paths}
\end{table*}
\begin{table*}
\center
    \begin{tabular}{|c|c|c|c|c|}
    \hline
    Instance & Mean Distance & Diameter & Distance standard deviation & Clustering coefficient\\ \hline
    Waxman 50 & 2.58 & 6 & 0.88 & 0.169 \\ \hline
    Waxman 60 & 2.45 & 4 & 0.76 & 0.108 \\ \hline
    Waxman 70 & 2.50 & 5 & 0.77 & 0.152 \\ \hline
    Waxman 80 & 2.41 & 5 & 0.73 & 0.142 \\ \hline
    Waxman 90 & 2.44 & 4 & 0.70 & 0.127 \\ \hline
    Waxman 100 & 2.27 & 4 & 0.66 & 0.160 \\ \hline
    Erdos-Renyi 0.03 & 4.33 & 9 & 1.60 & 0.032 \\ \hline
    Erdos-Renyi 0.04 & 3.66 & 8 & 1.27 & 0.042 \\ \hline
    Erdos-Renyi 0.06 & 3.27 & 7 & 1.09 & 0.019 \\ \hline
    Erdos-Renyi 0.08 & 2.74 & 6 & 0.87 & 0.054 \\ \hline
    Erdos-Renyi 0.11 & 2.34 & 4 & 0.68 & 0.109 \\ \hline
    Erdos-Renyi 0.16 & 2.01 & 3 & 0.55 & 0.142 \\ \hline
    Erdos-Renyi 0.2 & 1.89 & 3 & 0.49 & 0.182 \\ \hline
    PSS 1 & 1.93 & 4 & 0.62 & 0.385 \\ \hline
    PSS 2 & 1.82 & 4 & 0.61 & 0.476 \\ \hline
    PSS 3 & 1.91 & 4 & 0.59 & 0.376 \\ \hline
    PSS 4 & 1.94 & 4 & 0.59 & 0.358 \\ \hline
    PSS 5 & 1.83 & 4 & 0.56 & 0.425 \\ \hline
    PSS 6 & 1.98 & 4 & 0.62 & 0.413 \\ \hline
    PSS 7 & 1.92 & 4 & 0.60 & 0.398 \\ \hline
    PSS 8 & 1.91 & 4 & 0.58 & 0.360 \\ \hline
    \end{tabular}
    \caption{Statistics on example simulated topologies}
    \label{fig:stats_paths_sim}
\end{table*}

\end{document}

%% file: fig_mdp.tex
\begin{tikzpicture}[auto, thick]
\node[circle, draw, minimum size=1.5cm] (s0) at (0,0) {$s(0)$};
\node[circle, draw, minimum size=1.5cm] (s1) at (4,0) {$s(1)$};
\node[circle] (s2) at (8,0) {...};
\node[circle, draw, minimum size=1.5cm] (s3) at (13.5,0) {$s(k^{end})$};
\node[draw, rounded corners=0.2cm, text width=2.85cm] (s4) at (16.5,0) {Compute terminal reward (optimize link placement)};
\node[draw, rounded corners=0.2cm, text width=2.85cm] (s5) at (20,1) {Success: feasible embedding found};
\node[draw, rounded corners=0.2cm, text width=2.85cm] (s6) at (20,-1) {Failure due to link placement};
\node[] (s7) at (13.5,-1.2) {$s_a(k^{end})=\emptyset$};
\node[] (s7) at (13.5,-4.2) {$s_a(k^{end})\neq\emptyset$};

\draw[->] (s0) -- (s1) node [midway, above] {$R(s(0), a_0) = 0$} node [midway, below] {$a_0$};
\draw[->] (s1) -- (s2) node [midway, above] {$R(s(1), a_1) = 0$} node [midway, below] {$a_1$};
\draw[->] (s2) -- (s3) node[midway, above] {$R(s(k^{end}-1), a_{k^{end}-1})$} node[midway, below] (l1) {$a_{k^{end}-1}$};
\draw[->,dashed] (s3) -- (s4);
\draw[->,dashed] (s4) -- (s5);
\draw[->,dashed] (s4) -- (s6);
\draw[->,dashed] (s5) |- ++(-1, 1) -| (10.8, 0.65);

\node[] (a1) at (18.35-2, 2.3) {$R(s(k^{end}-1), a_{k^{end}-1}) = \frac{r_x}{c_x}$};
\node[] (a1) at (18.3-2, 3) {$R(s(k^{end}-1), a_{k^{end}-1}) = 0$};
\draw[->, dashed] (s6.east) -| ++(1, 4.3) -| (10.5, 0.65);

\node[circle, draw, minimum size=1.5cm] (s0b) at (0,-3) {$s(0)$};
\node[circle, draw, minimum size=1.5cm] (s1b) at (4,-3) {$s(1)$};
\node[circle] (s2b) at (8,-3) {...};
\node[circle, draw, minimum size=1.5cm] (s3b) at (13.5,-3) {$s(k^{end})$};
\node[draw, rounded corners=0.2cm, text width=2.85cm] (s4b) at (16.5,-3) {Failure due to node placement};

\draw[->] (s0b) -- (s1b) node [midway, above] {$R(s(0), a_0) = 0$} node [midway, below] {$a_0$};
\draw[->] (s1b) -- (s2b) node [midway, above] {$R(s(1), a_1) = 0$} node [midway, below] {$a_1$};
\draw[->] (s2b) -- (s3b) node [midway, above] {$R(s(k^{end}-1), a_{k^{end}-1})=0$} node [midway, below] {$a_{k^{end}-1}$};
\draw[->,dashed] (s3b) -- (s4b);

\end{tikzpicture}

%% file: root_tikz.tex
\begin{tikzpicture}[auto, thick]
  \node[cblue, label=above left:12] (a1) at (0,0) {\textit{1}};
  \node[cblue, label=below:8] (a2) at (1,-1) {\textit{2}};
  \node[cblue, label=above right:15] (a3) at (2,0) {\textit{3}};
  \node[cblue, label=below:10] (a4) at (1,1) {\textit{4}};
  

  \path[thin] (a1) edge node [draw=black, midway, yshift=2ex, below left=6pt]  {$9$}  (a2);
  \path[thin] (a2) edge node [draw=black, midway, yshift=2ex, below right=6pt]  {$8$}  (a3);
  \path[thin] (a4) edge node [draw=black, midway, yshift=2ex, below=2pt, right]  {$0$}  (a3);
  \path[thin] (a1) edge node [draw=black, midway, yshift=2ex, below=2pt, left]  {$9$}  (a4);
  \path[thin] (a1) edge node [draw=black, midway, yshift=2ex, below=9.5pt]  {$7$}  (a3);
  
  \node[qgre, label=10] (b1) at (0,1.5) {\textit{1}};
  \node[qgre, label=9] (b2) at (1,2) {\textit{2}};
  \node[qgre, label=14] (b3) at (2,1.5) {\textit{3}};
  
  \path[thin] (b1) edge node [draw=black, midway, yshift=2ex, above=-5pt]  {$1$}  (b2);
  \path[thin] (b2) edge node [draw=black, midway, yshift=2ex, above=-5pt]  {$2$}  (b3);
\end{tikzpicture}

%% file: l1_1.tex
\begin{tikzpicture}[auto, thick]
  \node[cblue, label=above left:12] (a1) at (0,0) {\textit{1}};
  \node[cblue, label=below:8] (a2) at (1,-1) {\textit{2}};
  \node[cblue, label=above right:15] (a3) at (2,0) {\textit{3}};
  \node[cblue, label=below:10] (a4) at (1,1) {\textit{4}};
  

  \path[thin] (a1) edge node [draw=black, midway, yshift=2ex, below left=6pt]  {$9$}  (a2);
  \path[thin] (a2) edge node [draw=black, midway, yshift=2ex, below right=6pt]  {$8$}  (a3);
  \path[thin] (a4) edge node [draw=black, midway, yshift=2ex, below=2pt, right]  {$0$}  (a3);
  \path[thin] (a1) edge node [draw=black, midway, yshift=2ex, below=2pt, left]  {$9$}  (a4);
  \path[thin] (a1) edge node [draw=black, midway, yshift=2ex, below=9.5pt]  {$7$}  (a3);
  
  \node[qgre, label=10] (b1) at (0,1.5) {\textit{1}};
  \node[qgre, label=9] (b2) at (1,2) {\textit{2}};
  \node[qgre, label=14] (b3) at (2,1.5) {\textit{3}};
  
  \path[thin] (b1) edge node [draw=black, midway, yshift=2ex, above=-5pt]  {$1$}  (b2);
  \path[thin] (b2) edge node [draw=black, midway, yshift=2ex, above=-5pt]  {$2$}  (b3);
  
  \path[rpath, line width=4] (b1) edge node [midway, yshift=2ex, above=-5pt]  {}  (a1);
\end{tikzpicture}

%% file: l2_1.tex
\begin{tikzpicture}[auto, thick]
  \node[cblue, label=above left:12] (a1) at (0,0) {\textit{1}};
  \node[cblue, label=below:8] (a2) at (1,-1) {\textit{2}};
  \node[cblue, label=above right:15] (a3) at (2,0) {\textit{3}};
  \node[cblue, label=below:10] (a4) at (1,1) {\textit{4}};
  

  \path[thin] (a1) edge node [draw=black, midway, yshift=2ex, below left=6pt]  {$9$}  (a2);
  \path[thin] (a2) edge node [draw=black, midway, yshift=2ex, below right=6pt]  {$8$}  (a3);
  \path[thin] (a4) edge node [draw=black, midway, yshift=2ex, below=2pt, right]  {$0$}  (a3);
  \path[thin] (a1) edge node [draw=black, midway, yshift=2ex, below=2pt, left]  {$9$}  (a4);
  \path[thin] (a1) edge node [draw=black, midway, yshift=2ex, below=9.5pt]  {$7$}  (a3);
  
  \node[qgre, label=10] (b1) at (0,1.5) {\textit{1}};
  \node[qgre, label=9] (b2) at (1,2) {\textit{2}};
  \node[qgre, label=14] (b3) at (2,1.5) {\textit{3}};
  
  \path[thin] (b1) edge node [draw=black, midway, yshift=2ex, above=-5pt]  {$1$}  (b2);
  \path[thin] (b2) edge node [draw=black, midway, yshift=2ex, above=-5pt]  {$2$}  (b3);
  
  \path[rpath] (b1) edge node [midway, yshift=2ex, above=-5pt]  {}  (a1);
  \path[rpath, line width=4] (b2) edge node [midway, yshift=2ex, above=-5pt]  {}  (a4);
\end{tikzpicture}

%% file: l3_1.tex
\begin{tikzpicture}[auto, thick]
  \node[cblue, label=above left:12] (a1) at (0,0) {\textit{1}};
  \node[cblue, label=below:8] (a2) at (1,-1) {\textit{2}};
  \node[cblue, label=above right:15] (a3) at (2,0) {\textit{3}};
  \node[cblue, label=below:10] (a4) at (1,1) {\textit{4}};
  

  \path[thin] (a1) edge node [draw=black, midway, yshift=2ex, below left=6pt]  {$9$}  (a2);
  \path[thin] (a2) edge node [draw=black, midway, yshift=2ex, below right=6pt]  {$8$}  (a3);
  \path[thin] (a4) edge node [draw=black, midway, yshift=2ex, below=2pt, right]  {$0$}  (a3);
  \path[thin] (a1) edge node [draw=black, midway, yshift=2ex, below=2pt, left]  {$9$}  (a4);
  \path[thin] (a1) edge node [draw=black, midway, yshift=2ex, below=9.5pt]  {$7$}  (a3);
  
  \node[qgre, label=10] (b1) at (0,1.5) {\textit{1}};
  \node[qgre, label=9] (b2) at (1,2) {\textit{2}};
  \node[qgre, label=14] (b3) at (2,1.5) {\textit{3}};
  
  \path[thin] (b1) edge node [draw=black, midway, yshift=2ex, above=-5pt]  {$1$}  (b2);
  \path[thin] (b2) edge node [draw=black, midway, yshift=2ex, above=-5pt]  {$2$}  (b3);
  
  \path[rpath] (b1) edge node [midway, yshift=2ex, above=-5pt]  {}  (a1);
  \path[rpath] (b2) edge node [midway, yshift=2ex, above=-5pt]  {}  (a4);
  \path[rpath, line width=4] (b3) edge node [midway, yshift=2ex, above=-5pt]  {}  (a3);
\end{tikzpicture}

%% file: l2_2.tex
\begin{tikzpicture}[auto, thick]
  \node[cblue, label=above left:12] (a1) at (0,0) {\textit{1}};
  \node[cblue, label=below:8] (a2) at (1,-1) {\textit{2}};
  \node[cblue, label=above right:15] (a3) at (2,0) {\textit{3}};
  \node[cblue, label=below:10] (a4) at (1,1) {\textit{4}};
  

  \path[thin] (a1) edge node [draw=black, midway, yshift=2ex, below left=6pt]  {$9$}  (a2);
  \path[thin] (a2) edge node [draw=black, midway, yshift=2ex, below right=6pt]  {$8$}  (a3);
  \path[thin] (a4) edge node [draw=black, midway, yshift=2ex, below=2pt, right]  {$0$}  (a3);
  \path[thin] (a1) edge node [draw=black, midway, yshift=2ex, below=2pt, left]  {$9$}  (a4);
  \path[thin] (a1) edge node [draw=black, midway, yshift=2ex, below=9.5pt]  {$7$}  (a3);
  
  \node[qgre, label=10] (b1) at (0,1.5) {\textit{1}};
  \node[qgre, label=9] (b2) at (1,2) {\textit{2}};
  \node[qgre, label=14] (b3) at (2,1.5) {\textit{3}};
  
  \path[thin] (b1) edge node [draw=black, midway, yshift=2ex, above=-5pt]  {$1$}  (b2);
  \path[thin] (b2) edge node [draw=black, midway, yshift=2ex, above=-5pt]  {$2$}  (b3);
  
  \path[rpath] (b1) edge node [midway, yshift=2ex, above=-5pt]  {}  (a1);
  \path[rpath, line width=4] (b2) edge node [midway, yshift=2ex, above=-5pt]  {}  (a3);
\end{tikzpicture}

%% file: l1_2.tex
\begin{tikzpicture}[auto, thick]
  \node[cblue, label=above left:12] (a1) at (0,0) {\textit{1}};
  \node[cblue, label=below:8] (a2) at (1,-1) {\textit{2}};
  \node[cblue, label=above right:15] (a3) at (2,0) {\textit{3}};
  \node[cblue, label=below:10] (a4) at (1,1) {\textit{4}};
  

  \path[thin] (a1) edge node [draw=black, midway, yshift=2ex, below left=6pt]  {$9$}  (a2);
  \path[thin] (a2) edge node [draw=black, midway, yshift=2ex, below right=6pt]  {$8$}  (a3);
  \path[thin] (a4) edge node [draw=black, midway, yshift=2ex, below=2pt, right]  {$0$}  (a3);
  \path[thin] (a1) edge node [draw=black, midway, yshift=2ex, below=2pt, left]  {$9$}  (a4);
  \path[thin] (a1) edge node [draw=black, midway, yshift=2ex, below=9.5pt]  {$7$}  (a3);
  
  \node[qgre, label=10] (b1) at (0,1.5) {\textit{1}};
  \node[qgre, label=9] (b2) at (1,2) {\textit{2}};
  \node[qgre, label=14] (b3) at (2,1.5) {\textit{3}};
  
  \path[thin] (b1) edge node [draw=black, midway, yshift=2ex, above=-5pt]  {$1$}  (b2);
  \path[thin] (b2) edge node [draw=black, midway, yshift=2ex, above=-5pt]  {$2$}  (b3);
  
  \path[rpath, line width=4] (b1) edge node [midway, yshift=2ex, above=-5pt]  {}  (a4);
\end{tikzpicture}

%% file: l2_3.tex
\begin{tikzpicture}[auto, thick]
  \node[cblue, label=above left:12] (a1) at (0,0) {\textit{1}};
  \node[cblue, label=below:8] (a2) at (1,-1) {\textit{2}};
  \node[cblue, label=above right:15] (a3) at (2,0) {\textit{3}};
  \node[cblue, label=below:10] (a4) at (1,1) {\textit{4}};
  

  \path[thin] (a1) edge node [draw=black, midway, yshift=2ex, below left=6pt]  {$9$}  (a2);
  \path[thin] (a2) edge node [draw=black, midway, yshift=2ex, below right=6pt]  {$8$}  (a3);
  \path[thin] (a4) edge node [draw=black, midway, yshift=2ex, below=2pt, right]  {$0$}  (a3);
  \path[thin] (a1) edge node [draw=black, midway, yshift=2ex, below=2pt, left]  {$9$}  (a4);
  \path[thin] (a1) edge node [draw=black, midway, yshift=2ex, below=9.5pt]  {$7$}  (a3);
  
  \node[qgre, label=10] (b1) at (0,1.5) {\textit{1}};
  \node[qgre, label=9] (b2) at (1,2) {\textit{2}};
  \node[qgre, label=14] (b3) at (2,1.5) {\textit{3}};
  
  \path[thin] (b1) edge node [draw=black, midway, yshift=2ex, above=-5pt]  {$1$}  (b2);
  \path[thin] (b2) edge node [draw=black, midway, yshift=2ex, above=-5pt]  {$2$}  (b3);
  
  \path[rpath] (b1) edge node [midway, yshift=2ex, above=-5pt]  {}  (a4);
  \path[rpath, line width=4] (b2) edge node [midway, yshift=2ex, above=-5pt]  {}  (a1);
\end{tikzpicture}

%% file: l3_2.tex
\begin{tikzpicture}[auto, thick]
  \node[cblue, label=above left:12] (a1) at (0,0) {\textit{1}};
  \node[cblue, label=below:8] (a2) at (1,-1) {\textit{2}};
  \node[cblue, label=above right:15] (a3) at (2,0) {\textit{3}};
  \node[cblue, label=below:10] (a4) at (1,1) {\textit{4}};
  

  \path[thin] (a1) edge node [draw=black, midway, yshift=2ex, below left=6pt]  {$9$}  (a2);
  \path[thin] (a2) edge node [draw=black, midway, yshift=2ex, below right=6pt]  {$8$}  (a3);
  \path[thin] (a4) edge node [draw=black, midway, yshift=2ex, below=2pt, right]  {$0$}  (a3);
  \path[thin] (a1) edge node [draw=black, midway, yshift=2ex, below=2pt, left]  {$9$}  (a4);
  \path[thin] (a1) edge node [draw=black, midway, yshift=2ex, below=9.5pt]  {$7$}  (a3);
  
  \node[qgre, label=10] (b1) at (0,1.5) {\textit{1}};
  \node[qgre, label=9] (b2) at (1,2) {\textit{2}};
  \node[qgre, label=14] (b3) at (2,1.5) {\textit{3}};
  
  \path[thin] (b1) edge node [draw=black, midway, yshift=2ex, above=-5pt]  {$1$}  (b2);
  \path[thin] (b2) edge node [draw=black, midway, yshift=2ex, above=-5pt]  {$2$}  (b3);
  
  \path[rpath] (b1) edge node [midway, yshift=2ex, above=-5pt]  {}  (a4);
  \path[rpath] (b2) edge node [midway, yshift=2ex, above=-5pt]  {}  (a1);
  \path[rpath, line width=4] (b3) edge node [midway, yshift=2ex, above=-5pt]  {}  (a3);
\end{tikzpicture}

%% file: l2_4.tex
\begin{tikzpicture}[auto, thick]
  \node[cblue, label=above left:12] (a1) at (0,0) {\textit{1}};
  \node[cblue, label=below:8] (a2) at (1,-1) {\textit{2}};
  \node[cblue, label=above right:15] (a3) at (2,0) {\textit{3}};
  \node[cblue, label=below:10] (a4) at (1,1) {\textit{4}};
  

  \path[thin] (a1) edge node [draw=black, midway, yshift=2ex, below left=6pt]  {$9$}  (a2);
  \path[thin] (a2) edge node [draw=black, midway, yshift=2ex, below right=6pt]  {$8$}  (a3);
  \path[thin] (a4) edge node [draw=black, midway, yshift=2ex, below=2pt, right]  {$0$}  (a3);
  \path[thin] (a1) edge node [draw=black, midway, yshift=2ex, below=2pt, left]  {$9$}  (a4);
  \path[thin] (a1) edge node [draw=black, midway, yshift=2ex, below=9.5pt]  {$7$}  (a3);
  
  \node[qgre, label=10] (b1) at (0,1.5) {\textit{1}};
  \node[qgre, label=9] (b2) at (1,2) {\textit{2}};
  \node[qgre, label=14] (b3) at (2,1.5) {\textit{3}};
  
  \path[thin] (b1) edge node [draw=black, midway, yshift=2ex, above=-5pt]  {$1$}  (b2);
  \path[thin] (b2) edge node [draw=black, midway, yshift=2ex, above=-5pt]  {$2$}  (b3);
  
  \path[rpath] (b1) edge node [midway, yshift=2ex, above=-5pt]  {}  (a4);
  \path[rpath, line width=4] (b2) edge node [midway, yshift=2ex, above=-5pt]  {}  (a3);
\end{tikzpicture}

%% file: l1_3.tex
\begin{tikzpicture}[auto, thick]
  \node[cblue, label=above left:12] (a1) at (0,0) {\textit{1}};
  \node[cblue, label=below:8] (a2) at (1,-1) {\textit{2}};
  \node[cblue, label=above right:15] (a3) at (2,0) {\textit{3}};
  \node[cblue, label=below:10] (a4) at (1,1) {\textit{4}};
  

  \path[thin] (a1) edge node [draw=black, midway, yshift=2ex, below left=6pt]  {$9$}  (a2);
  \path[thin] (a2) edge node [draw=black, midway, yshift=2ex, below right=6pt]  {$8$}  (a3);
  \path[thin] (a4) edge node [draw=black, midway, yshift=2ex, below=2pt, right]  {$0$}  (a3);
  \path[thin] (a1) edge node [draw=black, midway, yshift=2ex, below=2pt, left]  {$9$}  (a4);
  \path[thin] (a1) edge node [draw=black, midway, yshift=2ex, below=9.5pt]  {$7$}  (a3);
  
  \node[qgre, label=10] (b1) at (0,1.5) {\textit{1}};
  \node[qgre, label=9] (b2) at (1,2) {\textit{2}};
  \node[qgre, label=14] (b3) at (2,1.5) {\textit{3}};
  
  \path[thin] (b1) edge node [draw=black, midway, yshift=2ex, above=-5pt]  {$1$}  (b2);
  \path[thin] (b2) edge node [draw=black, midway, yshift=2ex, above=-5pt]  {$2$}  (b3);
  
  \path[rpath, line width=4] (b1) edge node [midway, yshift=2ex, above=-5pt]  {}  (a3);
\end{tikzpicture}

%% file: l2_5.tex
\begin{tikzpicture}[auto, thick]
  \node[cblue, label=above left:12] (a1) at (0,0) {\textit{1}};
  \node[cblue, label=below:8] (a2) at (1,-1) {\textit{2}};
  \node[cblue, label=above right:15] (a3) at (2,0) {\textit{3}};
  \node[cblue, label=below:10] (a4) at (1,1) {\textit{4}};
  

  \path[thin] (a1) edge node [draw=black, midway, yshift=2ex, below left=6pt]  {$9$}  (a2);
  \path[thin] (a2) edge node [draw=black, midway, yshift=2ex, below right=6pt]  {$8$}  (a3);
  \path[thin] (a4) edge node [draw=black, midway, yshift=2ex, below=2pt, right]  {$0$}  (a3);
  \path[thin] (a1) edge node [draw=black, midway, yshift=2ex, below=2pt, left]  {$9$}  (a4);
  \path[thin] (a1) edge node [draw=black, midway, yshift=2ex, below=9.5pt]  {$7$}  (a3);
  
  \node[qgre, label=10] (b1) at (0,1.5) {\textit{1}};
  \node[qgre, label=9] (b2) at (1,2) {\textit{2}};
  \node[qgre, label=14] (b3) at (2,1.5) {\textit{3}};
  
  \path[thin] (b1) edge node [draw=black, midway, yshift=2ex, above=-5pt]  {$1$}  (b2);
  \path[thin] (b2) edge node [draw=black, midway, yshift=2ex, above=-5pt]  {$2$}  (b3);
  
  \path[rpath] (b1) edge node [midway, yshift=2ex, above=-5pt]  {}  (a3);
  \path[rpath, line width=4] (b2) edge node [midway, yshift=2ex, above=-5pt]  {}  (a1);
\end{tikzpicture}

%% file: l2_6.tex
\begin{tikzpicture}[auto, thick]
  \node[cblue, label=above left:12] (a1) at (0,0) {\textit{1}};
  \node[cblue, label=below:8] (a2) at (1,-1) {\textit{2}};
  \node[cblue, label=above right:15] (a3) at (2,0) {\textit{3}};
  \node[cblue, label=below:10] (a4) at (1,1) {\textit{4}};
  

  \path[thin] (a1) edge node [draw=black, midway, yshift=2ex, below left=6pt]  {$9$}  (a2);
  \path[thin] (a2) edge node [draw=black, midway, yshift=2ex, below right=6pt]  {$8$}  (a3);
  \path[thin] (a4) edge node [draw=black, midway, yshift=2ex, below=2pt, right]  {$0$}  (a3);
  \path[thin] (a1) edge node [draw=black, midway, yshift=2ex, below=2pt, left]  {$9$}  (a4);
  \path[thin] (a1) edge node [draw=black, midway, yshift=2ex, below=9.5pt]  {$7$}  (a3);
  
  \node[qgre, label=10] (b1) at (0,1.5) {\textit{1}};
  \node[qgre, label=9] (b2) at (1,2) {\textit{2}};
  \node[qgre, label=14] (b3) at (2,1.5) {\textit{3}};
  
  \path[thin] (b1) edge node [draw=black, midway, yshift=2ex, above=-5pt]  {$1$}  (b2);
  \path[thin] (b2) edge node [draw=black, midway, yshift=2ex, above=-5pt]  {$2$}  (b3);
  
  \path[rpath] (b1) edge node [midway, yshift=2ex, above=-5pt]  {}  (a3);
  \path[rpath, line width=4] (b2) edge node [midway, yshift=2ex, above=-5pt]  {}  (a4);
\end{tikzpicture}

%% file: fig_nrpa.tex
\begin{tikzpicture}[auto, thick]
\node (l2) at (-2.5,-2) {\LARGE \textbf{Level l=2}};
\node (l1) at (-2.5,-4) {\LARGE \textbf{Level l=1}};
\node (l0) at (-2.5,-6) {\LARGE \textbf{Level l=0}};
\node[nblue] (l3) at (0.5,-0.5) {\textbf{initial policy}};
\node[bluebox] (l0_1) at (1,-6) {\textbf{simulation}};
\node[bluebox] (l0_2) at (3.5,-6) {\textbf{simulation}};
\node[] (l0_1s) at (1,-6.5) {\textbf{$R^{seq}=0.4$}};
\node[] (l0_2s) at (3.5,-6.5) {$R^{seq} = 0$};
\node (next) at (5.7,-6) {\LARGE \textbf{...}};
\node[bluebox] (l0_3) at (7.7,-6) {\textbf{simulation}};
\node[] (l0_3s) at (7.7,-6.5) {\textbf{$R^{seq} = 0.53$}};

\node[greenbox] (l1_0) at (2.2,-4) {\textbf{adapt}};
\node[greenbox] (l1_1) at (4.4,-4) {\textbf{adapt}};
\node[redbox] (l1_2) at (4.4,-4.8) {\textbf{~max~~}};

\node[greenbox] (l1_3) at (6.25,-4) {\textbf{adapt}};
\node[redbox] (l1_4) at (6.25,-4.8) {\textbf{~max~~}};

\node[redbox] (l1_5) at (8.2,-4.8) {\textbf{max}};

\node[greenbox] (l2_0) at (8.2,-1.75) {\textbf{adapt}};

\draw[<-, dashed] (l1_0.west) -- ++(-1,0);
\draw[->, dashed] (l3.south) -- (l3.south|-l0_1.north);
\draw[<-] (l1_0.south) -- ++(0,-0.5);
\draw[->, dashed] (l1_0) -| (l0_2.north);
\draw[->, dashed] (l1_0) -- (l1_1);
\draw[<-] (l1_2) -- ++(0, -1.);
\draw[->] (l1_2) -- (l1_1);
\draw[->, dashed] (l1_1) -- (l1_3);
\draw[->] (l1_4) -- (l1_3);

\draw[->] (l1_5) -- (l2_0);

\draw[->, dashed] (l1_1.east) -| (5.2,-5.7);
\draw[<-] (l1_4.south) -- ++(0,-0.7);

\draw[->] (l0_1.east) |- ++ (0.15,0) |- (l1_2);

\draw[->, dashed] (l1_3.east) -| (7.45,-5.7);

\draw[->] (l1_4.east) -| ++(0.1,0) |- (l1_5);

\draw[->, dashed] (l2_0.east) -| (10,-6+0.25);

\draw[<-] (l1_5) -- ++(0, -0.95);


\draw (-0.2, -5.3) -- (-0.2,-6.8);
\draw (-0.2+2.5, -5.3) -- (-0.2+2.5,-6.8);
\draw (-0.2,-6.8) -- (-0.2+5,-6.8);
\draw (-0.2+5, -5.3) -- (-0.2+5,-6.8);

\draw (6.55, -5.3) -- (6.55,-6.8);
\draw (6.35+2.5, -5.3) -- (6.35+2.5,-6.8);

\draw (-0.2, -5.3) -- (6.35+2.5,-5.3);

\draw (-0.2, -5.3) -- (-0.2,-3.5);
\draw (6.35+2.5, -3.2) -- (6.35+2.5, -5.3);

\draw (6.35+2.5, -6.8) -- (6.55, -6.8);


\node[bluebox] (l0_1b) at (1+9,-6) {\textbf{simulation}};
\node[bluebox] (l0_2b) at (3.5+9,-6) {\textbf{simulation}};
\node[] (l0_1sb) at (1+9,-6.5) {\textbf{$R^{seq}=0.8$}};
\node[] (l0_2sb) at (3.5+9,-6.5) {$R^{seq} = 0$};
\node (nextb) at (5.7+9,-6) {\LARGE \textbf{...}};
\node[bluebox] (l0_3b) at (7.7+9,-6) {\textbf{simulation}};
\node[] (l0_3sb) at (7.7+9,-6.5) {\textbf{$R^{seq} = 0.55$}};

\node[greenbox] (l1_0b) at (2.2+9,-4) {\textbf{adapt}};
\node[greenbox] (l1_1b) at (4.4+9,-4) {\textbf{adapt}};
\node[redbox] (l1_2b) at (4.4+9,-4.8) {\textbf{~max~~}};

\node[greenbox] (l1_3b) at (6.25+9,-4) {\textbf{adapt}};
\node[redbox] (l1_4b) at (6.25+9,-4.8) {\textbf{~max~~}};

\node[redbox] (l1_5b) at (8.2+9,-4.8) {\textbf{max}};

\node[greenbox] (l2_0b) at (8.2+9,-1.75) {\textbf{adapt}};

\node at (19.5, -4.3) {\Huge \textbf{...}};
\node[redbox] (l2_3) at (20.5, -2.5) {max};

\node[redbox] (l2_0b_max) at (8.2+9, -2.5) {max};

\draw[<-, dashed] (l1_0b.west) -- ++(-0.5,0);
\draw[<-] (l1_0b.south) -- ++(0,-0.5);
\draw[->, dashed] (l1_0b) -| (l0_2b.north);
\draw[->, dashed] (l1_0b) -- (l1_1b);
\draw[<-] (l1_2b) -- ++(0, -1.);
\draw[->] (l1_2b) -- (l1_1b);
\draw[->, dashed] (l1_1b) -- (l1_3b);
\draw[->] (l1_4b) -- (l1_3b);

\draw[->] (l1_5b) -- (l2_0b_max);
\draw[->] (l2_0b_max) -- (l2_0b);
\draw[<-] (l2_3) -- ++(0,-1.5);

\draw[->, dashed] (l1_1b.east) -| (5.2+9,-5.7);
\draw[<-] (l1_4b.south) -- ++(0,-0.7);

\draw[->] (l0_1b.east) |- ++ (0.15,0) |- (l1_2b);

\draw[<-] (l1_5b) -- ++(0, -0.95);

\draw[->, dashed] (l1_3b.east) -| (7.45+9,-5.7);

\draw[->] (l1_4b.east) -| ++(0.1,0) |- (l1_5b);

\draw[->, dashed] (l2_0b.east) -| (18.5,-4);

\node[greenbox] (l2_0d) at (20.5,-1.75) {\textbf{adapt}};

\draw[<-] (l2_0b_max) -- ++(-9,0);


\draw (-0.2+2.5+9, -5.3) -- (-0.2+2.5+9,-6.8);
\draw (-0.2+9,-6.8) -- (-0.2+5+9,-6.8);
\draw (-0.2+5+9, -5.3) -- (-0.2+5+9,-6.8);

\draw (6.55+9, -5.3) -- (6.55+9,-6.8);
\draw (6.35+2.5+9, -5.3) -- (6.35+2.5+9,-6.8);

\draw (-0.2+9, -5.3) -- (6.35+2.5+9,-5.3);

\draw (6.35+2.5+9, -3.2) -- (6.35+2.5+9, -5.3);

\draw[->] (l2_3) -- (l2_0d);

\draw (6.35+2.5+9, -6.8) -- (6.55+9, -6.8);


\node[bluebox] (l0_1c) at (1+21.5,-6) {\textbf{simulation}};
\node[bluebox] (l0_2c) at (3.5+21.5,-6) {\textbf{simulation}};
\node[] (l0_1sc) at (1+21.5,-6.5) {\textbf{$R^{seq}=0.78$}};
\node[] (l0_2sc) at (3.5+21.5,-6.5) {$R^{seq} = 0.89$};
\node (nextc) at (5.7+21.5,-6) {\LARGE \textbf{...}};
\node[bluebox] (l0_3c) at (7.7+21.5,-6) {\textbf{simulation}};
\node[] (l0_3sc) at (7.7+21.5,-6.5) {\textbf{$R^{seq} = 0.68$}};

\node[greenbox] (l1_0c) at (2.2+21.5,-4) {\textbf{adapt}};
\node[greenbox] (l1_1c) at (4.4+21.5,-4) {\textbf{adapt}};
\node[redbox] (l1_2c) at (4.4 +21.5,-4.8) {\textbf{~max~~}};

\node[greenbox] (l1_3c) at (6.25+21.5,-4) {\textbf{adapt}};
\node[redbox] (l1_4c) at (6.25+21.5,-4.8) {\textbf{~max~~}};

\node[redbox] (l1_5c) at (8.2+21.5,-4.8) {\textbf{max}};

\node[redbox] (l2_0c) at (8.2+21.5,-2.5) {\textbf{max}};

\draw[<-, dashed] (l1_0c.west) -- ++(-0.5,0);
\draw[<-] (l1_0c.south) -- ++(0,-0.5);
\draw[->, dashed] (l1_0c) -| (l0_2c.north);
\draw[->, dashed] (l1_0c) -- (l1_1c);
\draw[<-] (l1_2c) -- ++(0, -1.);
\draw[->] (l1_2c) -- (l1_1c);
\draw[->, dashed] (l1_1c) -- (l1_3c);
\draw[->] (l1_4c) -- (l1_3c);

\draw[->] (l1_5c) -- (l2_0c);

\draw[->, dashed] (l1_1c.east) -| (5.2+21.5,-5.7);
\draw[<-] (l1_4c.south) -- ++(0,-0.7);

\draw[->] (l0_1c.east) |- ++ (0.15,0) |- (l1_2c);

\draw[->, dashed] (l1_3c.east) -| (7.45+21.5,-5.7);

\draw[->] (l1_4c.east) -| ++(0.1,0) |- (l1_5c);


\draw[->, dashed] (l2_0d) -| (l0_1c);
\draw[<-] (l1_5c) -- ++(0, -0.95);

\draw[->] (l2_3) -- (l2_0c);

\draw (-0.2+21.5, -5.3) -- (-0.2+21.5,-6.8);
\draw (-0.2+2.5+21.5, -5.3) -- (-0.2+2.5+21.5,-6.8);
\draw (-0.2+21.5,-6.8) -- (-0.2+5+21.5,-6.8);
\draw (-0.2+5+21.5, -5.3) -- (-0.2+5+21.5,-6.8);

\draw (6.55+21.5, -5.3) -- (6.55+21.5,-6.8);
\draw (6.35+2.5+21.5, -5.3) -- (6.35+2.5+21.5,-6.8);

\draw (-0.2+21.5, -5.3) -- (6.35+2.5+21.5,-5.3);

\draw (-0.2+21.5, -5.3) -- (-0.2+21.5,-3.2);
\draw (6.35+2.5+21.5, -3.5) -- (6.35+2.5+21.5, -5.3);

\draw (6.35+2.5+21.5, -6.8) -- (6.55+21.5, -6.8);

\node[nred] (lfinal) at (8.2+21.5,-0.5) {\textbf{final sequence}};

\draw[->] (l2_0c) -- (lfinal);

\draw (-0.2, -1.4) -- (-0.2,-3.5);

\draw (21.5+6.35+2.5, -1.4) -- (21.5+6.35+2.5+0,-3.5);

\draw (21.5+6.35+2.5+0,-3.2) -- (-0.2,-3.2);

\draw (21.5+6.35+2.5, -1.4) -- (-0.2, -1.4);

\draw[->] (l1_2) -- (l1_4);
\draw[->] (l1_2b) -- (l1_4b);
\draw[->] (l1_2c) -- (l1_4c);

\end{tikzpicture}

%% file: fig_nepa.tex
\begin{tikzpicture}[auto, thick]
\node (l2) at (-2.5,-2) {\LARGE \textbf{Level l=2}};
\node (l1) at (-2.5,-4) {\LARGE \textbf{Level l=1}};
\node (l0) at (-2.5,-6) {\LARGE \textbf{Level l=0}};
\node[nblue] (l3) at (0.5,-0.5) {\textbf{initial policy}};
\node[bluebox] (l0_1) at (1,-6) {\textbf{simulation}};
\node[bluebox] (l0_2) at (3.5,-6) {\textbf{simulation}};
\node[] (l0_1s) at (1,-6.5) {\textbf{$R^{seq}=0.4$}};
\node[] (l0_2s) at (3.5,-6.5) {$R^{seq} = 0$};
\node (next) at (5.7,-6) {\LARGE \textbf{...}};
\node[bluebox] (l0_3) at (7.7,-6) {\textbf{simulation}};
\node[] (l0_3s) at (7.7,-6.5) {\textbf{$R^{seq} = 0.53$}};

\node[greenbox] (l1_0) at (2.2,-4) {\textbf{adapt}};
\node[greenbox] (l1_1) at (4.4,-4) {\textbf{adapt}};
\node[redbox] (l1_2) at (4.4,-4.8) {\textbf{~max~~}};

\node[greenbox] (l1_3) at (6.25,-4) {\textbf{adapt}};
\node[redbox] (l1_4) at (6.25,-4.8) {\textbf{~max~~}};

\node[redbox] (l1_5) at (8.2,-4.8) {\textbf{max}};

\node[greenbox] (l2_0) at (8.2,-1.5) {\textbf{adapt}};

\node[cyanbox] (l2_0r) at (8.2,-2.8) {\textbf{refine}};

\draw[->] (l1_2) -- (l1_4);
\draw[<-, dashed] (l1_0.west) -- ++(-1,0);
\draw[->, dashed] (l3.south) -- (l3.south|-l0_1.north);
\draw[<-] (l1_0.south) -- ++(0,-0.5);
\draw[->, dashed] (l1_0) -| (l0_2.north);
\draw[->, dashed] (l1_0) -- (l1_1);
\draw[<-] (l1_2) -- ++(0, -1.);
\draw[->] (l1_2) -- (l1_1);
\draw[->, dashed] (l1_1) -- (l1_3);
\draw[->] (l1_4) -- (l1_3);

\draw[->] (l1_5) -- (l2_0r);

\draw[->] (l2_0r) -- (l2_0);

\draw[->, dashed] (l1_1.east) -| (5.2,-5.7);
\draw[<-] (l1_4.south) -- ++(0,-0.7);

\draw[->] (l0_1.east) |- ++ (0.15,0) |- (l1_2);

\draw[->, dashed] (l1_3.east) -| (7.45,-5.7);

\draw[->] (l1_4.east) -| ++(0.1,0) |- (l1_5);

\draw[->, dashed] (l2_0.east) -| (10,-6+0.25);

\draw[<-] (l1_5) -- ++(0, -0.95);


\draw (-0.2, -5.3) -- (-0.2,-6.8);
\draw (-0.2+2.5, -5.3) -- (-0.2+2.5,-6.8);
\draw (-0.2,-6.8) -- (-0.2+5,-6.8);
\draw (-0.2+5, -5.3) -- (-0.2+5,-6.8);

\draw (6.55, -5.3) -- (6.55,-6.8);
\draw (6.35+2.5, -5.3) -- (6.35+2.5,-6.8);

\draw (-0.2, -5.3) -- (6.35+2.5,-5.3);

\draw (-0.2, -5.3) -- (-0.2,-3.5);
\draw (6.35+2.5, -3.2) -- (6.35+2.5, -5.3);

\draw (6.35+2.5, -6.8) -- (6.55, -6.8);


\node[bluebox] (l0_1b) at (1+9,-6) {\textbf{simulation}};
\node[bluebox] (l0_2b) at (3.5+9,-6) {\textbf{simulation}};
\node[] (l0_1sb) at (1+9,-6.5) {\textbf{$R^{seq}=0.8$}};
\node[] (l0_2sb) at (3.5+9,-6.5) {$R^{seq} = 0$};
\node (nextb) at (5.7+9,-6) {\LARGE \textbf{...}};
\node[bluebox] (l0_3b) at (7.7+9,-6) {\textbf{simulation}};
\node[] (l0_3sb) at (7.7+9,-6.5) {\textbf{$R^{seq} = 0.55$}};

\node[greenbox] (l1_0b) at (2.2+9,-4) {\textbf{adapt}};
\node[greenbox] (l1_1b) at (4.4+9,-4) {\textbf{adapt}};
\node[redbox] (l1_2b) at (4.4+9,-4.8) {\textbf{~max~~}};

\node[greenbox] (l1_3b) at (6.25+9,-4) {\textbf{adapt}};
\node[redbox] (l1_4b) at (6.25+9,-4.8) {\textbf{~max~~}};

\node[redbox] (l1_5b) at (8.2+9,-4.8) {\textbf{max}};

\node[greenbox] (l2_0b) at (8.2+9,-1.5) {\textbf{adapt}};

\node at (19.5, -4.3) {\Huge \textbf{...}};
\node[redbox] (l2_3) at (20.5, -2.8) {max};

\node[cyanbox] (l2_3br) at (20.5, -2.2) {refine};

\node[redbox] (l2_0b_max) at (8.2+9, -2.8) {max};

\node[cyanbox] (l2_0br) at (8.2+9, -2.2) {refine};

\draw[->] (l1_2b) -- (l1_4b);
\draw[<-, dashed] (l1_0b.west) -- ++(-0.5,0);
\draw[<-] (l1_0b.south) -- ++(0,-0.5);
\draw[->, dashed] (l1_0b) -| (l0_2b.north);
\draw[->, dashed] (l1_0b) -- (l1_1b);
\draw[<-] (l1_2b) -- ++(0, -1.);
\draw[->] (l1_2b) -- (l1_1b);
\draw[->, dashed] (l1_1b) -- (l1_3b);
\draw[->] (l1_4b) -- (l1_3b);

\draw[->] (l1_5b) -- (l2_0b_max);
\draw[->] (l2_0b_max) -- (l2_0br);
\draw[->] (l2_0br) -- (l2_0b);
\draw[<-] (l2_3) -- ++(0,-1.2);

\draw[->, dashed] (l1_1b.east) -| (5.2+9,-5.7);
\draw[<-] (l1_4b.south) -- ++(0,-0.7);

\draw[->] (l0_1b.east) |- ++ (0.15,0) |- (l1_2b);

\draw[<-] (l1_5b) -- ++(0, -0.95);

\draw[->, dashed] (l1_3b.east) -| (7.45+9,-5.7);

\draw[->] (l1_4b.east) -| ++(0.1,0) |- (l1_5b);

\draw[->, dashed] (l2_0b.east) -| (18.5,-4);

\draw[->] (l2_0br) -| ++(1,0) |- (l2_3.west);

\node[greenbox] (l2_0d) at (20.5,-1.5) {\textbf{adapt}};



\draw (-0.2+2.5+9, -5.3) -- (-0.2+2.5+9,-6.8);
\draw (-0.2+9,-6.8) -- (-0.2+5+9,-6.8);
\draw (-0.2+5+9, -5.3) -- (-0.2+5+9,-6.8);

\draw (6.55+9, -5.3) -- (6.55+9,-6.8);
\draw (6.35+2.5+9, -5.3) -- (6.35+2.5+9,-6.8);

\draw (-0.2+9, -5.3) -- (6.35+2.5+9,-5.3);

\draw (6.35+2.5+9, -3.2) -- (6.35+2.5+9, -5.3);

\draw[->] (l2_3) -- (l2_3br);
\draw[->] (l2_3br) -- (l2_0d);

\draw (6.35+2.5+9, -6.8) -- (6.55+9, -6.8);


\node[bluebox] (l0_1c) at (1+21.5,-6) {\textbf{simulation}};
\node[bluebox] (l0_2c) at (3.5+21.5,-6) {\textbf{simulation}};
\node[] (l0_1sc) at (1+21.5,-6.5) {\textbf{$R^{seq}=0.78$}};
\node[] (l0_2sc) at (3.5+21.5,-6.5) {$R^{seq} = 0.89$};
\node (nextc) at (5.7+21.5,-6) {\LARGE \textbf{...}};
\node[bluebox] (l0_3c) at (7.7+21.5,-6) {\textbf{simulation}};
\node[] (l0_3sc) at (7.7+21.5,-6.5) {\textbf{$R^{seq} = 0.68$}};

\node[greenbox] (l1_0c) at (2.2+21.5,-4) {\textbf{adapt}};
\node[greenbox] (l1_1c) at (4.4+21.5,-4) {\textbf{adapt}};
\node[redbox] (l1_2c) at (4.4 +21.5,-4.8) {\textbf{~max~~}};

\node[greenbox] (l1_3c) at (6.25+21.5,-4) {\textbf{adapt}};
\node[redbox] (l1_4c) at (6.25+21.5,-4.8) {\textbf{~max~~}};

\node[redbox] (l1_5c) at (8.2+21.5,-4.8) {\textbf{max}};

\node[redbox] (l2_0c) at (8.2+21.5,-2.8) {\textbf{max}};
\node[cyanbox] (l2_0cr) at (21.5+8.2, -2.2) {refine};

\draw[->] (l1_2c) -- (l1_4c);

\draw[<-, dashed] (l1_0c.west) -- ++(-0.5,0);
\draw[<-] (l1_0c.south) -- ++(0,-0.5);
\draw[->, dashed] (l1_0c) -| (l0_2c.north);
\draw[->, dashed] (l1_0c) -- (l1_1c);
\draw[<-] (l1_2c) -- ++(0, -1.);
\draw[->] (l1_2c) -- (l1_1c);
\draw[->, dashed] (l1_1c) -- (l1_3c);
\draw[->] (l1_4c) -- (l1_3c);

\draw[->] (l1_5c) -- (l2_0c);

\draw[->, dashed] (l1_1c.east) -| (5.2+21.5,-5.7);
\draw[<-] (l1_4c.south) -- ++(0,-0.7);

\draw[->] (l0_1c.east) |- ++ (0.15,0) |- (l1_2c);

\draw[->, dashed] (l1_3c.east) -| (7.45+21.5,-5.7);

\draw[->] (l1_4c.east) -| ++(0.1,0) |- (l1_5c);


\draw[->, dashed] (l2_0d) -| (l0_1c);
\draw[<-] (l1_5c) -- ++(0, -0.95);


\draw[->] (l2_3br) -| ++(1, 0) |- (l2_0c); 

\draw (-0.2+21.5, -5.3) -- (-0.2+21.5,-6.8);
\draw (-0.2+2.5+21.5, -5.3) -- (-0.2+2.5+21.5,-6.8);
\draw (-0.2+21.5,-6.8) -- (-0.2+5+21.5,-6.8);
\draw (-0.2+5+21.5, -5.3) -- (-0.2+5+21.5,-6.8);

\draw (6.55+21.5, -5.3) -- (6.55+21.5,-6.8);
\draw (6.35+2.5+21.5, -5.3) -- (6.35+2.5+21.5,-6.8);

\draw (-0.2+21.5, -5.3) -- (6.35+2.5+21.5,-5.3);

\draw (-0.2+21.5, -5.3) -- (-0.2+21.5,-3.2);
\draw (6.35+2.5+21.5, -3.5) -- (6.35+2.5+21.5, -5.3);

\draw (6.35+2.5+21.5, -6.8) -- (6.55+21.5, -6.8);

\node[nred] (lfinal) at (8.2+21.5,-0.5) {\textbf{final sequence}};

\draw[->] (l2_0c) -- (l2_0cr);
\draw[->] (l2_0cr) -- (lfinal);

\draw (-0.2, -1) -- (-0.2,-3.5);

\draw (21.5+6.35+2.5, -1) -- (21.5+6.35+2.5+0,-3.5);

\draw (21.5+6.35+2.5+0,-3.2) -- (-0.2,-3.2);

\draw (21.5+6.35+2.5, -1) -- (-0.2, -1);


\draw[->] (l2_0r) -- (l2_0b_max);
\end{tikzpicture}

%% file: fig5_1.tex
\begin{tikzpicture}[auto, thick, scale=0.7]
  \foreach \place/\x in {{(-0.25,0)/1}, 
    {(1.5,0)/2}}
  \node[cblue] (a\x) at \place {\small \x};
  
  \node[cblue] (a3) at (0.25,0.7) {\small 3};
  \node[cblue] (a4) at (0.75,-0.3) {\small 4};
  \node[cblue] (a5) at (2.5,0.4) {\small 5};
  
  \node[cblue] (a6) at (-1.5,0) {\small 6};
  
  \path[thin] (a1) edge (a4);
  \path[thick] (a3) edge[color=black] node [color=black, thin, near start, yshift=2ex, below=12pt]  {} (a4);
  \path[thick] (a3) edge[color=black] node [color=black, thin, midway, yshift=2ex, below=-4pt]  {}(a5);
  \path[thin] (a3) edge (a2) ;
  \path[thick] (a2) edge[color=black] node [color=black, thin, midway, yshift=2ex, below=12pt]  {} (a5);
  \path[thick] (a4) edge[color=black] node [thin, color=black, midway, yshift=2ex, right=34pt, below=10pt]  {} (a2);
  \path[thick] (a1) edge[color=black] node [thin, color=black, midway, yshift=2ex, right=34pt, below=10pt]  {} (a6);
 
    \node[qgre] (b3) at (0.25,2.7) {\small 1};
    \node[qgre] (b1) at (-0.25,1.7) {\small 3};
    \node[qgre] (b5) at (2.5,2.4) {\small 2};
 
  \path[thin] (b3) edge node [midway, yshift=2ex, below=10pt]  {}  (b1);
  \path[thin] (b5) edge node [midway, yshift=2ex, below=8pt]  {} (b1);
  \path[thin] (b5) edge node [midway, yshift=2ex, right]  {} (b3);
 
  \foreach \i in {3, 1, 5}
    \path[rpath] (a\i) edge (b\i);
\end{tikzpicture}

%% file: fig5_2.tex
\begin{tikzpicture}[auto, thick, scale=0.7]
  \foreach \place/\x in {{(-0.25,0)/1}, 
    {(1.5,0)/2}}
  \node[cblue] (a\x) at \place {\small \x};
  
  \node[cblue] (a3) at (0.25,0.7) {\small 3};
  \node[cblue] (a4) at (0.75,-0.3) {\small 4};
  \node[cblue] (a5) at (2.5,0.4) {\small 5};
  
  \node[cblue] (a6) at (-1.5,0) {\small 6};
  
  \path[thin] (a1) edge (a4);
  \path[thick] (a3) edge[color=black] node [color=black, thin, near start, yshift=2ex, below=12pt]  {} (a4);
  \path[thick] (a3) edge[color=black] node [color=black, thin, midway, yshift=2ex, below=-4pt]  {}(a5);
  \path[thin] (a3) edge (a2) ;
  \path[thick] (a2) edge[color=black] node [color=black, thin, midway, yshift=2ex, below=12pt]  {} (a5);
  \path[thick] (a4) edge[color=black] node [thin, color=black, midway, yshift=2ex, right=34pt, below=10pt]  {} (a2);
  \path[thick] (a6) edge[color=black] node [thin, color=black, midway, yshift=2ex, right=34pt, below=10pt]  {} (a1);
 
    \node[qgre] (b3) at (0.25,2.7) {\small 1};
    \node[qgre] (b1) at (-0.25,1.7) {\small 3};
    \node[qgre] (b5) at (2.5,2.4) {\small 2};
 
  \path[thin] (b3) edge node [midway, yshift=2ex, below=10pt]  {}  (b1);
  \path[thin] (b5) edge node [midway, yshift=2ex, below=8pt]  {} (b1);
  \path[thin] (b5) edge node [midway, yshift=2ex, right]  {} (b3);
 
  \foreach \i in {3, 5}
    \path[rpath] (a\i) edge (b\i);
\end{tikzpicture}

%% file: fig5_4.tex
\begin{tikzpicture}[auto, thick, scale=0.7]
          \foreach \place/\x in {{(-0.25,0)/1}, 
            {(1.5,0)/2}}
          \node[cblue] (a\x) at \place {\small \x};
          
          \node[cblue] (a3) at (0.25,0.7) {\small 3};
          \node[cblue] (a4) at (0.75,-0.3) {\small 4};
          \node[cblue] (a5) at (2.5,0.4) {\small 5};
          
          \node[cblue] (a6) at (-1.5,0) {\small 6};
          
          \path[thin] (a1) edge (a4);
          \path[thick] (a3) edge[color=black] node [color=black, thin, near start, yshift=2ex, below=12pt]  {} (a4);
          \path[thick] (a3) edge[color=black] node [color=black, thin, midway, yshift=2ex, below=-4pt]  {}(a5);
          \path[thin] (a3) edge (a2) ;
          \path[thick] (a2) edge[color=black] node [color=black, thin, midway, yshift=2ex, below=12pt]  {} (a5);
          \path[thick] (a4) edge[color=black] node [thin, color=black, midway, yshift=2ex, right=34pt, below=10pt]  {} (a2);
          \path[thick] (a1) edge[color=black] node [thin, color=black, midway, yshift=2ex, right=34pt, below=10pt]  {} (a6);
         
            \node[qgre] (b3) at (0.25,2.7) {\small 1};
            \node[qgre] (b2) at (1.5,1.7) {\small 3};
            \node[qgre] (b5) at (2.5,2.4) {\small 2};
         
          \path[thin] (b3) edge node [midway, yshift=2ex, below=10pt]  {}  (b2);
          \path[thin] (b5) edge node [midway, yshift=2ex, below=8pt]  {} (b2);
          \path[thin] (b5) edge node [midway, yshift=2ex, right]  {} (b3);
         
          \foreach \i in {3, 2, 5}
            \path[rpath] (a\i) edge (b\i);
    \end{tikzpicture}

%% file: fig5_3.tex
\begin{tikzpicture}[auto, thick, scale=0.7]
              \foreach \place/\x in {{(-0.25,0)/1}, 
                {(1.5,0)/2}}
              \node[cblue] (a\x) at \place {\small \x};
              
              \node[cblue] (a3) at (0.25,0.7) {\small 3};
              \node[cblue] (a4) at (0.75,-0.3) {\small 4};
              \node[cblue] (a5) at (2.5,0.4) {\small 5};
              
              \node[cblue] (a6) at (-1.5,0) {\small 6};
              
              \path[thin] (a1) edge (a4);
              \path[thick] (a3) edge[color=black] node [color=black, thin, near start, yshift=2ex, below=12pt]  {} (a4);
              \path[thick] (a3) edge[color=black] node [color=black, thin, midway, yshift=2ex, below=-4pt]  {}(a5);
              \path[thin] (a3) edge (a2) ;
              \path[thick] (a2) edge[color=black] node [color=black, thin, midway, yshift=2ex, below=12pt]  {} (a5);
              \path[thick] (a4) edge[color=black] node [thin, color=black, midway, yshift=2ex, right=34pt, below=10pt]  {} (a2);
              \path[thick] (a6) edge[color=black] node [thin, color=black, midway, yshift=2ex, right=34pt, below=10pt]  {} (a1);
             
                \node[qgre] (b3) at (0.25,2.7) {\small 1};
                \node[qgre] (b4) at (0.75,1.7) {\small 3};
                \node[qgre] (b5) at (2.5,2.4) {\small 2};
             
              \path[thin] (b3) edge node [midway, yshift=2ex, below=10pt]  {}  (b4);
              \path[thin] (b5) edge node [midway, yshift=2ex, below=8pt]  {} (b4);
              \path[thin] (b5) edge node [midway, yshift=2ex, right]  {} (b3);
             
              \foreach \i in {3, 4, 5}
                \path[rpath] (a\i) edge (b\i);
        \end{tikzpicture}